%% file: tether_survey.tex
\documentclass[manuscript]{acmart}

\usepackage{algorithmic}
\usepackage{array}
\usepackage{graphicx}
\usepackage{textcomp}
\usepackage{enumitem}

\usepackage{etoolbox} 

\usepackage{xcolor}
\usepackage{csvsimple}

\usepackage{subcaption} 
\usepackage{graphicx}

\usepackage{makecell}
\usepackage{multirow}

\usepackage{mdframed}
\usepackage{boxedminipage}
\usepackage{tikz}
\usepackage{balance}

\usepackage{enumitem}

\usepackage{tcolorbox}
\newenvironment{newbox1}{}{}

\input{custom.tex}

\AtBeginDocument{%
  }

\setcopyright{acmlicensed}
\copyrightyear{2024}
\acmYear{2024}
\acmDOI{XXXXXXX.XXXXXXX}

\acmConference[CHI '25]{Make sure to enter the correct
  conference title from your rights confirmation emai}{April 26--May 01,
  2025}{Yokohoma, Japan}
\acmISBN{978-1-4503-XXXX-X/18/06}

\begin{document}

\title{Understanding End-User Perception of Transfer Risks in Smart Contracts}

\author{Yustynn Panicker}
\affiliation{%
  \institution{Singapore University of Technology and Design}
  \country{Singapore}
}

\author{Ezekiel Soremekun}
\affiliation{%
  \institution{Singapore University of Technology and Design}
  \country{Singapore}
}
\author{Sudipta Chattopadhyay}
\affiliation{%
  \institution{Singapore University of Technology and Design}
  \country{Singapore}
}

\author{Sumei Sun}
\affiliation{%
  \institution{Institute for Infocomm Research A*STAR}
  \country{Singapore}
}

\renewcommand{\shortauthors}{Panicker et al.}

\input{abstract.tex}





\maketitle

\input{introduction}
\input{overview}
\input{related_work}

\input{evaluation_setup}

\input{evaluation}

\input{discussion}

\input{validity}
\input{conclusion_and_ft}

\bibliographystyle{ACM-Reference-Format}
\bibliography{tether_survey}

\input{appendix}
\end{document}

%% file: custom.tex
\newmdenv[innerlinewidth=0.5pt, roundcorner=4pt,linecolor=gray,innerleftmargin=4pt,innerrightmargin=4pt,innertopmargin=4pt,innerbottommargin=4pt]{note1}

\newenvironment{result}%
{\medskip\begin{note1}\centering\em}%
{\end{note1}\medskip}

\newcommand{\rev}[1]{\textcolor{black}{#1}}
\newcommand{\reva}[1]{\textcolor{black}{#1}}
\newcommand{\revb}[1]{\textcolor{black}{#1}}
\newcommand{\revc}[1]{\textcolor{black}{#1}}
\newcommand{\revise}[1]{\textcolor{black}{#1}}

\newcommand{\newrevise}[1]{\textcolor{black}{#1}}

\newcommand{\recheck}[1]{\textcolor{red}{#1}}

\newcommand{\newtodo}[1]{}%
\renewcommand{\newtodo}[1]{{\color{red} TODO: {#1}}}%

\newcommand*\circled[1]{\tikz[baseline=(char.base)]{
            \node[shape=circle,draw,inner sep=0.02cm] (char) {#1};}}

\newcommand*\emptycircle[1][0.75ex]{\tikz\draw[thick] (0,0) circle (#1);} 
\newcommand*\halfcircle[1][0.75ex]{%
  \begin{tikzpicture}
  \draw[fill] (0,0)-- (90:#1) arc (90:270:#1) -- cycle ;
  \draw[thick] (0,0) circle (#1);
  \end{tikzpicture}
}
\newcommand*\fullcircle[1][0.75ex]{
  \begin{tikzpicture}
  \draw[thick] (0,0) circle (#1);
  \draw[fill] (0,0) circle (#1);
  \end{tikzpicture}
}

\newcolumntype{C}[1]{>{\centering\arraybackslash}m{#1}}

%% file: abstract.tex
\begin{abstract}
    Blockchain smart contracts are increasingly used in critical use cases (e.g., financial transactions). Thus, it is pertinent to ensure that their end-users understand risks in attempting token transfers. Addressing this, we investigate end-user comprehension of five transfer risks (e.g. the end-user being blacklisted) in the most popular Ethereum contract, USD Tether (USDT), and their prevalence in other top ERC-20 contracts. First, we conducted a user study investigating end-user comprehension of transfer risks in USDT with 110 participants. Second, we performed source code analysis of the next top (78) ERC-20 smart contracts to identify the prevalence of these risks. Study results show that the majority of end-users do not comprehend some real risks, and confuse real and fictitious risks. This holds regardless of participants’ self-rated programming and Web3 proficiency. Source code analysis demonstrates that examined risks are prevalent in up to 19.2\% of the top ERC-20 contracts.
\end{abstract}

%% file: introduction.tex
\section{Introduction}
\label{sec:introduction}


Blockchain smart contracts have become 
increasingly important in our society.  Common 
applications of smart contracts include financial contracts,  and  gaming~\cite{contract_language_survey}.  
Notably,  
the top ten public blockchains
implement smart contracts~\cite{TopChains}, including Ethereum~\cite{EthereumWhitepaper}, 
Polygon~\cite{PolygonWhitepaper} and Binance Smart Chain~\cite{BSCWhitepaper}. Considering the critical use cases of  
smart contracts (e.g., in financial transactions), \textit{it is important to  understand 
the end-users' comprehension of the \rev{transfer} risks in smart contracts}.
Such risks 
(e.g.,  transaction risks) 
may have severe consequences,  e.g., 
financial loss and unexpected outcomes. 

%
%

\begin{figure}[h]
\centering
\includegraphics[width=0.75\linewidth]{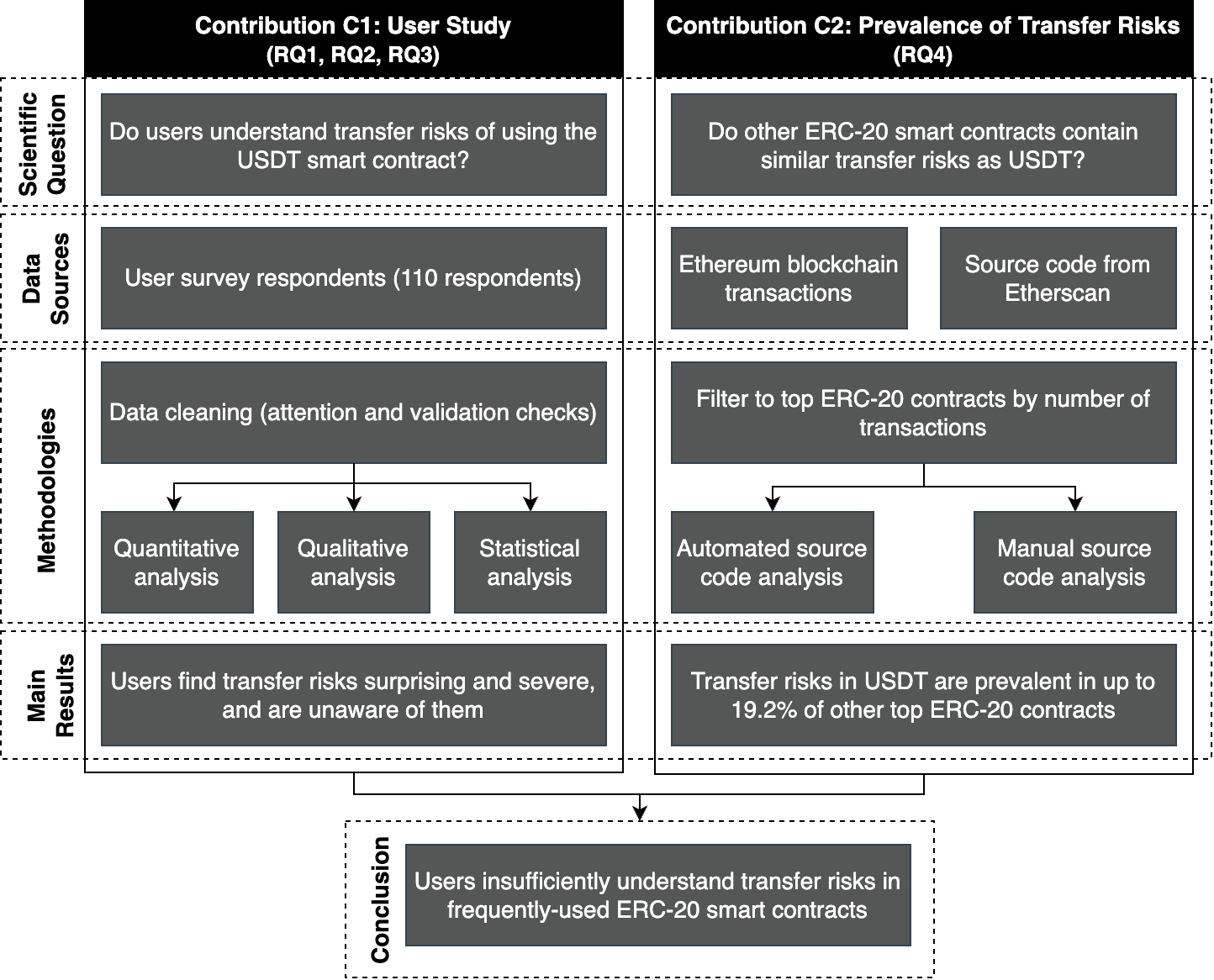}
\caption{\revc{
Overview of Research Methodology}}
\label{fig:methodology-overview}
\Description{A diagram with two parallel columns of nodes, with the columns labelled “User Study” and “Generalizability Analysis”. The elements are broken down into node groups, where each column has node(s) in the node group. This is the group breakdown, from node groups ordered top to bottom:

Research Question (User Study): A node with the text “Do users understand transfer risks of using the USDT smart contract?”
Research Question (Generalizability Analysis): A node with the text “Do other ERC-20 smart contracts contain similar transfer risks as USDT?”
Data Sources (User Study): A node with the text “User survey respondents (110 respondents)”
Data Sources (Generalizability Analysis): Two parallel nodes. The first contains the text “Ethereum blockchain transactions”. The second contains the text ”Source code from Etherscan”.
Methodologies (User Study): A parent node, with 3 children. The parent node contains the text “Data cleaning (attention and validation checks). The first child’s text is “Quantitative analysis”. The second child’s text is “Qualitative analysis”. The third child’s text is “Statistical analysis”.
Methodologies (User Study): A parent node, with 2 children. The parent node contains the text “Filter to top ERC-20 contracts by number of transactions”. The first child contains the text “Automated source code analysis”. The second child contains the text “Manual source code analysis”.
Main Results (User Study): A node with the text “Users find transfer risks surprising and severe, and are unaware of them”
Main Results (Generalizability Analysis): A node with the text “Transfer risks in USDT are prevalent in up to 19.2

The two columns then combine to a final node group labelled “Conclusion”. This node contains a node with the text “Users insufficiently understand transfer risks in frequently-used ERC-20 smart contracts”.
}
\end{figure}

\newrevise{
 To the best of our knowledge, this paper presents the first comprehensive study to investigate end-users' understanding of \rev{transfer} risks of smart contracts.}
Broadly, our research methodology comprises of two key components, namely {\em (i)} a \textit{user study }
examining smart contract \rev{transfer} risks with 110 end-users using the Etheruem-based USDT contract, 
and {\em (ii)} 
\textit{source code analysis} of (\rev{78}) Etheruem-based ERC-20 contracts 
investigating the generalization of
 the examined \rev{transfer} risks in \textit{(i)}.
\autoref{fig:methodology-overview} illustrates 
our research methodology. 
%

\newrevise{
In this work,  we investigate Ethereum-based contracts since 
Ethereum has the largest 
total value locked (TVL) \cite{TVLDefinition} by far (USD 25B~\cite{TopChains}) among the top ten public blockchains. 
Our source code analysis (in \autoref{fig:methodology-overview}) 
focuses on smart contracts that implement the ERC-20 standard \cite{ERC20TokenStandard} for fungible tokens. 
%
ERC-20 was chosen as 
contracts implementing the ERC-20 standard received the largest transaction volume 
during our analysis.  Concretely,  among the top 500 recipient addresses, \rev{USDT together with the other} 78 ERC-20 contract addresses  studied in this work,  accounted for 20.0\% of transaction volume.  
We design our \textit{user study} (in \autoref{fig:methodology-overview}) using the USDT contract since  
it is the most popular smart contract, 
accounting for 12.7\% of transaction volume to the top 500 recipient addresses in a three-month period.  
}

\begin{figure*}[t]
	\centering
	\includegraphics[width=\linewidth]{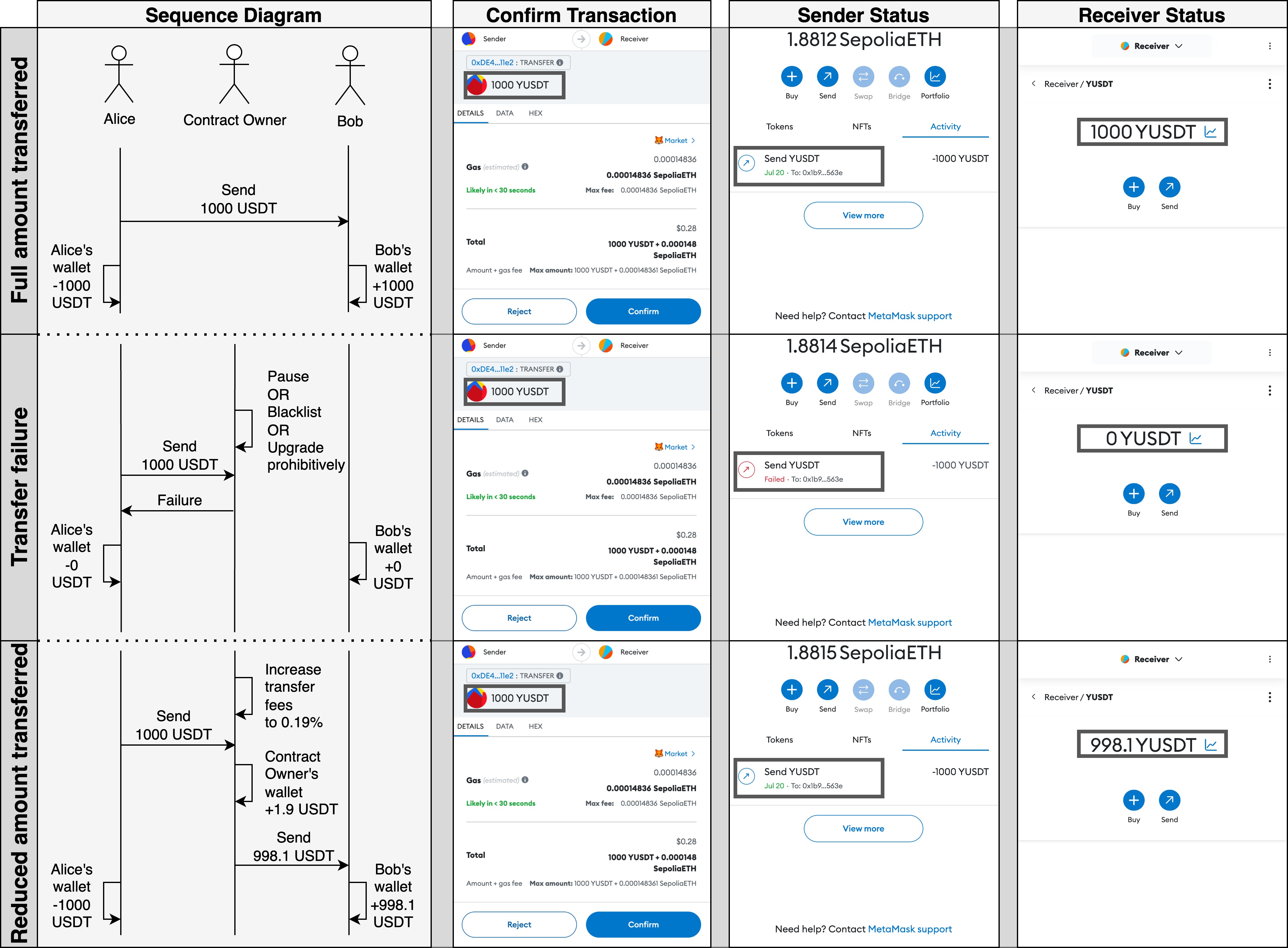}
	\caption{
		Sequence of events and corresponding MetaMask flow for \rev{a user's failure to make a USDT transfer}. 
		The MetaMask flow shows our functionally-equivalent clone of the USDT contract named YUSDT, with bounding boxes highlighting relevant aspects.
		\rev{All scenarios are based on the USDT source code, reflecting intentional features and not fraudulent behavior.}
	}
	\Description{The figure shows a sequence diagram and UI flow for three states. From top to bottom, these states are “Full amount transferred”, “Transfer failure” and “Reduced amount transferred”.

	The sequence diagram of each state contains the three actors (ordered left-to-right) “Alice”, “Contract Owner” and “Bob”. Each state also shows the MetaMask UI screenshots for the stages (ordered left-to-right) of “Confirm Transaction”, “Sender Status” and “Receiver Status” of the YUSDT, the functionally-equivalent clone of USDT.

	The “Full amount transferred” state diagram is as follows: Alice sends Bob 1000 USDT. Alice’s wallet decreases by 1000 USDT. Bob’s wallet increases by 1000 USDT.

	The “Full amount transferred” UI screenshots highlight the following:
	Confirm Transaction: Sender having 1000USDT in their account
	Sender Status: The date “Jul 20” in green, signalling success, along with the elided recipient address.
	Receiver Status: Receiver having 1000USDT in their account

	The “Transfer failure” state diagram is as follows: The contract owner pauses the contract, blacklists Alice, or upgrades the contract prohibitively. Alice attempts to send 1000USDT. Both Alice’s and Bob’s wallet balances remain unchanged.

	The “Transfer failure” UI screenshots highlight the following:
	Confirm Transaction: Sender having 1000USDT in their account
	Sender Status: The message “Failed” in red, signalling failure, along with the elided recipient address.
	Receiver Status: Receiver having 0USDT in their account

	The “Reduced amount transferred” state diagram is as follows: The Contract Owner increases the transfer fees to 0.19

	The “Reduced amount transferred” UI screenshots highlight the following:
	Confirm Transaction: Sender having 1000USDT in their account
	Sender Status: The date “Jul 20” in green, signalling success, along with the elided recipient address.
	Receiver Status: Receiver having 998.1USDT in their account}
	\label{fig:sequence-diagram}
\end{figure*}

Our inspection of the USDT contract revealed 
various potentially surprising 
features beyond the ERC-20 specification~\cite{ERC20TokenStandard}. These include the 
ability to {\em blacklist users}, {\em pause the contract}, {\em upgrade the contract arbitrarily} and 
{\em set fees on user transfers of USDT}. 
\autoref{fig:sequence-diagram} illustrates \reva{some of} these risks, capturing the sequence of 
events and the user interface (UI) flow (using MetaMask UI) for a
\reva{
\textit{failed transaction} due to the contract being 
{\em paused}, the user {\em blacklisted} or arbitrary contract {\em upgrade}. 
}
We focus on 
these risks 
because they affect the transfer outcomes, thus directly impacting
the financial objective of the smart contract end-user. To check if end-users are aware of these risks,
we conducted a user study with 110 smart contract users 
(``User Study" in  \autoref{fig:methodology-overview}). Through a series of quantitative and 
qualitative analysis, we discover that most users are {\em unaware} \reva{of} and {\em surprised} \reva{by} these risks. 
More importantly,  end-users perceived the \reva{studied risks (e.g.,  the three risks illustrated in \autoref{fig:sequence-diagram})} to be 
{\em severe}. 
Our statistical analysis \reva{further} shows 
that our findings hold for different user groups 
(e.g., \rev{self-rated} programmers vs. non-programmers.) 

\rev{To extend our analysis beyond the USDT contract,} (``Generalizability Analysis" in 
\autoref{fig:methodology-overview}), we show that 
\rev{the transfer risks examined in the USDT contract}
extend to  other popular ERC-20 smart contracts. 
To this end, we performed both 
manual and automated analysis of \rev{the next} 78 most frequently used ERC-20 contracts (excluding USDT, 
since we performed a manual analysis of USDT for the user study).  Moreover, 
manual 
analysis of these 78 ERC-20 contracts
revealed additional potential end-user risks beyond the 
%
%
ones considered in our user study. Overall, our findings concretely point to the insufficiency of 
user understanding in dealing with a variety of smart contract risks across the most popular contracts.

Our study is unique in that it deals with the end-users' perception directly. Several existing 
works have focused on programming languages~\cite{scilla,obsidian-toplas} and 
tools~\cite{fse20-fairness,osiris,david-tosem22,david-tse20} for smart contract {\em developers}, but 
these \reva{works} do not target {\em end-users}. Concurrently, existing user studies with smart contract aim to \reva{either} understand specific user preferences and misconceptions (e.g., cryptographic 
key)~\cite{key-icissp,mentalmodel-soups} or to validate certain technologies (e.g., user notice in the 
code~\cite{david-usernotice}). Our study is orthogonal to these approaches, as instead of focusing 
on a specific technology and user preference, we aim to broadly investigate the users' 
comprehension through a widely used smart contract and interface (USDT and MetaMask).


Concretely, we make the following contributions: 

\begin{enumerate}[leftmargin=*,label=(\textbf{C\arabic*)}]
	\item \textbf{User Study: } To the best of our knowledge,  we present the first user study to investigate the end-users' understanding of \rev{smart contract transfer} risks, 
	through USDT	(\autoref{sec:setup}).  
	\revc{
		We provide detailed analysis of the study responses, including statistical analysis where relevant.
		Notably, our analysis reveals that \textit{up to 71.8\% of 
		users believe that contract upgrade and blacklisting are 
		the most severe and most surprising transfer risks}. 
		Moreover, only up to 35.8\% of users found the MetaMask UI sufficient to understand transfer risks 
		(see  \autoref{fig:sequence-diagram}). 
		In comparison, more than twice as many 
		(82.7\% of) end-users understand the successful  
		transfer outcome (\autoref{sec:results}). 
		Statistical analysis reveals that neither self-rated programming nor Web3 proficiency significantly influence end-users' ratings 
		of transfer risks and their comprehension of MetaMask UI flows for successful and failing transfer 
		outcomes  (\autoref{sec:results}). 
	}
	
	\item \textbf{Prevalence of \rev{transfer} risks:}
	\revc{
		We investigate the prevalence of transfer risks in other ERC-20 contracts through automated and manual source code analysis
		of the top 78 contracts (excluding USDT) (\autoref{sec:setup}). This revealed that examined transfer risks are 19.2\% prevalent. Additionally, our manual analysis revealed three additional transfer risks (\autoref{sec:results}).
	}
\end{enumerate}
We discuss key takeaway points and future outlook based on the study results (\autoref{sec:discussion}) before 
highlighting possible threats in our study (\autoref{sec:validity}) and concluding (\autoref{sec:conclusion}).

%% file: overview.tex
\section{Background \protect\rev{and Terminology}}
\label{sec:overview}


\smallskip\noindent
\textbf{Ethereum and \rev{Smart Contracts}:}
\rev{A blockchain is a mechanism to arrive at \reva{an} agreement \reva{among} a set of decentralised actors about the order of events recorded in an append-only data structure ~\cite{DBLP:journals/csur/KolbAKC20}.}
\reva{{\em Ethereum}~\cite{EthereumWhitepaper} is a public blockchain 
with over}
 238M unique addresses \cite{etherscanEthereumUniqueAddresses} and a market capitalization of over USD 230B \cite{etherscanEtherMarketCapitalization}. 
\reva{In contrast to Bitcoin, 
Ethereum is able} to execute Turing-complete code 
known as smart contracts, a term coined prior to blockchain by 
Szabo~\cite{Szabo_1997}. Ethereum is known as the \rev{start} of the second generation of blockchains, which often have 
smart contract capabilities.


\smallskip\noindent
\textbf{Tokens:}
One major use case of Ethereum smart contracts is implementating auxiliary {\em virtual tokens}~\cite{ERC20TokenStandard, ERC721TokenStandard, ERC1155TokenStandard}. These tokens 
have a variety of uses 
e.g., for implementing currencies, for marking asset ownership (such as with non-fungible 
tokens - NFTs), or for conferring voting power in some community.


\smallskip\noindent
\textbf{ERC-20:}
The programmatic interfaces 
\reva{for tokens employ}
agreed-upon standards in the Ethereum community. 
\rev{ERC-20 is such an interface standard~\cite{ERC20TokenStandard} for fungible tokens}.
During our analysis period, the ERC-20 
standard~\cite{ERC20TokenStandard} for fungible tokens was the most used, accounting for 20.0\% 
of \reva{all} transactions \reva{involving} the top 500 recipient addresses.

\smallskip\noindent
\textbf{\rev{Transfer Risks}:}
\rev{By transfer risks, we refer to 
unexpected events which may not match users' expectations when transferring 
tokens from one Ethereum address to another}.

\smallskip\noindent
\textbf{USDT:}
The dominant share of ERC-20 transactions went to the USDT contract (12.7\% of transactions to the 
top 500 recipient addresses), which manages the USDT stablecoin. 
A stablecoin is a store of value in which its price relative to some fiat currency (i.e., the US dollar) is meant 
to stay stable. USDT is an asset-backed stablecoin which has a market capitalization of over USD 83B~\cite{TetherUSDTPrice}.

\smallskip\noindent
\textbf{Light Wallets and MetaMask:}
End-users typically use software known as {\em light wallets} to interact with blockchains. For Ethereum, a popular 
interface is the MetaMask wallet \cite{MetaMask}. The advantage of such wallets is their ease-of-use 
(the GUI removes the need for user code) and ease-of-setup (no need to download the 
entire blockchain).
The standardized interface of tokens allows end-user wallets to programmatically extract salient features of the 
contract (e.g., the token symbol, the function enabling fund transfer and the ability to retrieve a user’s balance) 
for presentation to the end-user.

\smallskip\noindent
\rev{
\textbf{\reva{Software Engineering Relevance}:} Our study is directly related to the end-users’ perception of risks associated with software systems, in particular, smart contracts. 
\reva{Since smart contracts are relatively new and complex,
there is a \textit{lack of 
empirical studies or automated methods to enable the 
understanding of 
end-users’ perception of transfer risks}. 
To this end, 
our work contributes to the broader field of \textit{(empirical) software engineering for smart contracts}}. Moreover, our study \textit{not only} highlights the concrete risks, but also provides a groundwork for 
\reva{automated detection of transfer risks in the source code of smart contracts. This also allows for the  }
development 
of better wallet interfaces to highlight the transfer risks to end users.  
\reva{Besides,
our study informs the need for 
end-user-oriented software pipeline for smart contracts.
For instance,  it motivates the development of 
(a) techniques (e.g., code/UI annotations)
to explain the associated transfer risks to end users and (b) automated testing and mitigation methods to discover, avoid and prevent such transfer risks. }
}

%% file: related_work.tex
\section{Related Work}
\label{sec:related_work}


\revc{Table~\ref{tab:related-work-comparison} illustrates the overall positioning of our work with respect to existing literature. 
In the following, we further categorize the current works based on their scope and objective.}

\noindent
\revc{
\textbf{User Perspectives on Blockchain Technology (C1):}
   Current works explore a variety of user perspectives, encompassing first-time users, experienced users, miners and people who do not use blockchain.
   The majority of these studies focus either on the perceptions and behaviors relating to trust, security and privacy
   ~\cite{Mining2019,Abramova2021,Sas2017,Ooi2021,Voskobojnikov2020,Froehlich2021Pressure,SecurityRisksWeb3}, or on wallet and key management perceptions ~\cite{Yu2024,Frhlich2020,Voskobojnikov2021,Mangipudi2023}.
   A minority of user studies engage in more focused topics.
      These include creator and holder perceptions of NFTs~\cite{Xiao2024}, user perceptions of third-party security audits ~\cite{Feng2023,huang2024exploring} and user perception of the category of sandwich attacks~\cite{wangSandwich}.
      Most related to our topic, two recent studies~\cite{Guan2024,guan2023examining} concern user perceptions of stablecoins,  but they study stablecoins 
      as a broad category rather than focusing on the implemented transfer risks of a particular stablecoin.
   Many of these works highlight incorrect user understanding, either by showcasing inaccuracies in user mental models~\cite{Saldivar2023,Mangipudi2023,mentalmodel-soups} 
   or highlighting misunderstandings which underpin some user statements~\cite{guan2023examining,Voskobojnikov2021}.
   While there has been work which encompasses perceptions on the breadth of Web3, including the smart contract layer, to the best of our knowledge, 
   there has been no work which studies user understanding of a particular, widely used smart contract.
   Our work thus distinguishes itself from the existing body of user studies in two ways: First, by focusing on a specific yet widely used smart contract (USDT), 
   and second, by examining the category of transfer risks. Additionally, we extend our findings through blockchain data analysis. This is methodologically 
   uncommon across existing studies, which are often centred on user perspectives alone.
}

\vspace{0.5mm}
\noindent
\textbf{\revc{Human Factors in} Smart Contracts \revc{(C1, C2)}:} 
%
%
Our findings are in line with 
 prior works that emphasized the user aspects in blockchain design and improving the design 
 of wallets accordingly~\cite{hcisurvey-dis,blockchain-tochi}. 
 However, our goal is orthogonal to approaches that involved users in the validation of smart 
 contracts, e.g., via 
 user experience design~\cite{ux-hci}, 
 generation of warnings~\cite{tls-metamask}, programming language 
 design~\cite{obsidian-toplas} and user notice generation~\cite{david-usernotice}.
 Specifically, unlike previous works,  which aim to validate a new technology,  our work 
 (\textit{see} \autoref{sec:discussion}) identifies the risks and gaps in smart contract usage using 
 a widely used smart contract (USDT/MetaMask). 
 Our work also complements 
 approaches that target users' specific behaviors 
 e.g.,  key management 
 preferences~\cite{key-icissp}, misconceptions 
 of cryptographic keys and blockchain encryption~\cite{mentalmodel-soups}.

\noindent
\textbf{Programming Languages and Tools for Smart Contracts \revc{(C2)}: } 
In the last few years, researchers have emphasized  the need for  
fixing defects in smart contract code~\cite{david-tse20}. 
To this end, prior works explore security practices of smart contract developers~\cite{practice_security_icse21},  
new programming language constructs for safe smart 
contract~\cite{wetseb-19,obsidian-toplas,scilla}, 
comprehension 
of smart contract code~\cite{ponzi-visual,tse-code-summary,comments-ist,self-destruct}, 
detection of API documentation errors~\cite{ase22-api}, verification of smart 
contract fairness~\cite{fse20-fairness} \reva{\revc{and verification of smart contracts written in a declarative language~\cite{icse24-declarative}. Other researchers have also conducted} oracle deviation analysis~\cite{icse24-oracle}}, testing based on symbolic execution~\cite{ase19-symbolic}, 
code clone detection via code embedding
~\cite{tse21-codeclone}, verification 
with the aid of specification tailored to smart contract~\cite{oopsla21-specification} and static analysis 
based bug hunting~\cite{osiris,verismart,icse-seip}.
\reva{Empirical studies have also analyzed the effectiveness of existing tools~\cite{icse24-fuzzers,icse24-tools}.}
\revc{Our work aligns with these works under the broader category of smart contract risks. However, it is distinguished in that the transfer risks we study are intended behaviors, and therefore unrelated to the implementation bugs that the majority of works in this category study.}


\vspace{0.5mm}
\noindent
\textbf{Security, Privacy and Fairness of Smart Contracts \revc{(C1, C2)}:}
%
To avoid exploitation of security and privacy concerns in smart contract~\cite{security_survey, exploitability-usenix}, 
several works have been 
proposed  including 
 threat modeling~\cite{threatmodeling-infocom19}, recommendation and validation 
of secure programming pattern~\cite{fse21-sse}, 
domain-specific and privacy-preserving services~\cite{medical-tsc23}, security incident response~\cite{ares22-response}, 
static analysis  for vulnerability detection~\cite{icbc-reentrant}\reva{,~\cite{icse24-gptscan}},  
dynamic-analysis-based online defense~\cite{aegis-ccs}\reva{,
investigating patterns of library misuse~\cite{icse24-library},
simulating user behavior with bots to unveil defects arising from multi-user interactions~\cite{bot}}
, smart contract code repair~\cite{david-tosem22,smartshield}, 
and fair, transparent use of smart contracts~\cite{flashboys-sp20}.  
Our study is orthogonal to these works, 
as we investigate 
{\em smart contract \rev{transfer} risks for end-users} instead of design or coding flaws. 


\vspace{0.5mm}
\noindent
\rev{
\textbf{Transfer Risks \revc{(C1, C2)}:}
There is substantial research work in detecting inconsistent and unexpected behaviors in ERC-20 tokens. 
TokenScope~\cite{TokenScope}  automatically detects inconsistencies by comparing the ERC-20 specification, events fired during usage and their effects on relevant data structures. 
Another work ~\cite{AdministratedTokens} proposes a classifier for identifying \textit{administration patterns} in existing tokens. 
There is an overlap in the transfer risks examined in our work and existing work (e.g., Arbitrary Mint and Destroy User Funds), 
but our work includes unexamined transfer risks (User Blacklist and Insufficient Funds). 
Besides, contrary to  prior works on detecting and avoiding transfer risks,  
%
we
examine the end-users’ perception of such risks. 
}



\begin{table}[h!]
   \centering
   \caption{\revc{Comparison of our work vs most relevant related works. $\fullcircle$: Full consideration, \halfcircle : Partial consideration, $\emptycircle$: Exclusion.
      The  comparison categories are derived based on our contributions (C1/C2):
      concerns smart contract transfer risks (``Transfer Risks''),
      concerns user perception (``User Perception''),
      concerns smart contract implementation bugs/vulnerabilities (``Bugs or Vulnerabilities''),
      utilizes user study (``User Study''),
      analyzes blockchain data or smart contract source code (``Blockchain Data or Code'').}
   }
   \label{tab:related-work-comparison}
   \begin{newbox1}
   \resizebox{\linewidth}{!} {
   \begin{tabular}{|C{3cm}|C{2cm}|C{2cm}|C{2cm}|C{2cm}|C{2cm}|}
      \hline
         \textbf{Works}
         & \textbf{Transfer Risks (C1, C2)}
         & \textbf{User Perception (C1)}
         & \textbf{Bugs or Vulnerabilities}
         & \textbf{User Study (C1)}
         & \textbf{Blockchain Data or Code (C2)}
      \\\hline
         This Work & $\fullcircle$ & $\fullcircle$ & $\emptycircle$ & $\fullcircle$ & $\fullcircle$ \\\hline
         ~\citet{Froehlich2021Pressure} & $\halfcircle$ & $\fullcircle$ & $\fullcircle$ & $\fullcircle$ & $\emptycircle$ \\\hline
         ~\citet{SecurityRisksWeb3} & $\halfcircle$ & $\fullcircle$ & $\halfcircle$ & $\fullcircle$ & $\emptycircle$ \\\hline
         ~\citet{TokenScope} & \halfcircle & \fullcircle & \halfcircle & \emptycircle & \fullcircle \\\hline
         ~\cite{Xiao2024,Guan2024} & $\halfcircle$ & $\fullcircle$ & $\emptycircle$ & $\fullcircle$ & $\emptycircle$ \\\hline
         ~\citet{icse24-tools} & $\halfcircle$ & $\emptycircle$ & $\fullcircle$ & $\emptycircle$ & $\fullcircle$ \\\hline
         ~\citet{AdministratedTokens} & \halfcircle & \emptycircle & \emptycircle & \emptycircle & \fullcircle \\\hline
         ~\citet{icse24-fuzzers} & \emptycircle & \fullcircle & \fullcircle & \fullcircle & \fullcircle \\\hline
         ~\citet{self-destruct} & \emptycircle & \fullcircle & \halfcircle & \fullcircle & \fullcircle \\\hline
         ~\citet{huang2024exploring} & $\emptycircle$ & $\fullcircle$ & $\halfcircle$ & $\fullcircle$ & $\emptycircle$ \\\hline
         ~\cite{wangSandwich,tse-code-summary,david-usernotice,comments-ist} & \emptycircle & \fullcircle & \emptycircle & \fullcircle & \fullcircle \\\hline
         ~\cite{Yu2024,Abramova2021,Saldivar2023,Sas2017,Frhlich2020,Froehlich2021Dont,Ooi2021,guan2023examining,Gao2016,Voskobojnikov2020,Mangipudi2023,mentalmodel-soups,threatmodeling-infocom19} & $\emptycircle$ & $\fullcircle$ & $\emptycircle$ & $\fullcircle$ & $\emptycircle$ \\\hline
         ~\citet{Voskobojnikov2021} & $\emptycircle$ & $\fullcircle$ & $\emptycircle$ & $\emptycircle$ & $\emptycircle$ \\\hline
         ~\cite{Zhou2023,icse24-library,bot} & $\emptycircle$ & $\emptycircle$ & $\fullcircle$ & $\emptycircle$ & $\fullcircle$ \\\hline
         ~\citet{ponzi-visual} & \emptycircle & \emptycircle & \halfcircle & \emptycircle & \fullcircle \\\hline
         ~\citet{flashboys-sp20} & \emptycircle & \emptycircle & \emptycircle & \emptycircle & \fullcircle \\\hline

   \end{tabular}}
\end{newbox1}
\end{table}

%% file: evaluation_setup.tex
\section{Research Methodology}
\label{sec:setup}

In this work,  we pose the following research questions:

\noindent
\textbf{RQ1 Risk Perception:} 
How do end-users perceive the 
transaction risks of USDT smart contracts?

\noindent
\textbf{RQ2 \revc{Understanding} of User Interface:} 
Does the MetaMask wallet effectively inform users of possible transfer outcomes?


\noindent
\textbf{RQ3 Smart Contract Understanding:} 
\revc{How do 
smart contract end-users educate themselves? }
How do they perceive their
own comprehension of how smart contracts 
work? 

\noindent
\textbf{RQ4 Generalizability of Risks:} 
Do the transfer risks
of USDT contract 
occur in other ERC-20 contracts?



\revc{The first three research questions lead to contribution \textbf{C1}, while \textbf{RQ4} leads to contribution \textbf{C2}}.
\textbf{RQ1} \revc{describes} the central question of transfer risk perception in the most used Ethereum smart contract (USDT). 
\revc{Building on this, }\textbf{RQ2} investigates \revc{how the MetaMask UI may lead to incorrect transfer risk perception}.
\revc{In a similar vein, \textbf{RQ3} aims to deepen our understanding of factors that mediate transfer risk perception by investigating how users comprehend smart contracts, as well as their confidence in their comprehension}. Finally, \textbf{RQ\revc{4}} studies the generalization of these transfer risks to other ERC-20 contracts, extending our findings beyond USDT. 

\vspace{-0.1in}
\subsection{User Study Design}
\label{ssec:setup:user_study_design}

\noindent
\textbf{Survey Questionnaire:}
The user study questionnaire contains 
196 questions
implemented on Google Forms~\cite{googleforms}.  It begins with a consent form explaining 
study goals and obtaining participants' consent to \textit{anonymously} collect response data.  In the first 
part of the survey,  we posed questions about the demography of respondents (e.g., age and profession) and their 
knowledge/expertise level.  For instance,  we asked about the participants' knowledge of smart contracts, USDT,  
stablecoins and programming. The second part of the survey gathers information about USDT user behavior 
and risk awareness. It contains questions about the users perception of transaction outcomes and transfer risks,  
the usabilty of the MetaMask user interface,  and user preferences for alternative descriptions of smart contract 
behavior. We provide a copy of the survey questionnaire for scrutiny and reuse (see \autoref{sec:conclusion}).


\vspace {0.5mm}
\noindent
\textbf{Recruitment and Compensation:} 
The study was conducted using Prolific \cite{prolific}, a well-known platform for conducting industrial and research surveys.  
We chose this platform because it allows to pre-screen for participants with experience or expertise in specific domains,  
in our case smart contracts. Specifically,  we screened for participants that are fluent in English language,  and completed 
at least secondary school education.  We also require that participants have high quality answers (approval rating between 
98 to 100, with at least 100 prior submissions on the platform) in prior surveys and have a knowledge of cryptocurrency 
or cryptocurrency exchanges.  
\rev{All participants gave consent prior to participation, and were provided with multiple avenues for feedback during and 
after the survey. 
After the study, we sent 
a draft of the paper
to all participants 
explicitly informing them about true and fictitious transfer risks.}
Overall, we recruited 110 respondents for the final study and paid each participant £7.04.
The main study took 44 days to complete starting from 30th January 2023.

\vspace {0.5mm}
\noindent
\textbf{Pilot 
Study:} 
We conducted the first pilot study for three days (from 11th to 13th January 2023) with four researchers and librarians 
in order to obtain feeedback on the design and identify unclear questions in the study questionnaire.  
The feedback enabled us to add several free-text questions to elicit reasons 
for users' responses and to allow participants to provide additional information about risk comprehension.  Using the 
revised questionnaire, we conducted a second pilot study for a day with 10 participants from Prolific (paid £7.04) who 
had used smart contracts before.
Analyzing these responses informed the inclusion of new questions, e.g.,  
questions on comprehension of the MetaMask user interface (UI) flow.

\vspace {0.5mm}
\noindent
\textbf{Demographics:} The survey was conducted on 110 respondents. Participants' age is from 18 
to 64 years,  most (54.6\%) being 25-34 years old and least (4.6\%) being between 55 to 64 years old.  
Participants are from 
over 21 different sectors, the top three being Computing or IT (23.6\%), 
Engineering or Manufacturing (11.8\%), and Students (8.2\%).  Geographically,  the respondents are from four continents 
and 18 countries. The top three countries were the United States (26.4\%), South Africa (22.7\%) and the United Kingdom (17.3\%).  

\vspace {0.5mm}
\noindent
\revc{
	\textbf{Validating Smart Contract Usage}:
	The main threat to construct validity is that respondents may falsely claim to be smart contract
	users, e.g., because they transact on intermediary cryptocurrency exchanges. We mitigate this by (a) pre-screening for
	users (on Prolific) who were knowledgeable about cryptocurrency, (b) asking validation questions about smart contract
	usage, (c) notifying respondents that transferring from an exchange to their wallet is not smart contract usage, and (d)
	asking respondents to rate themselves on their Web3 proficiency. We note that most (94\%) of the respondents have at least
	some Web3 proficiency and passed our validation questions.
}

\vspace{0.5mm}
\noindent
\revc{
\textbf{Proficiency Validation:} 
The self-rated Web3 and programming proficiencies of participants
were verified with validation questions based on verifying the participants' knowledge of common concepts (e.g.,  \textit{Please give an example of a low-level programming language}). 
However, we note that these validation questions do not provide full confidence in self-rated scores which are high, particularly for programming --  a more skill-based proficiency. 
}


\vspace {0.5mm}
\noindent
\textbf{Quantitative Response Data Analysis: } 
We report the number of respondents that chose each option (e.g., severity level score on Likert scale) 
as well as the percentage of respondents that chose the option
(\textit{see} \autoref{tab:risk-perception}, \autoref{tab:fake-risk-perception} and \autoref{tab:outcome-perception}). 
We further categorised each likert scale response into three levels, e.g.,  Aware (score 1-2), Somewhat Aware (3) and Unaware (4-5). 
The total percentages may be below 100\% (\textit{in} \autoref{tab:risk-perception-reasons} and \autoref{tab:outcome-perception-reasons}) 
since some  responses were discarded due to clear misunderstandings of the question by the participant,  confusing reasons or  incorrect assessments 
of the presented scenario.  Such responses were not categorized, but are mentioned when worth highlighting 
(\autoref{sec:discussion}, \autoref{tab:risk-perception-reasons} and \autoref{tab:outcome-perception-reasons}). 

\vspace {0.5mm}
\noindent
\textbf{Qualitative Coding Protocol:} 
To analyze free-text questions for qualitative results, we use a coding protocol~\cite{charmaz2006constructing}
 involving at least two researchers.  We extracted qualitative results for four (4) real transaction risks (\textbf{RQ1}) and all MetaMask flow evaluations (\textbf{RQ2}).  Our coding protocol involved one researcher \rev{manually} deriving the initial response categorizations, which was then validated by another researcher and conflicts were resolved in the coding to agree on the categorization of responses.
 \rev{This process took approximately 30 hours in total}.
 \rev{In general, we categorized reasons with associated scores above three (3) as positive, and below three (3) as negative.}
 The presented results (in \autoref{sec:results}, e.g.,  \autoref{tab:risk-perception-reasons} and  \autoref{tab:outcome-perception-reasons}) are the consensus after coding and validation.


\vspace {0.5mm}
\noindent
\textbf{Examined Real Risks:} We examined \revc{all five transfer risks in the USDT contract} to investigate users' 
perception\rev{s} of real risks. 
The first four risks stop users from transferring USDT.  

\noindent
\textit{(1) Contract Pause: } 
The contract being \textit{paused}.

\noindent
\textit{(2) User Blacklist: } 
The user being \textit{blacklisted}.

\noindent
\textit{(3) Contract Upgrade: } 
The contract being \textit{upgraded} to a new, arbitrary  contract,  \revise{contrary to the original implementation}.

\noindent
\textit{(4)  Insufficient Funds: } 
\textit{insufficient funds} in the user's account.

\noindent
\textit{(5) Transfer Fee Increase: } \newrevise{Increased fee parameters (currently zero) reducing the amount the receiver obtains from the sender.}

Except for \revc{the} insufficient funds risk, these identified risks represent ways in which a USDT transfer could fail beyond the ERC-20 specification.

\vspace {0.5mm}
\noindent
\textbf{Fake/Fictitious Risks:} 
To further establish users' misunderstandings of smart contracts, 
we constructed \revc{the following} five fake risks based on misconceptions \revb{that we suspected users may hold} about the design and implementation of Ethereum and USDT\revc{:}

\noindent
\textit{(1) Consortium Reject: } A user's inability to transfer USDT  due to a consortium of Tether \textit{users voting to reject} the transfer.
\revc{This risk is based on the possibility that users may have incorrect concepts of decentralization, potentially due to misunderstandings ~\cite{mentalmodel-soups} or centralization-decentralization tradeoffs ~\cite{Xiao2024}, which may lead them to believe that majority vote for acceptance of token transfers is plausible.}
This feature is not in any ERC-20 source code we analyzed.

\noindent
\textit{(2) Government Block: } 
The transaction fails due to a \textit{government blocking} it. 
\revc{The intuition for this risk is that users might incorrectly anchor expectations on the centralized financial systems ~\cite{mentalmodel-soups} which they may be more familiar with or assume cooperation with regulatory bodies ~\cite{guan2023examining} that goes beyond what is possible on the smart contract layer}.
This is not possible as Ethereum's consensus algorithm~\cite{EthereumWhitepaper} was designed \revc{to prevent} 
centralized control. 

\noindent
\textit{(3) Receiver Reject: } 
The \textit{receiver rejects} the transfer.
\revc{The risk is designed to check if users might believe they have a level of control over their wallets which they do not possess, possibly due to misunderstandings about permissions and decentralization ~\cite{mentalmodel-soups}.}
This \revc{ability} was not present in any ERC-20 source code we analyzed.

\noindent
\textit{(4) Partial Funds: }
Sending a larger amount (e.g., 10 USDT) than a sender owns (e.g., 5 USDT) results in the receiver receiving only the sender's wallet amount (i.e., 5 USDT, not 10 USDT).
\revc{The intuition behind this risk is a check for whether, perhaps due to lack of knowledge, users might have incorrect assumptions about the implementation of the transfer function. While such an implementation contradicts the ERC20 specification ~\cite{ERC20TokenStandard}, prior work ~\cite{TokenScope} has noted many instances in which this specificaion is not well-implemented.}
In reality, \revc{the Tether implementation would correctly cause} the transaction \revc{to} simply fail. 

\noindent
\textit{(5) Gas Fee Increase: } Receivers \revc{receive} less USDT due to \textit{fluctuations in gas  fees}. 
\revc{This risk is in line with numerous prior works ~\cite{Voskobojnikov2021,mentalmodel-soups,Saldivar2023} which demonstrate that users often find gas fees difficult to understand.}
This is misconceived \revc{since} gas fees, paid in ETH (not USDT), cannot reduce the USDT transferred.


\revb{In order to avoid biasing participants during the survey, participants were only informed that fake risks were included after the survey.
They were then informed about which risks were real and which were fictitious.}

\vspace {0.5mm}
\noindent
\textbf{Collecting User Interface (UI) Flows:}
In our user study,  we used the MetaMask light wallet 
to present users with screenshots of a standard transfer flow for USDT.  
To achieve this,  \revise{we cloned the USDT smart contract and deployed it on 
the Ethereum Sepolia test network~\cite{SepoliaTestnet} under the token name YUSDT. Our ownership over the 
deployed contract allowed us to tweak YUSDT parameters in the same manner possible 
in the actual USDT contract by its owner. This, in turn, enabled observation of  
the MetaMask interface differences under varied contract parameters.
To avoid bias, we discard responses that may be due to incomplete UI flows,  e.g.,  missing final screenshot or intermittent pop-ups.}

\noindent
\textbf{Evaluating Transaction Outcomes:} We examined three main transaction outcomes from the MetaMask UI flow:

\noindent
\textit{(1) Full amount transferred} -- 
the full transfer amount reaching the recipient.

\noindent
\textit{(2) Reduced amount transferred}  --
a reduced transfer amount reaching the recipient.

\noindent
\textit{(3) Failure} of the transfer --  no change occurring in both the sender and recipient's wallet.

\noindent
For each outcome, we asked the following questions: 
  
\noindent
\textit{(1)}  Was it (the UI) sufficient to \textbf{discover} the possibility of the outcome (with one - \textit{Not sufficient at all (I am completely uninformed about the possible outcomes)}, five - \textit{Fully sufficient (I fully understand the possible outcomes)})
  
 \noindent
\textit{(2)} Was it (the UI) sufficient to \textbf{understand} reasons behind those outcomes arising (one - \textit{Not sufficient at all (I am completely uninformed about the possible outcomes)}, five - \textit{Fully sufficient (I fully understand the possible outcomes)})

Each question was accompanied with a text field asking the respondent to describe the reason for their choice.

\begin{figure}[t!]
	\centering
	\includegraphics[width=0.7\linewidth]{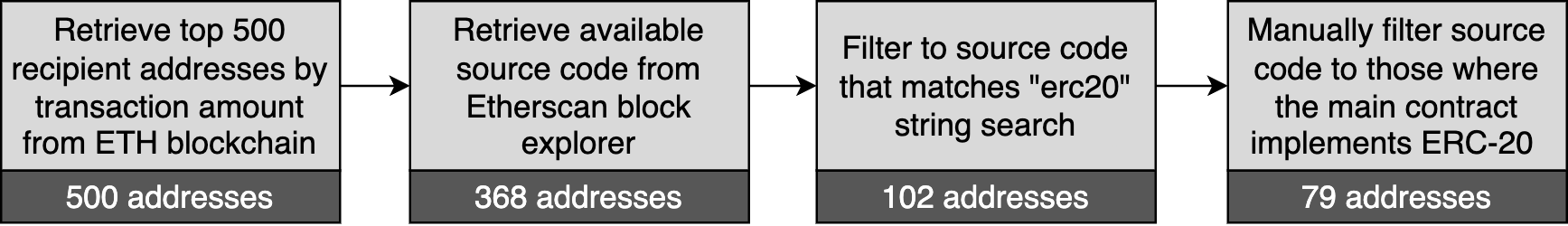}
	\caption{Source code retrieval process}
	\label{fig:source-code-methodology}
	\Description{The figure shows a flowchart of 4 nodes connected in sequence. Each node represents a processing stage, containing text describing the processing and a count of the number of smart contract address remaining after its processing.

	Node 1 text: “Retrieve top 500 recipient addresses by transaction amount from ETH blockchain”.
	Node 1 number of addresses: 500 addresses
	
	Node 2 text: “Retrieve available source code from Etherscan block explorer”.
	Node 2 number of addresses: 368 addresses
	
	Node 3 text: “Filter to source code that matches “erc20” string search”.
	Node 3 number of addresses: 102 addresses
	
	Node 4 text: “Manually filter source code to those where the main contract implements ERC-20”.
	Node 4 number of addresses: 79 addresses
	}
\end{figure}

\noindent
\textbf{Metrics and Measures:}
%
For each examined (real or fake) risk, we asked participants to evaluate their {\em awareness}, {\em surprise} and perceived {\em severity} 
of the risk on Likert scales as follows:

\noindent
\textit{(a) Unawareness} 
\noindent\footnote{Note that in the user study,  we examined the inverse, i.e.,  
 	``awareness'' rather than ``unawareness''. However,  this was inverted in the presentation of 
 	results (\autoref{sec:results}) for consistent analysis of negative 
 	reasons (unawareness)}
\textit{: }
score ``one (1)''  as \textit{I fully knew this could happen} to ``five (5)'' as
\textit{I had no idea that this could happen}.
\revc{We choose to evaluate unawareness as it addresses perception from a knowledge angle. More precisely, we check if the user knows 
	about a risk prior to being affected by it.}

\noindent
\textit{(b) Surprising: }  
score ``one (1)''  as \textit{Completely unsurprising} to ``five (5)'' as \textit{Completely surprising}.
\revc{Moving beyond knowledge, surprisingness gives a sense of the ``interestingness" of a risk to a user. 
	Indeed, prior work in highlighting interesting correlations within personal informatics \cite{Interesting2016} found 
	surprisingness to be a statistically significant (p < 0.01) predictor of interestingness to a user.}

\noindent
\textit{(c) Severity: }  
score ``one (1)'' as \textit{Not severe at all (it does not bother you at all)} to ``five (5)'' as \textit{Extremely severe 
	(you would not use USDT because of this possibility)}. 
\revc{Going further than knowledge and interestingness, severity is an action-oriented indication of the user's perceived impact of a risk on their own usage behavior. 
Severity is also a typical feature of Web3 audit reports, which are a common practice in the smart contract ecosystem that enhance user trust~\cite{huang2024exploring,Feng2023}.}

Similar to \revc{the} user interface evaluation, we followed each score with a text field asking the respondent to explain their choice.
\revc{Additionally, we note that the measurements of surprisingness and awareness in \textbf{RQ1} are in danger of being undifferentiated by users due to their semantic similarity.
To mitigate this, we added additional descriptive text to each question.
For surprisingness, the description was ``\textit{Does it surprise you that this could happen, or did it happening make sense to you?}''.
For unawareness, the description was ``\textit{Prior to reading this reason, did you know that your transaction could be rejected for this reason?}''
}
%


\vspace {0.5mm}
\noindent
\textbf{Statistical Analysis: } Our analysis includes statistical significance tests using 
the Mann-Whitney U test for unpaired analysis,  and the Wilcoxon signed rank test for paired analysis. We also reported normality and Levene tests,  
median values, p-values and U-statistics\footnote{Statistical analysis data and results are publicly available (\textit{see} \autoref{sec:conclusion}).}. \newrevise{These tests were performed on the following three pairs of groups of 
respondents, each rated on a Likert scale (one: \textit{not proficient}, five: \textit{extremely proficient})}: 

\noindent
\textit{(a) Programmers vs. Non-Programmers: }
We classified respondents with a self-rated programming proficiency of at least two\footnote{The programmer threshold differs from the other two thresholds to ensure that each group in the statistical test has $\ge$ 
20 participants - with a threshold of four, there would only be 15 participants in the programmer group.} out of five (85 respondents, 77.3\%)  
as programmers (see \textbf{RQ1}, \textbf{RQ2}, \textbf{RQ4}).
 
\noindent
\textit{(b) High vs. Low Web3 Proficiency: }
We classified respondents with a self-rated Web3 proficiency at least four out of five (32 respondents, 29.1\%) as high Web3 proficiency
(see \textbf{RQ1}, \textbf{RQ2}, \textbf{RQ4}).
 
\noindent
\textit{(c) High vs. Low Behavior Anticipation of Users:} 
We classified respondents with a self-rated ability to anticipate smart contract behavior of at least four out of five (87 users, 79.1\%) 
as confident in \textbf{RQ4}. 

\vspace{-0.1in}
\subsection{ERC-20 Source Code Analysis}
\vspace {0.5mm}
\noindent
\revc{\textbf{Source Code Collection:}}
We analyzed all Ethereum transactions over 94 days from 22nd March 2022 (block 14434001 to block 15012398).
From the available source code on Etherscan~\cite{etherscan},
we collected the top 500 (out of 11M) recipient addresses by transaction volume,
and filtered to ERC-20 contracts (see \autoref{fig:source-code-methodology}). 
%
The USDT contract had the most transactions by far, accounting for over 12.7\% of the transactions sent to those top 500 addresses. 
We thus based our analysis around it as the focal point.

\vspace {0.5mm}
\noindent
\revc{\textbf{Automated Analysis:}}
We conducted 
automated analysis of the source code of the remaining (78) ERC-20 contracts 
for only three risks (\textit{Contract Pause}, \textit{User Blacklist} and \textit{Contract Upgrade}). 
The \textit{Insufficient Funds} and \textit{Transfer Fee Increase} risks were excluded as they were not 
amenable to our approach of string matching of function names.
The strings used in our automated analysis, for case-insensitive string matching on function names, 
\reva{
are ``\textit{paus}'' for ``\textit{Contract Pause}'', ``\textit{blacklist}'' for ``\textit{User Blacklist}'' and ``\textit{deprecat}'' for ``\textit{Contract Upgrade}''.
}
These are based on the main components of the function names used by USDT 
for the respective features (e.g., we stem \textit{"pause"} to \textit{"paus"} for better generalizability). 
\revc{This methodology fails to detect  
code semantics 
e.g., it will not detect the ``Contract Pause'' feature implemented with a function named {\em stop} 
(thus failing the match term \textit{paus}). 
However, 
after manually evaluating the results for the examined ERC-20 contracts, 
the string matching was found effective in practice.
}

Our \textit{automated 
analysis} is implemented in about 1.1 KLOC of Python and JavaScript code. 
The detection experiment took $\approx$3 minutes, using a single thread on a 16-inch Macbook Pro (2021 model, 
M1 Max CPU, 32GB RAM, 1TB SSD).

\vspace {0.5mm}
\noindent
\revc{\textbf{Manual \revc{Evaluation}:}}
\revb{To validate and evaluate the automated detections, we indepedently}
conducted \textit{manual analysis} to identify instances of \textit{Insufficient Funds} and \textit{Transfer Fee Increase}, 
determine additional risks and identify the true/false positives
for the automated analysis.
\revb{To this end, a single researcher inspected all source code to identify presence of risky features. For verification, a second researcher inspected a 10\% sample of the source code to identify the same risky features, independent of the first researcher's results.}
Our data is publicly available (\autoref{sec:conclusion}).



%% file: evaluation.tex
\section{Evaluation Results}
\label{sec:results}
%
%
%


\input{evaluation-rq1}
\input{evaluation-rq2}
\input{evaluation-rq3}

\input{evaluation-rq3ii}

\input{evaluation-rq4}

%% file: evaluation-rq1.tex
\input{evaluation-rq1-real-qualitative-table.tex}



{
\renewcommand{\arraystretch}{1} 
\setlength{\tabcolsep}{5pt} 

\input{evaluation-rq1-real-quantitative-table}

\begin{figure}[t!]
	\centering
	\includegraphics[width=0.8\linewidth]{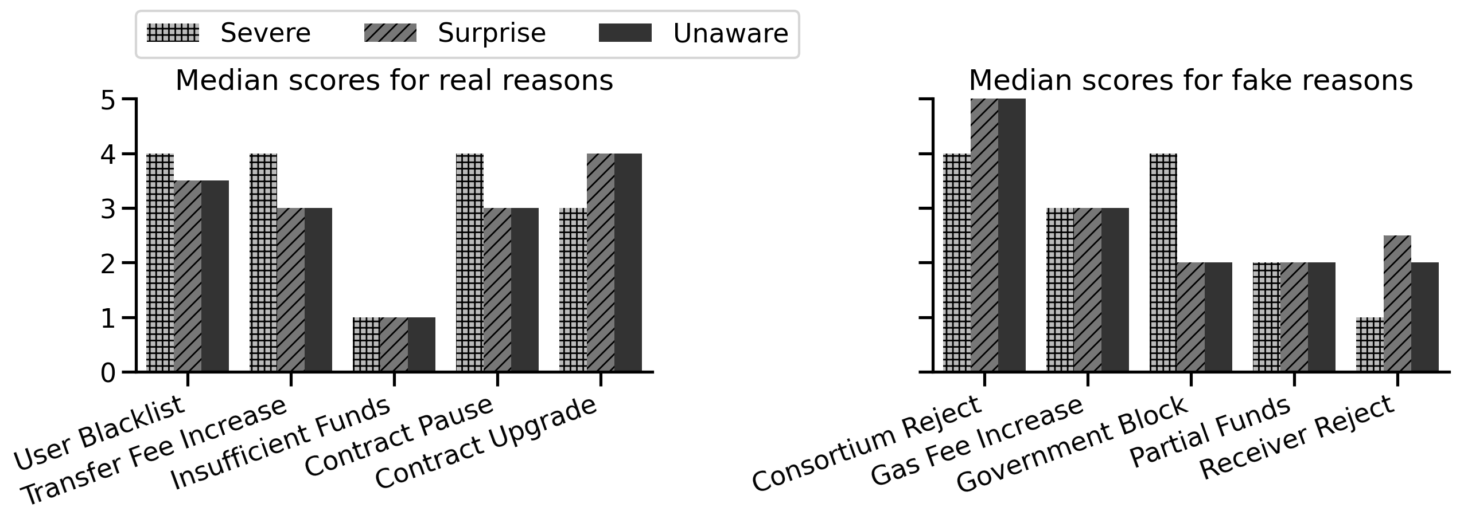}
	\caption{Median scores for rejection and reduction reasons}
	\label{fig:risk-perception}
	\Description{The figure shows two graphs, one titled “Median scores for real reasons” and the other “Median scores for fake reasons”.The reasons are placed on the X axes, while the scores are placed on the Y axes. Three bar graphs are drawn for each reason, representing the median scores for “Severe”, “Surprise” and “Unaware”.

	For the real reasons, all median scores save for Insufficient Funds are between 3 and 4. Insufficient Funds has all scores at 1.
	
	For the fake reasons, the median scores have more diversity.
	}
\end{figure}

\noindent
\textbf{RQ1 Risk Perception: }
%
%
We evaluate end-users' unawareness,  surprisingness 
and severity of transfer risks when using the USDT smart contract.  Given a description of 
transactions affected by each risk,  we asked participants to rate the level of each aforementioned 
property using Likert scales between one (e.g., least surprising) to five (e.g., most surprising).  We also collect and analyze the reasons for participants ratings.     
This includes both the real transfer risks, and the fictitious (fake) ones 
(see \autoref{sec:setup}).
Our findings are shown in 
\autoref{fig:risk-perception}, \autoref{tab:risk-perception} and \autoref{tab:fake-risk-perception}.

%
%
%

\input{evaluation-rq1-fake-quantitative-table}

\vspace{0.5mm}
\noindent
\textbf{\textit{Real Transfer Risks: }} We found that \textit{most (71.8\%) end-users see blacklisting as the most severe risk}, e.g., because it leads to loss of control over their USDT assets (\textit{see} \autoref{tab:risk-perception-reasons}). 
\revc{As an example, one respondent gave the reason that ``Being able to blacklist somebody from transferring their own cryptocurrency is an extremely severe issue, that is basically the same thing as blocking somebody's bank account with money in it''.}
We also observed that  \textit{more than half (55.5\% and 51.8\%) of end-users are surprised and unaware of contract upgrading}.  Qualitative analysis (i.e., coding, \revc{\textit{see} \autoref{tab:risk-perception-reasons}}) of free-text responses shows this to be largely due to user unawareness of the contract owner's ability to upgrade contracts
\revc{(e.g., ``I had no idea that this could happen'')},
to upgrading being an unexpected behavior
\revc{(e.g., ``It’s surprising as I assumed there would need to be a solid reason for rejecting the contract, especially after the company making the choice to upgrade it'')}
or to a lack of user notice prior to upgrade
\revc{(e.g., ``They should inform customers of these things so that it doesn't come as a shock to them when it happens'')}.
Pausing is the least severe and surprising because end-users believe it is temporary
\revc{(e.g., ``I dont think this pause is permanent'')},
infrequent
\revc{(e.g., ``It doesn’t bother me unless it’s constant '')}
or happens for legitimate reasons
(e.g., \revc{``It was just temporary for funds safety''}).

Insufficient funds and transfer fee increase are considered not risky 
by most users: 
most users (81.8\% and 60.9\%) are aware,  not surprised or unconcerned about the severity of both risks
(\textit{see} \autoref{fig:risk-perception}, \autoref{tab:risk-perception-reasons} and \autoref{tab:risk-perception}).  
\revc{Rejection due to insufficient funds is generally seen as commonsensical, with some anchoring on their experience with traditional banking systems
(e.g., ``It's like a bank so not too sursprising'')}.
}
Most end-users also believe increases in fees are normal
\revc{(e.g., ``Most companies take a fee for money transfers so this is completely unsurprising'')},
for profit
\revc{(e.g., ``The company would want to profit from transactions'')},
or are necessary for maintenance 
\revc{(e.g., ``As a result of the system maintenance fees, it is not a serious problem'')},
and it does not deter their usage of the smart contract, especially if agreed to ahead of time
\revc{(e.g., ``Not a severe issue, that is why it's important to read the terms and conditions.'')}.

\newrevise{These results imply that most end-users  
believe blacklisting and contract upgrade are highly risky,  pausing is somewhat 
risky,  but 
insufficient funds,  and fee increase are low risks}. 

\begin{result}
\newrevise{
Most (up to 71.8\%) end-users believe contract upgrade and user blacklist to be the most severe and surprising transfer risks. 
}
\end{result}



\noindent
\textbf{\textit{Real versus Fake Transfer Risks: }} 
We note that \textit{users rated fake risks  similarly as real risks across all three metrics},  i.e.,  surprisingness, unawareness and severity
(\textit{see} \autoref{fig:risk-perception}, \autoref{tab:risk-perception} and \autoref{tab:fake-risk-perception}).
  For instance,  the majority (up to 40\%) of users \textit{wrongly} claim to be highly aware (score one) of \textit{fake} risks -- receiver's rejecting a transaction or receiving less USDT due to insufficient funds.  Notably, this rating is more than their awareness for all \textit{real} risks (except insufficient funds) where
the highest awareness scores (score one)
are for pausing and fee increase with only 27\% and 30.9\%, respectively.   
\revb{These numeric ratings, validated by our \revc{qualitative} analysis of the free-text responses},  suggest that \textit{end-users are as surprised, unaware and  concerned about the severity of real risks as much as fictitious,  fake risks. } Users confusing and similarly rating real and fake risks shows they are
uninformed about the risks inherent in USDT transfers.

\begin{result}
End-users are uninformed (unaware) about real transfer risks.
They confuse real and fake risks,  
rating both as similarly surprising and severe,  \newrevise{while being incorrectly more informed (aware) about fake risks.}
%
\end{result}

\noindent
\textbf{\textit{Statistical Analysis and Differences in Distributions: }} 
We tested for statistically significant differences in risk rating, 
differentiated by users' \rev{self-rated} proficiency in 
programming 
 and Web3 separately (\textit{see} \autoref{sec:setup} for setup details).
\textit{There was no statistically significant difference 
between tested groups 
across all metrics,  i.e.,  unawareness,  surprisingness and severity.}
This 
suggests that neither \rev{self-rated} programming nor high Web3 proficiency 
are related to
risk perception of end-users. Notably, 
none of the proficiencies yielded better identification of fake risks.
\revc{We also note that even with both thresholds changed to three, programming and Web3 proficiency still show no statistically significant differences.}

\begin{figure}[t!]
\begin{newbox1}
	\centering
	\includegraphics[width=0.8\linewidth]{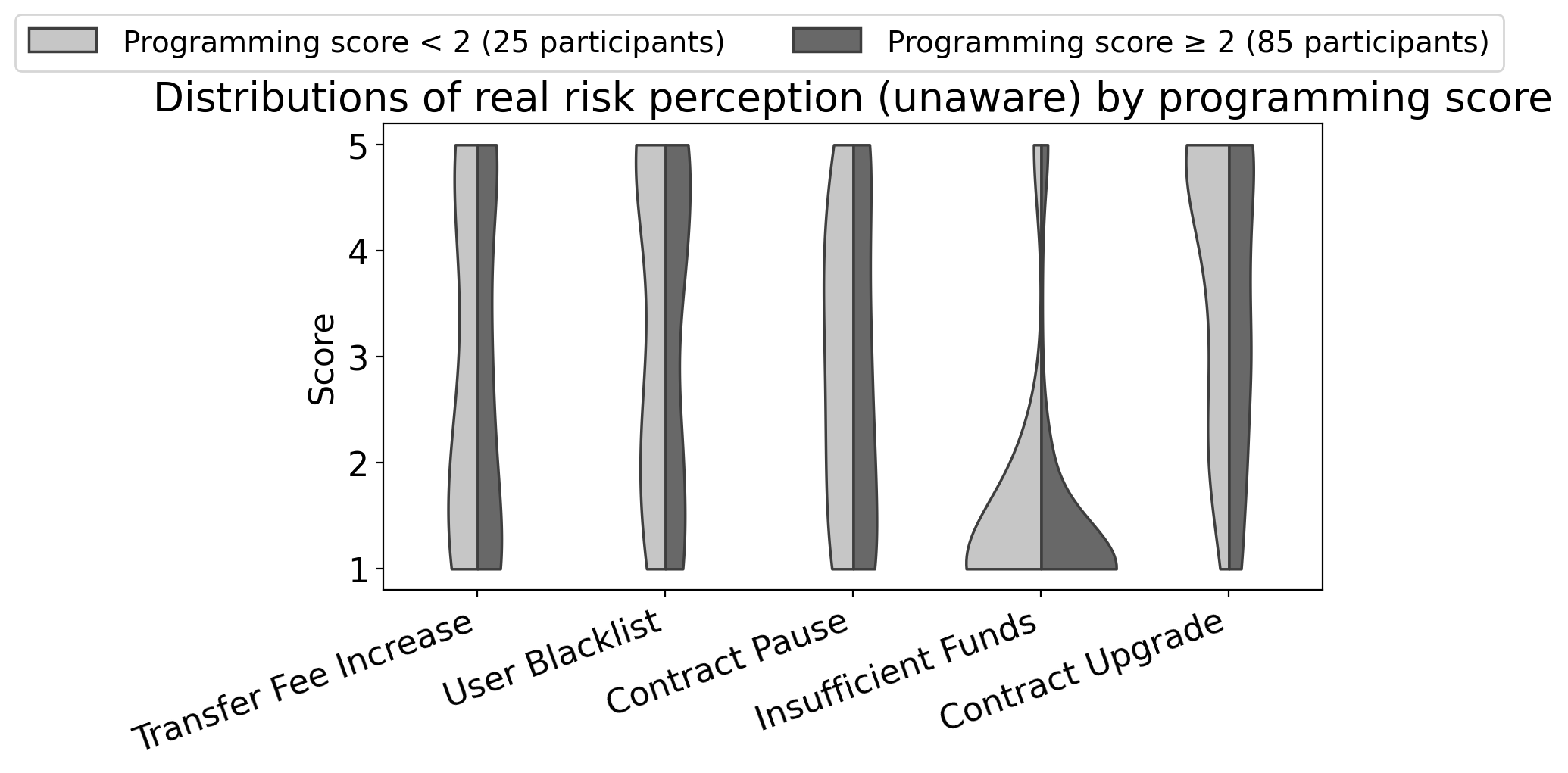}
	\caption{
		\revc{
			A violin plot comparing distributions of risk perception response scores, segmented by programming skill.
			Exhaustive plots for the facets of risk perception questions (unawareness, surprisingness, severity), realness (real, fake) and skill level (programming and Web3) at two different thresholds are found in \autoref{sec:appendix:distributions}.
		}
	}
	\label{fig:violin-plot}
	\Description{
		This showcases the distribution of scores for awareness of real risks, split by self-rated programming proficiency into a proficient group and a non-proficient group. Generally speaking, the distributions look similar despite different programiming proficiencies.
	}
\end{newbox1}
\end{figure}

\revc{Additionally, in order to better intuit the (lack of) differences between the groups, we provide violin plots of risk perception distributions across two different thresholds for both programming and Web3 proficiency  skill. These are found in \autoref{sec:appendix:distributions}, with \autoref{fig:violin-plot} serving as a representative example.}


\begin{result}
\rev{Self-rated} programming/Web3 proficiency
does not significantly influence 
end-users rating 
of real and fake transfer risks.
\end{result}

\input{evaluation-rq2-table}

%% file: evaluation-rq1-real-qualitative-table.tex
{
\begin{table*}[tb!]
 \begin{center}
 \caption{\centering Reasons for end-users' perception of
\textit{real} transfer risks,  highest number/percentage of participants and reasons are marked in \textbf{bold text} and \textit{wrong reasons} are in 
\textbf{\textit{italicized and bold text}} 
(``\#'' = ``Number of Respondents'', ``\%'' = ``Percentage of Respondents'' ). \newrevise{Exclusions for ambiguous/misanswered/incorrect answers and multiply-categorized answers lead to percentage totals and absolute totals deviating from 100\% and 110 responses.}
}
 \resizebox{\linewidth}{!}{
 \begin{tabular}{@{}|l|l|l|c|c|l|c|c|}
 \hline

\multirow{3}{*}{\rotatebox[origin=c]{90}{\textbf{Metric}}} & & \multicolumn{6}{c|}{\textbf{Most Prevalent Reasons for Users' Perception of Risks}} \\

& &  \multicolumn{3}{c|}{\textbf{Negative Reasons}} & \multicolumn{3}{c|}{\textbf{Positive Reasons}} \\  

 & \textbf{Outcome}  & 
\multicolumn{1}{c|}{\textbf{Examples} (Why Unaware,  Surprising or Severe?)}  & \textbf{\#} & \textbf{\%} & 
\multicolumn{1}{c|}{\textbf{Examples} (Why Aware,  Unsurprising or Not Severe?)}  & \textbf{\#} & \textbf{\%} \\
\hline


\multirow{4}{*}{\rotatebox[origin=c]{90}{\textbf{Unaware}}} 
& \textbf{Contract Pause} & 
\textbf{surprising capability} \&  no personal experience & \textbf{40} &  \textbf{36.4} &  maintenance  \& centralization  & 14 & 12.7
\\ \cline{2-8}

& \textbf{User Blacklist} &\textbf{no personal experience},  
\textit{\textbf{it is impossible}} 
or a bug & \textbf{55} & \textbf{50} & 
used to stop bad actors \& centralization & 40 & 36.4
\\  \cline{2-8}

& \textbf{Contract Upgrade} & 
\textbf{no knowledge of capability} \& unexpected behavior
& \textbf{57} & \textbf{51.8} & 
prior belief that contract/policy is alterable 
\& sensible 
& 35 & 31.8 \\  
\cline{2-8}

& \textbf{Transfer Fee Increase} & 
fee is unexpected and unspecified prior to the transfer
& 41 & 37.3 & 
\newrevise{\textbf{expected financial institution charges} \& profit motive}
& \textbf{67} & \textbf{60.9} 
\\
\hline 

\multirow{4}{*}{\rotatebox[origin=c]{90}{\textbf{Surprise}}} 
& \textbf{Contract Pause} &
no knowledge of pausing or that it affects all users & 41 & 37.3 & \textbf{users know of similar incident} 
\& due to centralization & \textbf{44} & \textbf{40}
\\  \cline{2-8}

& \textbf{User Blacklist} & 
\textbf{no justification provided} 
\& it is unfair  & \textbf{56} & \textbf{50.9} & 
 prior knowledge of capability \& it is sensible 
& 37 & 33.6 
\\  \cline{2-8}

& \textbf{Contract Upgrade} & \textbf{not aware of capability},  
no prior notice  and unexpected 
& \textbf{61} &\textbf{55.5} &
belief capability is possible/anticipated \& prior knowledge 
& 31 & 28.2 \\  \cline{2-8}

& \textbf{Transfer Fee Increase} &
less money received, 
\& unaware of 
capability 
& 23 & 20.9 &
\textbf{fees are normal},  fees are for profit or maintenance
& \textbf{36} & \textbf{32.7}
 \\
\hline

\multirow{4}{*}{\rotatebox[origin=c]{90}{\textbf{Severity}}}  & \textbf{Contract Pause} &  
loss of control of assets, impacts all users \& centralization & 
32 & 29.1 & \textbf{it is temporary},  
infrequent 
or for reasonable purposes
&  \textbf{63} & \textbf{57.3} 
\\ \cline{2-8}

& \textbf{User Blacklist} & 
\textbf{loss of control of assets }
\& not decentralized 
&  \textbf{79} & \textbf{71.8} & 
justifable,  benefits all users, 
\& against fraudulent users
& 16 & 14.5
\\  \cline{2-8}

& \textbf{Contract Upgrade} & 
\textbf{arbitrary power} \& potential for abuse &  \textbf{44} & \textbf{40}  
& the issue is likely fixable \& (only) inconvenient & 29 & 26.4
\\  \cline{2-8}

& \textbf{Transfer Fee Increase} & \textbf{users will switch to alternatives}, fee is disliked/unfair
& \textbf{45} & \textbf{40.9} & 
fee is known prior to transfer \& it does not deter usage
& 41 & 37.3 
\\ 
\hline


   \end{tabular}}
 \label{tab:risk-perception-reasons}   
\end{center}
\end{table*}
}

%% file: evaluation-rq1-real-quantitative-table.tex
{
\begin{table*}[h]
 \begin{center}
 \caption{\centering End-Users' perception 
of \textit{real} transfer risks.  The highest values 
are marked in \textbf{bold text}  (``\#'' = ``Number of Respondents'', ``\%'' = ``Percentage of Respondents'',  ``\textbf{\protect\circled{U}}'' = ``Unawareness'',   ``\textbf{\protect\circled{S}}'' = ``Surprising'',  ``\textbf{\protect\circled{SV}}'' = ``Severity'') 
}
  \resizebox{\linewidth}{!}{
 {\scriptsize
 \begin{tabular}{@{}|l|c|c|c|c|c|c|c|c|c|c|c|c|c|c|c|c|c|c|c|c|c|c|c|c|c|c|c|}
 \hline

\multirow{3}{*}{\rotatebox[origin=c]{90}{\textbf{Score}}} 
& \multicolumn{3}{c|}{\textbf{Contract Pause}}  & \multicolumn{3}{c|}{\textbf{User Blacklist}}  & \multicolumn{3}{c|}{\textbf{Contract Upgrade}} & \multicolumn{3}{c|}{\textbf{Transfer Fee Increase}}  & \multicolumn{3}{c|}{\textbf{Insufficient Funds}}   \\

& \textbf{\circled{U}} & \textbf{\circled{S}} & \textbf{\circled{SV}} 
& \textbf{\circled{U}} & \textbf{\circled{S}} & \textbf{\circled{SV}} 
& \textbf{\circled{U}} & \textbf{\circled{S}} & \textbf{\circled{SV}} 
& \textbf{\circled{U}} & \textbf{\circled{S}} & \textbf{\circled{SV}} 
& \textbf{\circled{U}} & \textbf{\circled{S}} & \textbf{\circled{SV}} 
\\


&  \multicolumn{1}{c|}{\textbf{\#/\%}} & \multicolumn{1}{c|}{\textbf{\#/\%}} & \multicolumn{1}{c|}{\textbf{\#/\%}}
&  \multicolumn{1}{c|}{\textbf{\#/\%}} & \multicolumn{1}{c|}{\textbf{\#/\%}} & \multicolumn{1}{c|}{\textbf{\#/\%}}
&  \multicolumn{1}{c|}{\textbf{\#/\%}} & \multicolumn{1}{c|}{\textbf{\#/\%}} & \multicolumn{1}{c|}{\textbf{\#/\%}}
&  \multicolumn{1}{c|}{\textbf{\#/\%}} & \multicolumn{1}{c|}{\textbf{\#/\%}} & \multicolumn{1}{c|}{\textbf{\#/\%}}
&  \multicolumn{1}{c|}{\textbf{\#/\%}} & \multicolumn{1}{c|}{\textbf{\#/\%}} & \multicolumn{1}{c|}{\textbf{\#/\%}}
\\
\hline 

1 & \textbf{27}/\textbf{24.5} & \textbf{32}/\textbf{29.1 }& 17/15.5 & 22/20.0 & 21/19.1 & 4/3.6 & 12/10.9 & 22/20.0 & 21/19.1 & \textbf{33}/\textbf{30.0} & \textbf{39}/\textbf{35.5} & 25/22.7 & \textbf{88}/\textbf{80.0 }&  \textbf{90}/\textbf{81.8} & \textbf{85}/\textbf{77.3} 
\\ \hline

2 & 23/20.9 & 16/14.5 & 11/ 10.0 & 22/20.0 & 18/16.4 & 12/10.9 & 22/20.0 &  18/16.4 & 12/10.9 & 21/19.1 & 11/10.0 & 14/12.7 & 12/10.9 & 11/10.0 & 10/9.1
\\ \hline

3 & 20/18.2 & 17/15.5  & 22/20.0 & 11/10.0 & 16/14.5 & 19/17.3 & 11/10.0 &  16/14.5 & 19/17.3 & 14/12.7 & 13/11.8 & 13/11.8 & 1/0.9 & 2/1.8 & 7/6.4 
\\ \hline

4 & 19/17.3 & 19/17.3 & 26/23.6 & 21/19.1 & 23/20.9 & 38/34.5 & 21/19.1 &  23/20.9 & \textbf{38}/\textbf{34.5} & 13/11.8 & 15/13.6 & 19/17.3 & 1/0.9  &   1/0.9 & 5/4.5 
\\ \hline

5 & 21/19.1 & 26/23.6 & \textbf{34}/\textbf{30.9} & \textbf{34}/\textbf{30.9} & \textbf{32}/\textbf{29.1} & \textbf{37}/\textbf{33.6} & \textbf{34}/\textbf{30.9} &  \textbf{32}/\textbf{29.1} & 37/33.6 & 29/26.4 & 32/29.1 & \textbf{39}/\textbf{35.5} & 8/7.3 & 6/5.5 & 3/2.7
\\ \hline

\textbf{Mean} & 2.85 & 2.92 & \textbf{3.45} & 3.21 & 3.25 & \textbf{3.84} & \textbf{3.44} & 3.28 & 3.04 & 2.85 & 2.91 & \textbf{3.3} & 1.45 & 1.38 & \textbf{1.46}
\\ \hline
\revc{\textbf{Std Dev}} & \revc{1.46} & \revc{\textbf{}1.56} & \revc{1.42} & \revc{\textbf{}1.55} & \revc{1.5} & \revc{1.12} & \revc{1.39} & \revc{\textbf{}1.46} & \revc{1.41} & \revc{1.6} & \revc{\textbf{}1.68} & \revc{1.6} & \revc{\textbf{}1.1} & \revc{1.0} & \revc{0.99}
\\ \hline
\textbf{Median}  & 3 & 3 & \textbf{4} & 3.5 & 3.5 & \textbf{4} & \textbf{4} & \textbf{4} & 3  &  3 &  3 &  \textbf{4} & \textbf{1} &  \textbf{1} &  \textbf{1} 
\\ \hline

   \end{tabular}}
  } 
 \label{tab:risk-perception}   
\end{center}
\end{table*}
}

%% file: evaluation-rq1-fake-quantitative-table.tex
{
\begin{table*}[tb!]
 \begin{center}
 \caption{\centering End-Users' perception 
of 
\textit{fictitious} (fake) transfer risks. The highest values 
are marked in \textbf{bold text}  (``\#'' = ``Number of Respondents'', ``\%'' = ``Percentage of Respondents'',  ``\textbf{\protect\circled{U}}'' = ``Unawareness'',   ``\textbf{\protect\circled{S}}'' = ``Surprising'',  ``\textbf{\protect\circled{SV}}'' = ``Severity'') 
}

\resizebox{\linewidth}{!}{
 {\scriptsize
 \begin{tabular}{@{}|l|c|c|c|c|c|c|c|c|c|c|c|c|c|c|c|c|c|c|c|c|c|c|c|c|c|c|c|}
 \hline

\multirow{3}{*}{\rotatebox[origin=c]{90}{\textbf{Score}}} 
& \multicolumn{3}{c|}{\textbf{Consortium Reject}}  & \multicolumn{3}{c|}{\textbf{Government Block}}  & \multicolumn{3}{c|}{\textbf{Receiver Reject}} & \multicolumn{3}{c|}{\textbf{Partial Funds}}  & \multicolumn{3}{c|}{\textbf{Gas Fee Increase}}   \\

& \textbf{\circled{U}} & \textbf{\circled{S}} & \textbf{\circled{SV}} 
& \textbf{\circled{U}} & \textbf{\circled{S}} & \textbf{\circled{SV}} 
& \textbf{\circled{U}} & \textbf{\circled{S}} & \textbf{\circled{SV}} 
& \textbf{\circled{U}} & \textbf{\circled{S}} & \textbf{\circled{SV}} 
& \textbf{\circled{U}} & \textbf{\circled{S}} & \textbf{\circled{SV}} 
\\

&  \multicolumn{1}{c|}{\textbf{\#/\%}} & \multicolumn{1}{c|}{\textbf{\#/\%}} & \multicolumn{1}{c|}{\textbf{\#/\%}}
&  \multicolumn{1}{c|}{\textbf{\#/\%}} & \multicolumn{1}{c|}{\textbf{\#/\%}} & \multicolumn{1}{c|}{\textbf{\#/\%}}
&  \multicolumn{1}{c|}{\textbf{\#/\%}} & \multicolumn{1}{c|}{\textbf{\#/\%}} & \multicolumn{1}{c|}{\textbf{\#/\%}}
&  \multicolumn{1}{c|}{\textbf{\#/\%}} & \multicolumn{1}{c|}{\textbf{\#/\%}} & \multicolumn{1}{c|}{\textbf{\#/\%}}
&  \multicolumn{1}{c|}{\textbf{\#/\%}} & \multicolumn{1}{c|}{\textbf{\#/\%}} & \multicolumn{1}{c|}{\textbf{\#/\%}}
\\
\hline 

1 & 9/8.2 & 10/9.1 & 14/12.7 & \textbf{32}/\textbf{29.1} & \textbf{40}/\textbf{36.4} & 12/10.9  & \textbf{40}/\textbf{36.4} &  \textbf{44}/\textbf{40.0} & \textbf{69}/\textbf{62.7} & \textbf{44}/\textbf{40.0} & \textbf{50}/\textbf{45.5} & \textbf{48}/\textbf{43.6} & 31/28.2 & \textbf{34}/\textbf{30.9} & 18/16.4 
\\ \hline

2 & 9/8.2 & 9/8.2 & 8/7.3 & 24/21.8 & 20/18.2 &  12/10.9  &  18/16.4 & 11/10.0 & 13/11.8 & 12/10.9  & 9/8.2 & 21/19.1 & 19/17.3 & 16/14.5 & 19/17.3
\\ \hline

3 & 13/11.8 & 10/9.1 & 15/13.6 & 13/11.8 & 13/11.8 & 16/14.5 & 12/10.9 & 14/12.7 & 17/15.5 & 12/10.9 &  11/10.0 & 15/13.6 & 15/13.6 & 16/14.5  & 20/18.2
\\ \hline

4 & 23/20.9 & 19/17.3 & 32/29.1 & 11/10.0 & 7/6.4 & 24/21.8 & 15/13.6 & 20/18.2 & 7/6.4 & 14/12.7 & 13/11.8 & 11/10.0 & 12/10.9 & 10/9.1 & 20/18.2
\\ \hline

5 & \textbf{56}/\textbf{50.9} & \textbf{62}/\textbf{56.4} & \textbf{41}/\textbf{37.3} & 30/27.3 & 30/27.3 & \textbf{46}/\textbf{41.8} & 25/22.7 & 21/19.1 & 4/3.6 & 28/25.5 & 27/24.5 & 15/13.6 & \textbf{33}/\textbf{30.0} & \textbf{34}/\textbf{30.9} & \textbf{33}/\textbf{30.0}
\\ \hline

\textbf{Mean} & 3.98 & \textbf{4.04} & 3.71 & 2.85 & 2.7 & \textbf{3.73} & \textbf{2.7} & 2.66 & 1.76 & \textbf{2.73} & 2.62 & 2.31 & 2.97 & 2.95 & \textbf{3.28}
\\ \hline

\revc{\textbf{Std Dev}} & \revc{1.31} & \revc{1.35} & \revc{\textbf{1.37}} & \revc{1.6} & \revc{\textbf{1.65}} & \revc{1.39} & \revc{\textbf{1.61}} & \revc{1.6} & \revc{1.15} & \revc{1.68} & \revc{\textbf{1.7}} & \revc{1.46} & \revc{1.62} & \revc{\textbf{1.65}} & \revc{1.47}
\\ \hline

\textbf{Median}  & \textbf{5} & \textbf{5}  & 4 & 2 & 2 & \textbf{4} & 2 & \textbf{2.5} & 1 & \textbf{2} & \textbf{2} & \textbf{2} & \textbf{3} & \textbf{3} & \textbf{3}
\\ \hline

   \end{tabular}}
} 
 \label{tab:fake-risk-perception}   
\end{center}
\end{table*}
}

%% file: evaluation-rq2-table.tex
{
\setlength{\tabcolsep}{1.8pt} 
\begin{table}[tb!]
\centering
 \caption{\centering \rev{End-Users' perception of discoverability (``Discov.'') and understandability (``Underst.'') 
 		of USDT transaction outcomes using the MetaMask UI.} 
 	The highest number/percentage of participants and scores are marked in \textbf{bold}
	(``\#'' = ``Number of Respondents'', ``\%'' = ``Percentage of Respondents'') 
}
 {
 \begin{tabular}{@{}|l|c|c|c|c|c|c|c|c|c|@{}}
 \hline

& \multicolumn{2}{c|}{\textbf{\makecell{Full amount \\ transferred}}} 
& \multicolumn{2}{c|}{\textbf{\makecell{Reduced amount \\ transferred}}} 
& \multicolumn{2}{c|}{\textbf{Transfer failure}}
\\

& \textbf{Discov.} & \textbf{Underst.}
& \textbf{Discov.} & \textbf{Underst.}
& \textbf{Discov.} & \textbf{Underst.}
\\

\textbf{Score}
&  \revc{\textbf{\#/\%}} & \revc{\textbf{\#/\%}} &  \revc{\textbf{\#/\%}} & \revc{\textbf{\#/\%}} &  \revc{\textbf{\#/\%}} & \revc{\textbf{\#/\%}} 
\\
\hline

1 & \rev{11/10.0} & \rev{5/4.5} & \rev{\textbf{40}/\textbf{36.4}} & \rev{\textbf{36}/\textbf{32.7}} & \rev{\textbf{42}/\textbf{38.2}} & \rev{\textbf{43}/\textbf{39.1}}
\\ \hline

2 & \rev{3/2.7} & \rev{5/4.5} & \rev{18/16.4} & \rev{20/18.2} & \rev{9/8.2} & 10/9.1
\\ \hline

3 & \rev{7/6.4} & \rev{9/8.2} & \rev{15/13.6} & \rev{19/17.3} & \rev{14/12.7} & \rev{18/16.4}
\\ \hline

4 & \rev{29/26.4} & \rev{17/15.5} & \rev{15/13.6} & \rev{21/19.1} & \rev{19/17.3} & \rev{16/14.5}
\\ \hline

5 & \rev{\textbf{60}/\textbf{54.5}} & \rev{\textbf{74}/\textbf{67.3}} & \rev{22/20.0} & \rev{14/12.7} & \rev{26/23.6} & \rev{23/20.9}
\\ \hline

\textbf{Mean} & 4.13 & 4.36 & 2.65 & 2.61 & 2.8 & 2.69
\\ \hline

\revc{\textbf{Std Dev}} & \revc{1.27} & \revc{1.11} & \revc{1.57} & \revc{1.43} & \revc{1.65} & \revc{1.6}
\\ \hline

\textbf{Median} &   5 &   5 &   2 &   2 &   3 &   3 
\\ \hline

   \end{tabular}}
 \label{tab:outcome-perception}   
\end{table}
}

%% file: evaluation-rq2.tex

{
	\begin{table*}[tb!]
		\begin{center}
			\caption{
				\centering 
				Reasons for end-users' perception of discoverability (``Disc.'') and understandability (``Underst.'') \reva{USDT} transaction outcomes using the \reva{MetaMask UI}.  The  highest number/percentage of participants and reasons are marked in \textbf{bold text} and \textit{wrong reasons} are in 
				\textbf{\textit{italicized and bold text}}
				(``\#'' =  ``Number of Respondents'', ``\%'' = ``Percentage of Respondents'' ).
				\newrevise{Exclusions for ambiguous/misanswered/incorrect answers and multiply-categorized answers lead to percentage totals and absolute totals deviating from 100\% and 110 responses.}
			}
			{\scriptsize
				
				\renewcommand{\arraystretch}{1} 
				\setlength{\tabcolsep}{3pt} 
				\begin{tabular}{@{}|l|l|l|c|c|l|c|c|}
					\hline
					
					& & \multicolumn{6}{c|}{\textbf{Most Prevalent Reasons for Users' Perception of Transaction Outcomes}} \\
					
					&  &  \multicolumn{3}{c|}{\textbf{Negative Reasons}} & \multicolumn{3}{c|}{\textbf{Positive Reasons}} \\

					\textbf{Metric} 
					& \textbf{Outcome}  & 
					\multicolumn{1}{c|}{\textbf{Examples} (Why \textit{Not} Discoverable/Understandable?)}  & \textbf{\#} & \textbf{\%} & 
					\multicolumn{1}{c|}{\textbf{Examples} (Why Discoverable/Understandable?)}  & \textbf{\#} & \textbf{\%} \\
					\hline{}

						
						\multirow{3}{*}{\rotatebox[origin=c]{0}{\shortstack[c]{\textbf{Disc.}}}}
						& \textbf{Full} & 
						N/A (no meaningful subcategories emerged)
						& 2 & 1.8 &
						\textbf{understandable flow},  \&
						displayed information
						& \textbf{87} & \textbf{79.1}
						\\ \cline{2-8}
						
						& \textbf{Reduced} & 
						\textbf{misleading/unexplained user flow}
						\& unexpected/
\textit{\textbf{impossible fee}}
						& \textbf{50} & \textbf{45.5} & 
						fee is expected \& the UI flow is understandable
						& 21  & 19.1
						\\ \cline{2-8}

						& \textbf{Failure} & 
						\textbf{contradictory flow} \&
						no failure reason 
						&  \textbf{47} & \textbf{42.7} & 
						failure report after occurrence \& no failure cost 
						&  34 & 30.9 
						\\ \hline
						
						\multirow{3}{*}{\rotatebox[origin=c]{0}{\shortstack[c]{\textbf{Underst.}}}}
						& \textbf{Full} & 
						\revc{m}ore details needed in the user flow &  9 & 8.2 & 
						\textbf{sufficient funds } \& \revc{u}ser flow is explanatory  
						& \textbf{92} & \textbf{83.6}
						\\ \cline{2-8}
						
						& \textbf{Reduced} &
						\textbf{user flow is uninformative, contradictory},  or unexpected
						& \textbf{51} & \textbf{46.4} & 
						expected fee \& contract parameters knowledge
						& 14 & 12.7
						\\ \cline{2-8}
						
						& \textbf{Failure} & 
						\textbf{no failure reasons}  \&  no indication of possibility,
						& 
						\textbf{50} & \textbf{45.5} & 
						possibly blacklisted \& parameters not met  &  24 & 21.8
						\\  \hline
						
				\end{tabular}}
				\label{tab:outcome-perception-reasons}   
				
			\end{center}
		\end{table*}
	}

\begin{figure}[t!]
	\begin{newbox1}
	\centering
	\includegraphics[width=0.5\linewidth]{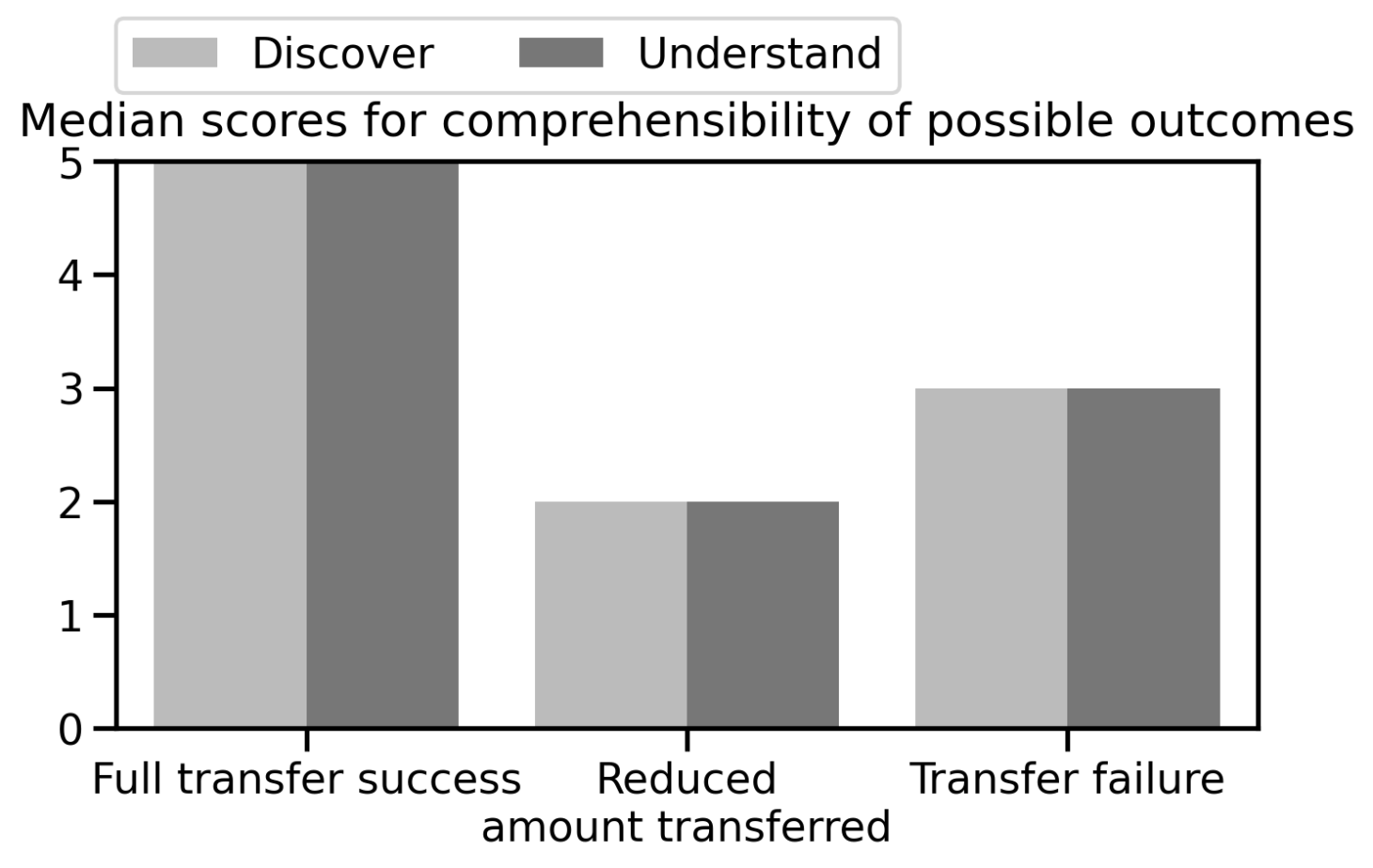}
	\caption{End-user knowledge of smart contracts}
	\label{fig:metamask-comprehensibility}
    \Description{The figure shows a grouped bar plot titled "Median scores for comprehensibility of possible outcomes.
		The groups are for the "Discover" metric and "Understand" metric.
		The "Full transfer success" bars report a median of 5 for both metrics.
		The "Reduced amount transferred" bars report a median of 2 for both metrics.
		The "Transfer failure" bars report a median of 3 for both metrics.
    }
	\end{newbox1}
\end{figure}

\noindent
\textbf{RQ2 \revc{Understanding} of User Interface:} We investigate \revc{end-user understanding} of the MetaMask (USDT) smart contract user interface 
in communicating transaction outcomes i.e.,  
\textit{full amount transferred},  a \textit{reduced amount transfered} and \textit{transfer failure} (no amount transferred). 
We inspect if these outcomes are discoverable  and understandable 
for end-users.   
\revc{\autoref{fig:metamask-comprehensibility},} \autoref{tab:outcome-perception} and \autoref{tab:outcome-perception-reasons} highlight our results. 

\vspace{0.5mm}
\noindent
\textbf{\textit{Risky Transaction Outcomes: }}
\revc{\autoref{tab:outcome-perception}} shows that \textit{end-users find the UI flow of MetaMask to be insufficient for discovering or understanding 
risky transaction outcomes}
(i.e.,  reduced amount transferred and transfer failure)\revc{, with reasons highlighted in \autoref{tab:outcome-perception-reasons}}.  
For instance,  
up to 46.4\% (51)  users find it difficult to 
comprehend 
the ``reduced amount'' transferred outcome 
because the flow is misleading
\revc{(e.g., ``The transaction should be 5 usdt plus fees, no reason for it be just 5 usdt'')},
uninformative
\revc{(e.g., ``It does not give a warning about this'')}
or contradictory to the outcome
\revc{(e.g., ``The transactions actually shows that 5 USDT will be sent'')}.
In contrast, over twice as many (2X)
users could comprehend 
the full amount transferred outcome: 
82.7\% (91) 
of users understand the successful outcome,  while only
31.8\% (35) 
of users understand the reduced amount transferred outcome\footnote{The actual absolute numbers are provided in \autoref{tab:outcome-perception}.  Note that \autoref{tab:outcome-perception-reasons} only reports the coded qualitative responses,  hence the percentage totals and absolute totals may deviate from 100\% and 110 responses due to the exclusion of ambiguous/incorrect answers. 
}.
%

\revc{\autoref{fig:metamask-comprehensibility}} also shows that the full amount transferred flow has a high comprehensibility score (median score of five),  
while comprehensibility of the other outcomes is low (median scores of two and three).  
\revc{
	As seen in the overview of reasons provided in \autoref{tab:outcome-perception-reasons}, we note that
	for the transfer failure outcome, this is primarily due to the user flow not explaining the reasons behind the failure
	(e.g., ``There were no prior knowledge shared as to whether an upgrade was being undertaking, also the account had sufficient balance to ensure that the transaction goes through successfully and finally there were no signs of the account being blocked.'')
}


\begin{result}
Most (82.7\% of) end-users find the MetaMask UI to be sufficient 
to comprehend \revc{the}
full amount transferred outcome, 
while fewer users could comprehend the risky transaction outcomes (31.8\% and 35.5\%).
\end{result}

%
%
%
%
%

\vspace{0.5mm}
\noindent
\textbf{\textit{Statistical Analysis: }}
We found \textit{no statistically significant differences in end-user ability to discover or understand potential outcomes from the MetaMask UI}, 
across tested groups of experts (\rev{self-rated} programmers and Web3 proficiency) versus non-experts.  
The poor 
comprehension is thus not due to the lack of \rev{self-rated} proficiency. 

\begin{result}
\rev{Self-rated} programming/Web3 proficiency does not affect end-user (in)comprehension of 
transfer outcomes in the UI. 
\end{result}

%

%
%
%
%
%



%
%

%% file: evaluation-rq3.tex

\begin{figure}[tbp!]
	\centering
	\begin{subfigure}{0.35\textwidth}
		\includegraphics[width=\textwidth]{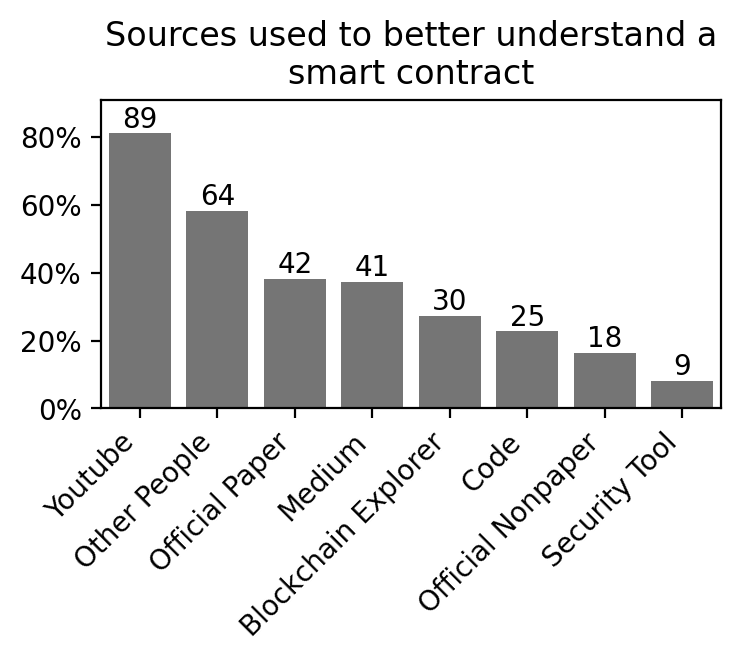}
		\caption{Information sources for users}
		\label{fig:sources-overview}
	\end{subfigure}
	\begin{subfigure}{0.35\textwidth}
		\includegraphics[width=\textwidth]{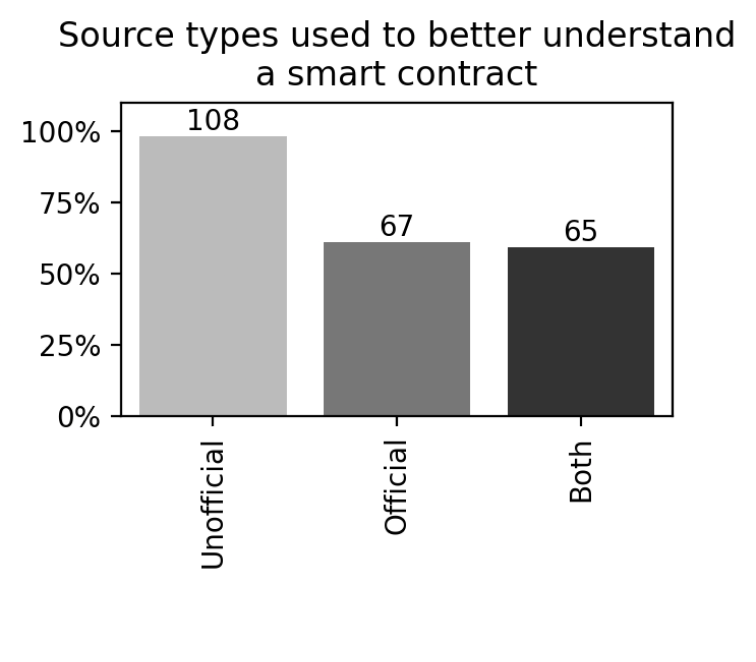}
		\caption{Type of information source}
		\label{fig:sources-by-type}
	\end{subfigure}
	\caption{End-users' information sources}
	\Description{The figure shows two bar chart graphs..

	Graph (a) is titled “Sources used to better understand a smart contract. The Y-axis is the percentage score for the number of users, and the X axis is the information source. Each bar has the number of participants labelled on top of it. The scores for graph (a) are as follows:
	Youtube: 81\% (89 participants)
	Other People: 58\% (64 participants)
	Official Paper: 38\% (42 participants)
	Medium: 37\% (41 participants)
	Blockchain Explorer: 27\% (30 participants)
	Code: 23\% (25 participants)
	Official Nonpaper: 16\% (18 participants)
	Security Tool: 8\% (9 participants)
	
	Graph (b) is titled “Sources types used to better understand a smart contract. The Y-axis is the percentage score for the number of users, and the X axis is the information source type. Each bar has the number of participants labelled on top of it. The scores for graph (b) are as follows:
	Unofficial: 98\% (108 participants)
	Official: 61\% (67 participants)
	Both: 59\% (65 participants)
	}
\end{figure}\

\noindent 
\revc{
\textbf{RQ3 Smart Contract Understanding:}
In order to investigate how end-users understand smart contracts, we examine sources of information they use (\textit{see} ~\autoref{fig:sources-overview},~\autoref{fig:sources-by-type}). 
In particular, we ask participants to provide the sources of information they use to educate themselves about the behavior 
of smart contracts they interact with. 
We also ask them to rate their perceived levels of knowledge before and after using a smart contract (\textit{see} ~\autoref{fig:perception-informed}), 
their trust that smart contracts will behave as they expect (\textit{see} ~\autoref{fig:perception-trust}) and their perceived 
ability to anticipate the behavior of a smart contract (\textit{see} ~\autoref{fig:perception-anticipate}).
}

\noindent 
\textbf{\textit{Sources of Information:}} 
\textit{All respondents employ external sources,  besides the smart contract wallet UI, to educate themselves on the behavior of the smart contract}.  
\revc{\autoref{fig:sources-by-type}
shows that while}
almost all (98.2\%, 108) respondents used an unofficial source of information (e.g., YouTube videos \revc{(\textit{see} \autoref{fig:sources-overview}))}, \revc{under} two-thirds (60.9\%, 67)  of 
respondents employ an official source (e.g., source code or whitepaper). 
%
Overall,  
59.1\% (65) of respondents employ both official and unofficial information sources. 
This implies that end-users prefer unofficial sources 
to learn about smart contracts.  Thus, we recommend that unofficial modes (e.g., videos and blogposts) are also used to communicate smart contract 
implementation. 

Additionally, 
respondents with relevant \rev{self-rated} proficiencies used official sources more frequently: 52\% of non-programmers, compared to 63.5\% of programmers. Likewise, 
52.6\% of users without high Web3 proficiency used official sources, compared to 79.4\% of users with Web3 proficiency. 


\begin{result}
\newrevise{
Most (98.2\% of) end-users 
employ unofficial sources of information (e.g., Youtube) for self-education on smart contract behaviors. 
}
\end{result}

%% file: evaluation-rq3ii.tex

\begin{figure}[t!]
	\centering
	\includegraphics[width=0.6\linewidth]{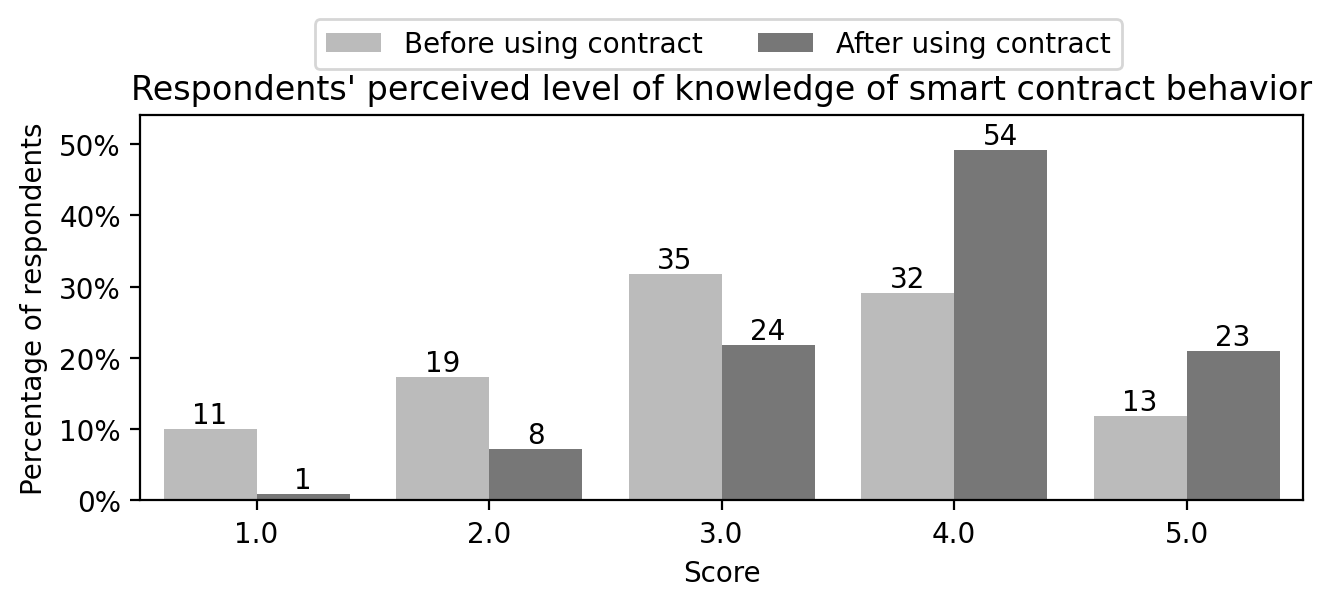}
	\caption{End-user knowledge of smart contracts}
	\label{fig:perception-informed}
    \Description{The figure shows a bar chart with the title “Respondents’ perceived level of knowledge of smart contract behavior”. The Y axis represents the percentage of respondents. The X axis represents the score given, ranging from 1 to 5.  Each bar has the number of participants labelled on top of it. Each score has 2 bar charts in its group. A legend describes the first bar in the group as being “Before using contract”, and the second bar in each group as being “After using contract”.

    The scores are as follows.
    Score 1: 11 (10\%) before, 1 (1\%) after
    Score 2: 19 (17\%) before, 8 (7\%) after
    Score 3: 35 (32\%) before, 24 (22\%) after
    Score 4: 32 (29\%) before, 54 (49\%) after
    Score 5: 13 (12\%) before, 23 (21\%) after
    }
\end{figure}

\vspace{0.5mm}
\noindent
\textbf{\textit{Level of Knowledge:}} 
We found that \textit{respondents are (19\%) more informed about a smart contract's behavior 
after using the  contract versus before using it, on average (3.8 vs. 3.2 mean scores). }
\autoref{fig:perception-informed} shows 
the level of 
informedness (least to most informed) of users based on a five-point Likert scale. 
The difference in the knowledge level \textit{before} and \textit{after}  contract usage is 
statistically significant (W = 384, p = 6.33E-8 under the Wilcoxon signed rank test).
This suggests that users become more knowledgeable through 
experience. 

\begin{figure}[t!]
\centering
\begin{subfigure}{0.35\textwidth}
    \includegraphics[width=\textwidth]{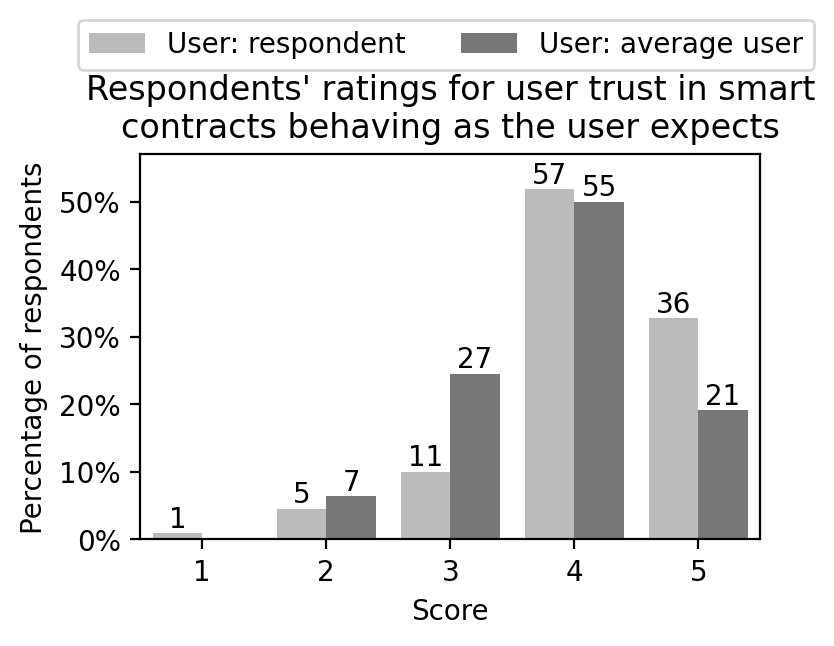}
    \caption{Trust in Smart Contracts}
    \label{fig:perception-trust}
\end{subfigure}
\begin{subfigure}{0.35\textwidth}
    \includegraphics[width=\textwidth]{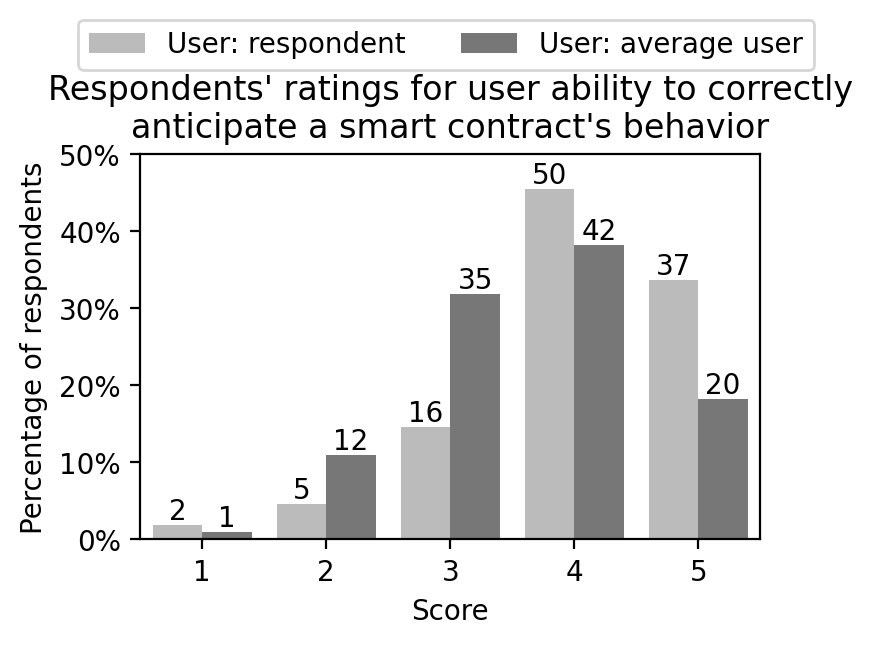}
    \caption{Anticipate Contract Behavior}
    \label{fig:perception-anticipate}
\end{subfigure}
\caption{\centering End-user trust and anticipation in smart contract behavior\revc{, rating both themselves and their perceptions of what they see as the average user.}}
\Description{The figure shows two bar charts. The Y axes for both represent the percentage of respondents. The X axes for both represent the score given, ranging from 1 to 5. Each score has 2 bar charts in its group.  Each bar has the number of participants labelled on top of it. In both bar charts, legend describes the first bar in the group as being for “User: respondent”, and the second in each group as being for “User: average user”.

Graph (a) is titled “Respondents’ ratings for user trust in smart contracts behavirng as the user expects”. The scores are as follows:
Score 1: 1 (1\%) for respondent, 0 (0\%) for average user
Score 2: 5 (5\%) for respondent, 7 (6\%) for average user
Score 3: 11 (10\%) for respondent, 27 (25\%) for average user
Score 4: 57 (52\%) for respondent, 55 (50\%) for average user
Score 5: 36 (33\%) for respondent, 21 (19\%) for average user

Graph (b)  is titled “Respondents’ ratings for user ability to correctly antiicpate a smart contract’s behavior”. The scores are as follows:
Score 1: 2 (2\%) for respondent, 1 (1\%) for average user
Score 2: 5 (5\%) for respondent, 12 (11\%) for average user
Score 3: 16 (15\%) for respondent, 35 (32\%) for average user
Score 4: 50 (45\%) for respondent, 42 (38\%) for average user
Score 5: 37 (34\%) for respondent, 20 (18\%) for average user
}
\end{figure}

\vspace{0.5mm}
\noindent
\textbf{\textit{Statistical Analysis (Level of Knowledge): }}
We analyzed the effects of \rev{self-rated} Web3 and programming proficiency (separately) on levels of knowledge before and after smart contract use. \textit{There were 
no statistically significant differences for either proficiency}. Users 
in general seem to benefit from 
experiences with the smart contract. 



\vspace{0.5mm}
\noindent
\textbf{\textit{Trust:
}}
We found that \textit{users 
trusted their own expectation of smart contract behavior (8\%) more than they believed the average smart contract user would trust hers, on average
(mean 4.1 versus 3.8, see \autoref{fig:perception-trust}).} 
This difference is statistically significant (W = 547, p = 0.0022 under the Wilcoxon signed rank test). 
Additionally, \rev{self-rated} programming or high Web3 proficiency had no statistically significant effect on these scores.
This suggests that 
none of the 
proficiencies influences end-user trust
in smart contract behavior.


\vspace{0.5mm}
\noindent
\textbf{\textit{Behavior Anticipation:}} 
\textit{
Users believe they are able to anticipate smart contract's behavior 11\% better than the average user (mean 4.0 vs. 3.6 in \autoref{fig:perception-anticipate})}.
This difference was also statistically significant (W = 593.5, p = 1.9E-4 under Wilcoxon's signed rank test) and \autoref{fig:perception-anticipate} shows that the difference holds across most scores.
Similar to trust scores, \rev{self-rated} programming proficiency and high Web3 proficiency had no statistically significant effect.
This suggests that neither programming nor Web3 proficiency influence users' confidence in their ability to anticipate smart contract behaviors.


\begin{result}
\newrevise{Most end-users believe they are (up to 11\%) more  
anticipatory and trusting of smart contracts  behaviors 
than the average user.}
\end{result}

\vspace{0.5mm}
\noindent
\textbf{\textit{Statistical Analysis (Behavior Anticipation and Risk Perception): }}
%
\textit{End-users' belief in their ability to anticipate smart contract behavior
does not imply their awareness of 
USDT transfer risks.}
There was no significant difference in high and low behavior anticipation users for 
USDT transfer risks.  

\begin{result}
A high self-assessment in anticipating smart contract behavior 
does not imply that a user is more aware of USDT
transfer risks than other users.
\end{result}




%% file: evaluation-rq4.tex

\noindent
\textbf{RQ4 Generalizability of Risks:} 
We extracted all known ERC-20 contracts (78, excluding USDT) out of the top 500 Ethereum recipient addresses (
see \autoref{fig:source-code-methodology}). 
We analyzed their source code for (a) the existence/prevalence of the five (5) real transfer risks examined in \textbf{RQ1} as well as the existence of a transfer fee,
(b) the existence and prevalence of other (new) risks,  and (c) the effectiveness of our experimental string-matching heuristic in automatically detecting three 
of the five risks in these 78 contracts (see \autoref{tab:generalize-features} and \autoref{tab:auto-detection-scores}). 

\vspace{0.5mm}
\noindent
\textbf{\textit{Prevalence of Risks: }}
\newrevise{We found occurrences of all (five) transfer risks across examined smart contracts}
(\textit{see} \autoref{tab:generalize-features}).  
Some of the least prevalent risks are the most severe and surprising for users. 
In particular, contract upgrade and user blacklist are the least prevalent (1.3\%) 
\reva{but rated most severe and surprising in \textbf{RQ1}.  }
In contrast,  the most prevalent transfer risk (contract pause: 19.2\%)
is the least severe and least surprising transfer risk 
(\textit{see} \textbf{RQ1}). 
Overall, these results imply 
the need to expose and 
\reva{clearly}
explain rare risks to end-users.

\begin{result}
The (five) transfer risks 
in USDT are present in the examined top ERC-20 contracts and up to 19.2\% prevalent.
\end{result}

\vspace{0.5mm}
\noindent
\textbf{\textit{Other Risks: }} \textit{In addition to the examined transfer risks, we discovered 
three (3) other 
risky and unexpected features with potentially significant impact} (see  \autoref{tab:generalize-features}). 
These risks are either beyond a user's ability to transfer tokens, or else not present in the USDT contract.
\textit{Arbitrary Mint} allows one or more authorized addresses to arbitrarily add new tokens to the total token supply. This 
\reva{may be misused for 
price manipulation via supply increase.}
\textit{Transfer Limit Change} allows an authorized address to limit the amount of tokens transferred by other users per transaction. This
\reva{potentially limits}
users from selling tokens during time-sensitive periods,
\reva{making} them more vulnerable to
Pump and Dump
scheme\reva{s}
~\cite{PumpAndDump}.
\textit{Destroy(ing) User Funds} allows an authorized address to destroy any particular user's token holdings without justification.
These risks may reduce trust
and \revc{understanding} of
contract\reva{s}
\reva{as}
they are not
exposed via the \reva{UI}.
These results motivate the need to further study users' risks in smart contract usage.


\begin{result}
We found three additional risks in ERC-20  smart contracts with potentially significant impact on \revc{end-user understanding} and trustworthiness.
\end{result}

\vspace{0.5mm}
\noindent
\textbf{\textit{Effectiveness of Automated Detection: }}
\autoref{tab:auto-detection-scores} shows that 
\textit{our proposed risk detection approach is effective especially in detecting Contract Pause.} 
We recall that our detection is based on partial string matching of function names (\textit{see} \autoref{sec:setup}). 
\textit{Contract Pause} detection (F1=80\%) performed well, but 
\textit{User Blacklist} (F1=100\%) and 
\textit{Contract Upgrade} (F1 not applicable) tests were only evaluated on one positive case. 
Overall, this result shows that the automatic detection of transactions risks is feasible.
It also motivates developing more effective methods, e.g.,  using program analysis techniques.


\begin{result}
Our automated risk detection 
was effective in detecting two out of three attempted transfer risks.   
\end{result}

{
\begin{table}[tb!]
  \centering
  \caption{
    \centering
Prevalence of Risky Features in  
  the top 
  ERC-20 contracts.
  (``\#'' =  ``Number of Contracts'',
    ``P": Risk identified \textit{prior} to analysis,  ``D": Risk identified \textit{during} analysis) 
  }
  {
  \begin{tabular}{|l|c|c|l|l|}
  \hline

  \textbf{Feature} & \textbf{\#} & \textbf{\%} & \textbf{Example} & \textbf{P / D}  \\ \hline
  Insufficient Funds    & 78   & 100\%   & MaticToken    & P  \\ \hline
  Contract Pause        & 15   & 19.2\%  & MANAToken     & P  \\ \hline
  User Blacklist        & 1    & 1.3\%   & KOKContract   & P  \\ \hline
  Contract Upgrade      & 1    & 1.3\%   & AVINOCToken   & P  \\ \hline
  Transfer Fee Increase & 1    & 1.3\%   & Shibnobi      & P  \\ \hline 
  Transfer Fee          & 4    & 5.1\%   & SaitamaInu    & P  \\ \hline
  Arbitrary Mint        & 21   & 26.9\%  & Stronger      & D  \\ \hline
  Transfer Limit Change & 2    & 2.6\%   & KishuInu      & D  \\ \hline
  Destroy User Funds    & 1    & 1.3\%   & SmartToken    & D  \\ \hline
  \end{tabular}
}  
  \label{tab:generalize-features}
\end{table}
}



{
\setlength{\tabcolsep}{5pt} 
\begin{table}[tb!]
  \caption{\centering Effectiveness of automated transfer risk detection}
  \begin{tabular}{|l|l|l|l|l|l|l|l|l|}
  \hline
  \textbf{Feature} & \textbf{Accuracy} & \textbf{F1 score} & \textbf{Precision} & \textbf{Recall} & \textbf{True} & \textbf{False} & \textbf{True} & \textbf{False}              \\ 
   &   & &  &  & \textbf{Positive} & \textbf{Positive} & \textbf{Negative} & \textbf{Negative}              \\ \hline
  Contract Pause   & 0.92 & 0.80 & 0.80 & 0.80 & 12 & 3 & 60 & 3                  \\ \hline
  User Blacklist   & 1    & 1    & 1    & 1    & 1  & 0 & 77 & 0                   \\ \hline
  Contract Upgrade & 0.99 & N/A & N/A & 0 & 0 & 0 & 77 & 1                  \\ \hline
  \end{tabular}
  \label{tab:auto-detection-scores}
\end{table}
}

%% file: discussion.tex
\section{Discussion and Future Work}
\label{sec:discussion}

\revc{
In sections \ref{sec:discussion:user-edu} and \ref{sec:discussion:erroneous-beliefs},
we address takeaways which are relevant to both end-users and designers of smart contracts and wallet interfaces.
	In section \ref{sec:discussion:future-work},
	we detail some potential directions for future research, centered around addressing the core issue of end-user explainability of smart contracts.
	The majority of these directions are relevant for designers.
}


%

\revc{\subsection{User Education and Risk Perception}
\label{sec:discussion:user-edu}}
\noindent
\textbf{Insufficient information sources:}
{\em While all respondents in the study attempted to educate themselves on smart contracts they use, they still  
lacked sufficient risk comprehension} (\textit{see} \textbf{RQ3)}.
We observed that 
most resources focus on communicating the goals of the respective project, 
but they 
do not provide the concrete implementation of the smart contract, which dictates the 
rules of interaction with end-users. For instance, the official Tether whitepaper~\cite{TetherWhitepaper} made no 
reference to 
blacklisting, pausing and contract upgrade. 
\rev{
Additionally,  previous work \cite{TokenScope} investigating inconsistent token behavior noted that of 752 whitepapers of inconsistent tokens, only 31 (4.1\%) included detailed token behavior descriptions.
}
\rev{
Some developer-facing resources make transfer risk features explicit, e.g., 
OpenZeppelin 
\cite{OpenZeppelinContracts} and 
Alchemy 
\cite{AlchemySmartContracts}. However, 
\reva{these resources do not target 
end-users (e.g., non-developers). }
}

To improve the users'  comprehension of smart contracts, 
we recommend  the source code should be made publicly available.  At present, only bytecode is available on the 
Ethereum blockchain and not all projects upload their source code for public inspection.
Most (430 of the top 500) recipient addresses have bytecode available, but 62 (14.4\%) of those do not have verified source code on Etherscan.
End-users desire to 
understand the source code of the smart contracts: 
Almost one in every four (19 /85 = 22.4\%)  respondents
with programming ability read smart contract 
source code
(\textbf{RQ3}).

\vspace{0.5mm}
\noindent
\textbf{Confidence/skills vs. Risk comprehension:}
{\em Neither \rev{self-rated} programming/Web3 proficiency nor high ability to anticipate smart contract behavior 
	have statistical significance on risk comprehension (see \textbf{RQ1})}. 
We thus suggest that many users may 
\reva{have}
an inflated sense of confidence, as 79.1\% (87) users rated themselves 
as having high behavior anticipation ability yet generally performed badly in risk awareness.
Similarly, users do no  seem to be effectively engaging relevant skills (\rev{self-rated} programming ability, Web3 proficiency) 
to better comprehend these risks (\textit{see} statistical tests in \textbf{RQ1}).
We believe further research is needed to uncover the reasons behind inflated self-confidence, and why users  
may ineffectively use relevant skills  for smart contract comprehension.

\revc{\subsection{Common Erroneous Beliefs}
\label{sec:discussion:erroneous-beliefs}}
\noindent
\revc{\textbf{Tether can communicate directly to users:}}
{\em Many respondents erroneously believed that Tether would be able to communicate changes to all affected users}. 
For instance, regarding the upgrading capability, one user believed that ``{\em ...the company Tether Limited was supposed 
to inform me about upgrading the smart contract}".
The ability for Tether to directly notify a user by their MetaMask wallet does not exist in the USDT contract. Respondents, however, frequently implied 
otherwise. For example, 15 (13.6\%)  respondents noted that a fee increase would only be surprising if no notice was given. 
Besides, 
12 (10.9\%) respondents claimed that the blacklisting was surprising due to no reason 
given 
for being 
blacklisted. 
\reva{We note it is not possible,  at present, }to communicate such reasons via the wallet interface. 
We thus expect that measuring and improving user comprehension of smart contract is a fertile ground for \reva{future research}.

\vspace{0.5mm}
\noindent
\revc{\textbf{Tether is decentrally governed:}}
{\em In the absence of appropriate tutorial, users may believe that blockchain properties, such as decentralization, 
transfer over to the projects held on the blockchain.} For instance, 58 (52.7\%) study respondents incorrectly 
believed that USDT is ``governed in a decentralized manner". Of the 58 (52.7\%) respondents who themselves 
owned USDT, 28 (48.3\%) held the same incorrect belief. Alternatively, many users seem 
to anchor expectations on their experiences with centralized institutions. For example, 19 (17.3\%) respondents 
rated themselves as aware of a potential USDT fee increase as ``{\em companies/financial institutions increasing 
transaction charges is normal}".
\revc{Interestingly, this anchoring in some of our participants is consistent with the ``bank bias" found for non-users 
	in prior work~\cite{mentalmodel-soups}.} We expect that tutorial is necessary for smart contract comprehension.

\revc{
	\subsection{Future Research Directions in Explainable Smart Contracts}
	\label{sec:discussion:future-work}
	This work demonstrates 
the \textit{explainability gap} between smart contract end-users and designers.  
In particular, this gap refers to how end-users poorly understand smart contracts, despite using them. As such, we lay out the following research directions for potential future work in both understanding and bridging this gap.
}

\vspace{0.5mm}
\noindent
\revc{
	\textbf{The explainability gap in other smart contracts.} \textit{More work is needed to confirm that this gap is present for other smart contracts, beyond the ERC-20 smart contracts.} Due to the standardization of the ERC-20 specification, we were able to generalize our insights from USDT.
However,  
some heavily-used smart contract 
are not as standardized or well-specified as the ERC-20 smart contracts (e.g., the MakerDAO Vat contract~\cite{MakerDAOVat}). 
Thus,  there is a need to investigate end-user understanding of 
such heterogenous smart contracts.
}


\vspace{0.5mm}
\noindent
\revc{
	\textbf{Understanding end-users and factors which affect their understanding.}
	\textit{
Smart conctract understanding is not homogeneous among end-users. 
Hence, more research is needed 
		to determine the level of user understanding and factors influencing understanding level.} 
	As discussed, our results reveal a tension wherein some users incorrectly attributed centralized properties 
	to USDT while others incorrectly attributed decentralized properties to it. This is consistent with prior 
	work~\cite{mentalmodel-soups}, although our focus is on actual end-users. 
	In follow-up work, we would examine 
the distinct factors causing different levels 
	of end-user understanding. 
	Above shedding descriptive light on the space of end-users, such work may also help designers to take informed 
	decisions 
	in order to increase overall end-user understanding.
}

\vspace{0.5mm}
\noindent
\revc{
	\textbf{Increasing explainability in light wallet user interfaces}.
	\textit{Through identification of common patterns in source code, wallet interfaces might be able to present additional salient information to the end-user.}
	As seen through both our work and TokenScope~\cite{TokenScope}, a significant portion of ERC-20 contracts extend or deviate from the specification in common ways.
	While this deviation is observable in the source code, we note that even among our participants with programming experience, more than three quarters (77.6\%) have never read any smart contract source code.
	Thus, a potential research direction is increasing the scope of explanations downstream in the wallet interfaces. 
	While our study is focused on MetaMask due to its overwhelming market dominance, we also 
	inspected two other high-usage wallets 
	(Trust Wallet~\cite{TrustWallet} and OKX Wallet~\cite{OKXWallet},
	each with one million users in the Chrome Web Store~\cite{ChromeOKXWallet,googleTrustWallet}) \revc{through} the same YUSDT-based procedure (\textit{see} \autoref{ssec:setup:user_study_design}).
	All three wallets failed to explain these transfer risks in a manner similar to that shown in \autoref{fig:sequence-diagram}.
	In \textbf{RQ4}, we show that a simple string matching approach is effective at detecting some of these patterns, which suggests that more sophisticated approaches may achieve even greater success.
}

\vspace{0.5mm}
\noindent
\revc{
	\textbf{Increasing explainability of smart contract source code.}
		\textit{Enhancing the explainability of source code may facilitate end user understanding.}
		Smart contract programming languages are generally Turing-complete, and this expressiveness vastly hinders the ability to derive explanations from them.
		However, the source code is the root of the smart contract and therefore any improvements in explainability here is likely to make explainability easier 
		downstream (e.g., in the UI).
		To this end, a pattern-based approach for identifying common patterns and extracting explanations from them might be useful. 
		Such an approach has indeed been used in the context of security to identify administrative patterns~\cite{AdministratedTokens} and 
		to aid in loop summarization~\cite{loops2021}.
		An alternative approach 
		is to modify the smart contract programming language. 
		While smart contract programming languages have emerged both in industry and academia~\cite{contract_language_survey}, 
		few of them focus on enhancing the explainability of the source code to end-users.
}

%% file: validity.tex
\section{Limitations and Threats to Validity}
\label{sec:validity}


\vspace{0.5mm}
\noindent
\revc{
	We note three limitations regarding our user study.
	First, while our user study is done with a large number (110) of respondents with varying demographics (e.g., industries and countries), it was focused on the USD Tether smart contract and recruited users solely from the Prolific platform. 
	Additionally, users from some countries may not be able to access Google Forms, which we used for our survey.
	As discussed in \autoref{sec:discussion:future-work}, more work is needed to generalize our findings to other users and smart contracts.
}
\revc{Second, both
our transfer risk evaluation metrics (surprisingness,  awareness and severity) and UI evaluation metrics (discoverability and understandability) are measured through self-assessment rather than an investigation of actual user behavior.  
}
Thus,  our findings may differ from 
behavior-oriented studies (e.g. ,  observational study).
Mitigating these threats, we employed attention-checking, validation and knowledge checking questions. 
We also conducted pilot studies 
to  
revise confusing questions, add 
follow-up questions, and identify users' misunderstanding or contradictory responses. 
\revc{Third, w}e note some internal inconsistencies in participant response:
e.g., five participants claimed to simultaneously own no stablecoins and yet own USDT, despite being informed prior that USDT is a stablecoin. We conjecture that this is caused by participants failing to make the connection between their experience and new knowledge provided in the survey.

\revc{Regarding o}ur source code analysis (\textbf{RQ\revc{4}}), 
results may not generalize to other periods (e.g., before 2022), or non-ERC-20 smart contracts. 
\revise{However,  we note that during the period of our search, ERC-20 implementations were the most used standard (25.6\% of transaction volume) 
	for the top recipient addresses with published source code (368 addresses out of the top 500)}.
To further mitigate this threat, we have provided our experimental data.

%% file: conclusion_and_ft.tex
\section{Conclusion}
\label{sec:conclusion}

This paper investigates end-users' 
comprehension of smart contract \rev{transfer} risks
\reva{using the most popular smart contract (USDT),  a widely used smart contract interface (MetaMask) and 78 frequently used ERC-20 contracts.}
We observed that 
respondents are unaware 
of transfer risks,  irrespective of 
their \rev{self-rated} programming ability or Web3 proficiency. 
Users also consider the current USDT/MetaMask UI flow to be 
insufficient in communicating transaction outcomes. 
We further analyzed the 78 \recheck{\rev{next}} most frequently used ERC-20 contracts, \rev{after USDT}, to 
show that the transfer risks considered in our study are prevalent beyond USDT. 
This analysis also discovered additional \rev{transfer} risks beyond the transfer risks considered 
in our study. Our research points to the need for \textit{explainable smart contracts}.
\rev{Additionally, we hope that this work will motivate 
policy-makers to make informed decisions regarding \textit{\revc{end-user understanding}} of smart contracts.}
This work demonstrates that significant research is required in user interface design to explain the risky transaction 
outcomes to smart contract users. We hope this work provides a foundation for 
further research in improving end-user comprehension of smart contract \rev{transfer} risks. 
For reproduction and further research, 
our research data and code are available in the following: 

\begin{center}
	\rev{\url{https://zenodo.org/communities/tether-study}}
\end{center}

%% file: appendix.tex
%

\newpage
\appendix
\revc{\section{Appendix}}
\label{sec:appendix:distributions}

\begin{figure}[h!]
    \begin{newbox1}
    \centering
    \begin{subfigure}{0.48\textwidth}
      \includegraphics[width=\textwidth]{img/violin-plots/unaware-programming-2-real.png}
      \caption{Unawareness for real risks}
    \end{subfigure}
    \begin{subfigure}{0.48\textwidth}
      \includegraphics[width=\textwidth]{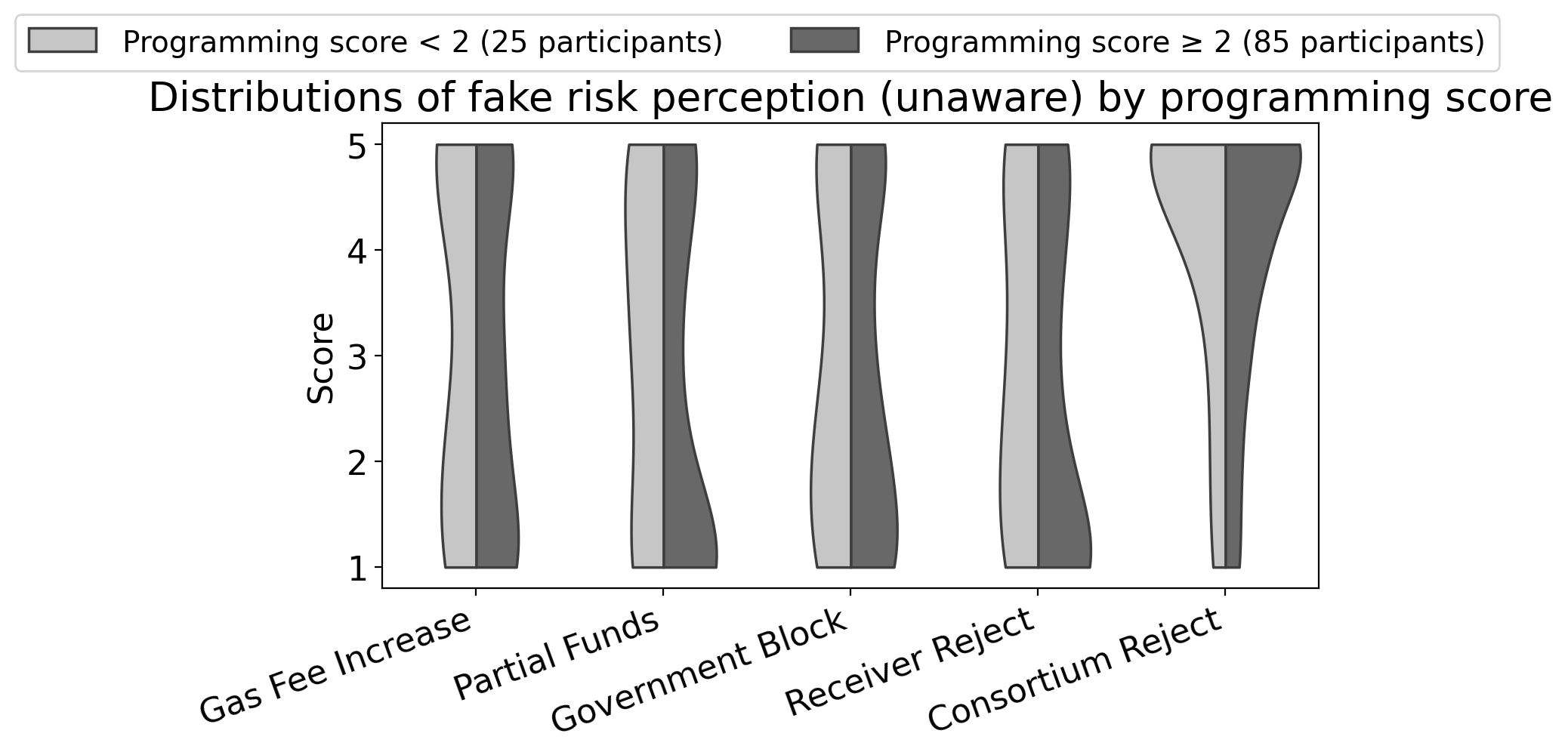}
      \caption{Unawareness for fake risks}
    \end{subfigure}
    \begin{subfigure}{0.48\textwidth}
      \includegraphics[width=\textwidth]{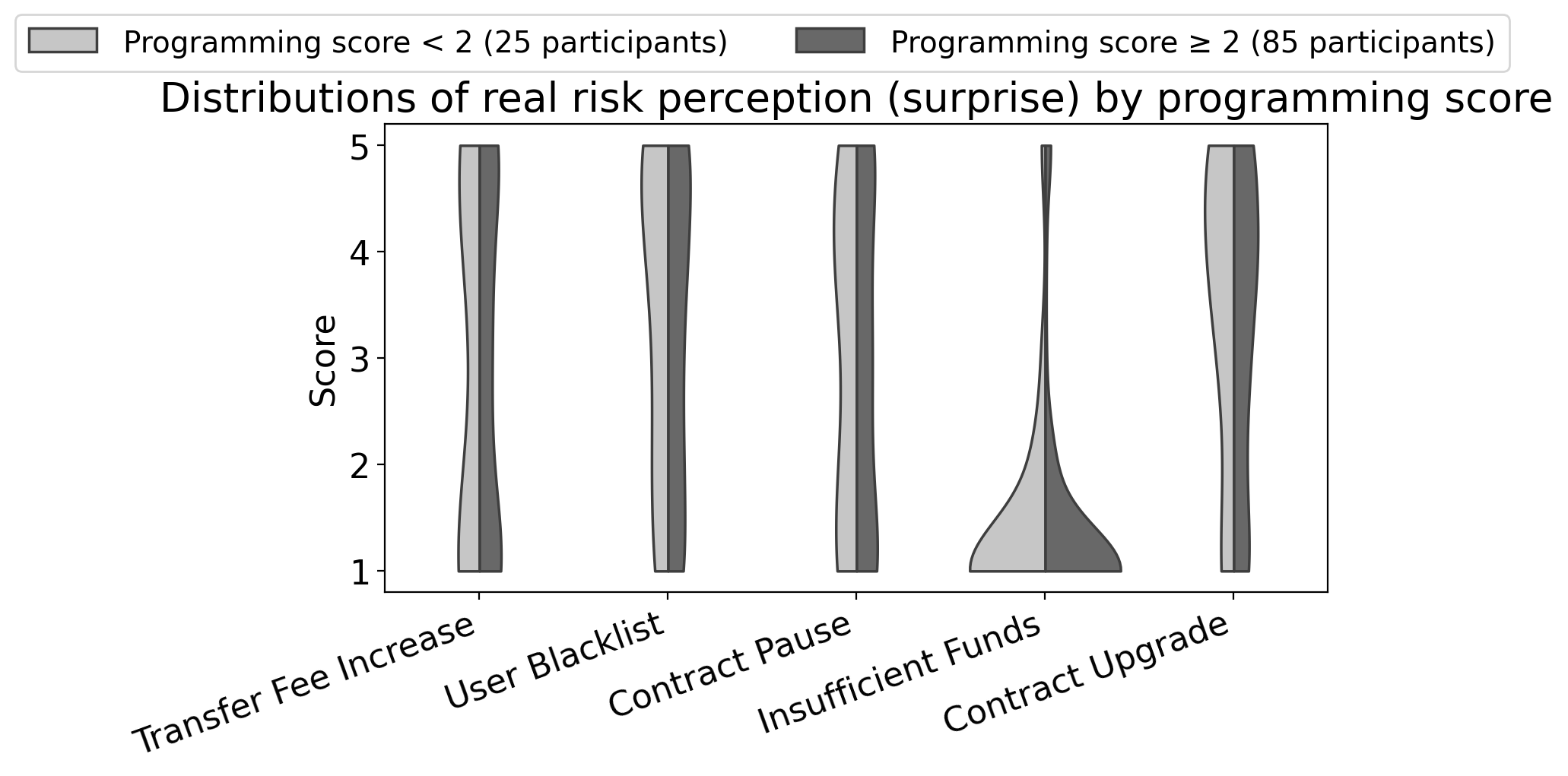}
      \caption{Surprisingness for real risks}
    \end{subfigure}
    \begin{subfigure}{0.48\textwidth}
      \includegraphics[width=\textwidth]{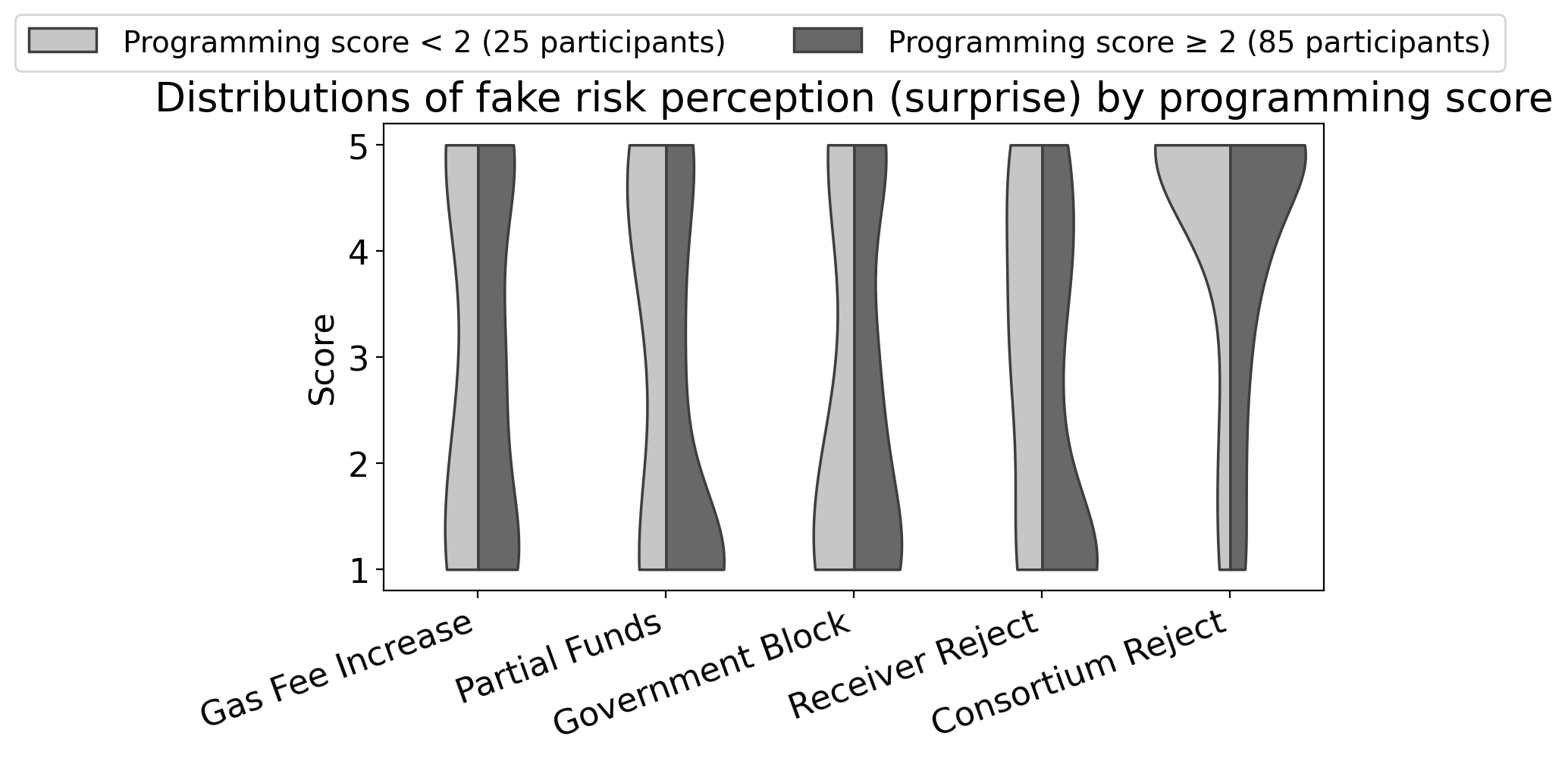}
      \caption{Surprisingness for fake risks}
    \end{subfigure}
    \begin{subfigure}{0.48\textwidth}
      \includegraphics[width=\textwidth]{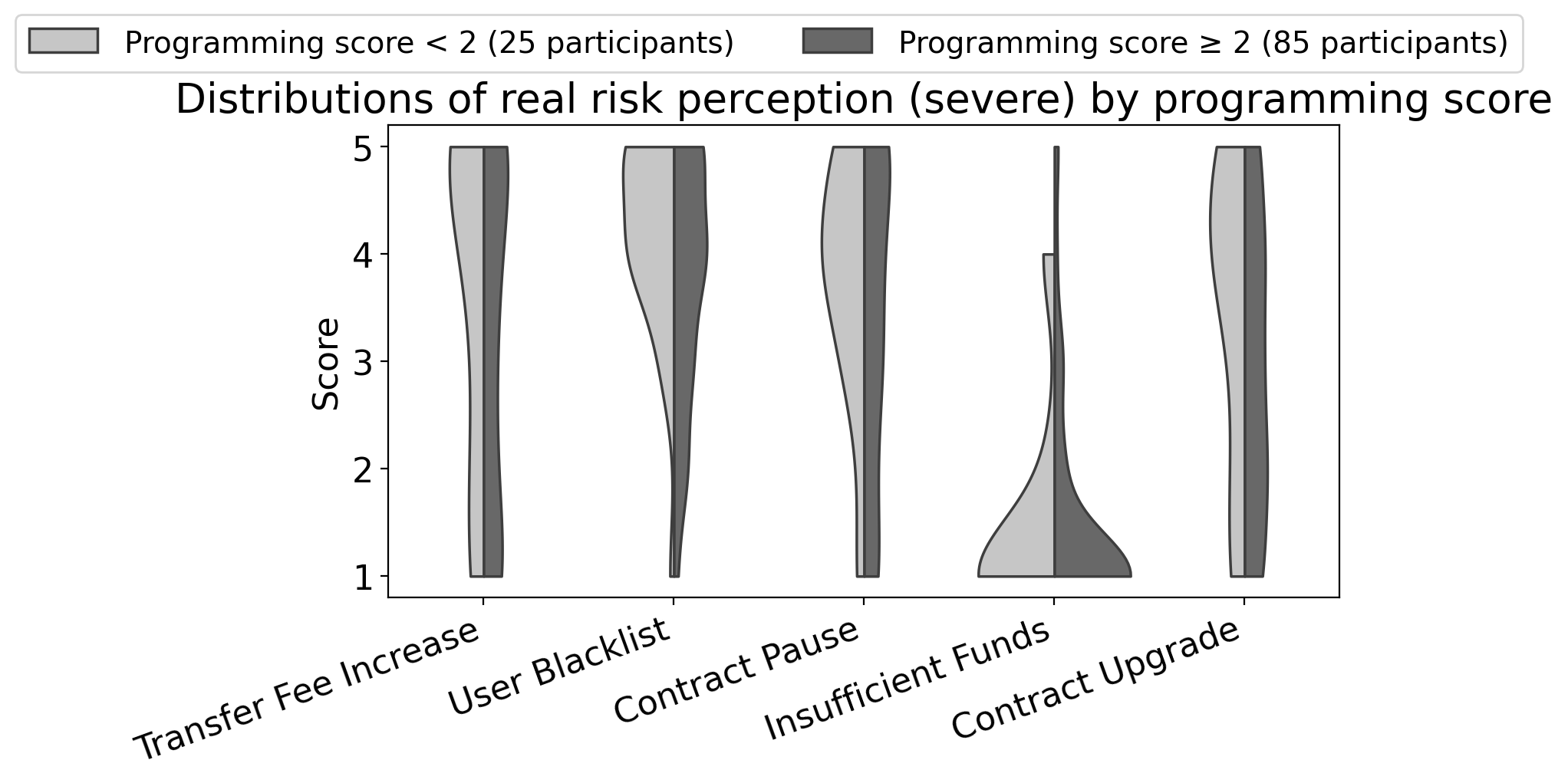}
      \caption{Severity for real risks}
    \end{subfigure}
    \begin{subfigure}{0.48\textwidth}
      \includegraphics[width=\textwidth]{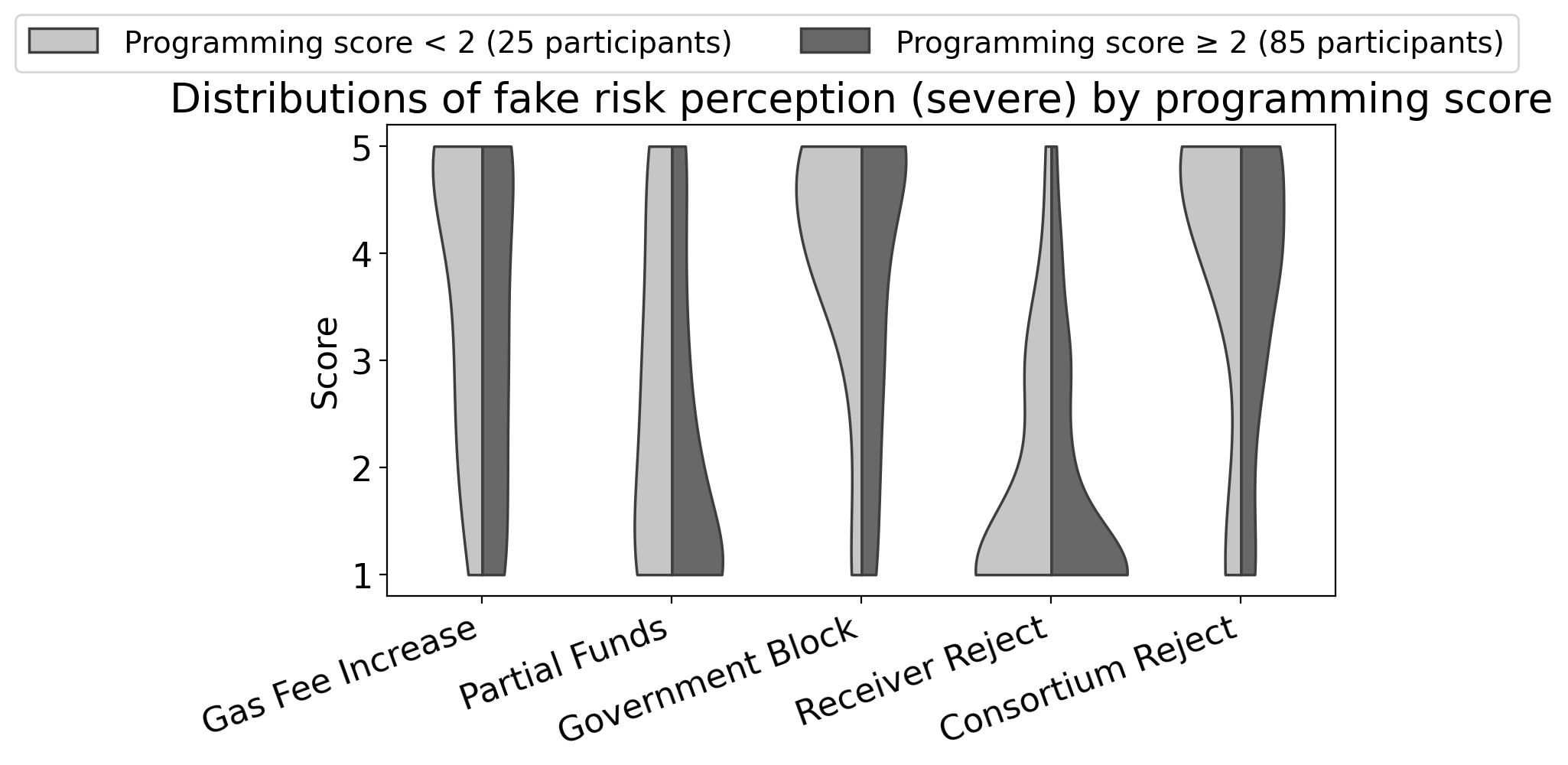}
      \caption{Severity for fake risks}
    \end{subfigure}
    \caption{\centering }Risk perception distributions split by programming proficiency, with the self-rated skill threshold set to two out of five.
    \Description{This details several plots which showcase the distribution of scores for risk perception of real and fake risks, split by self-rated skill proficiency into a proficient group and a non-proficient group. Generally speaking, the distributions look similar despite different skill proficiencies.}
      
    \end{newbox1}
    \end{figure}

    \begin{figure}[h!]
      \begin{newbox1}
    \centering
    \begin{subfigure}{0.48\textwidth}
      \includegraphics[width=\textwidth]{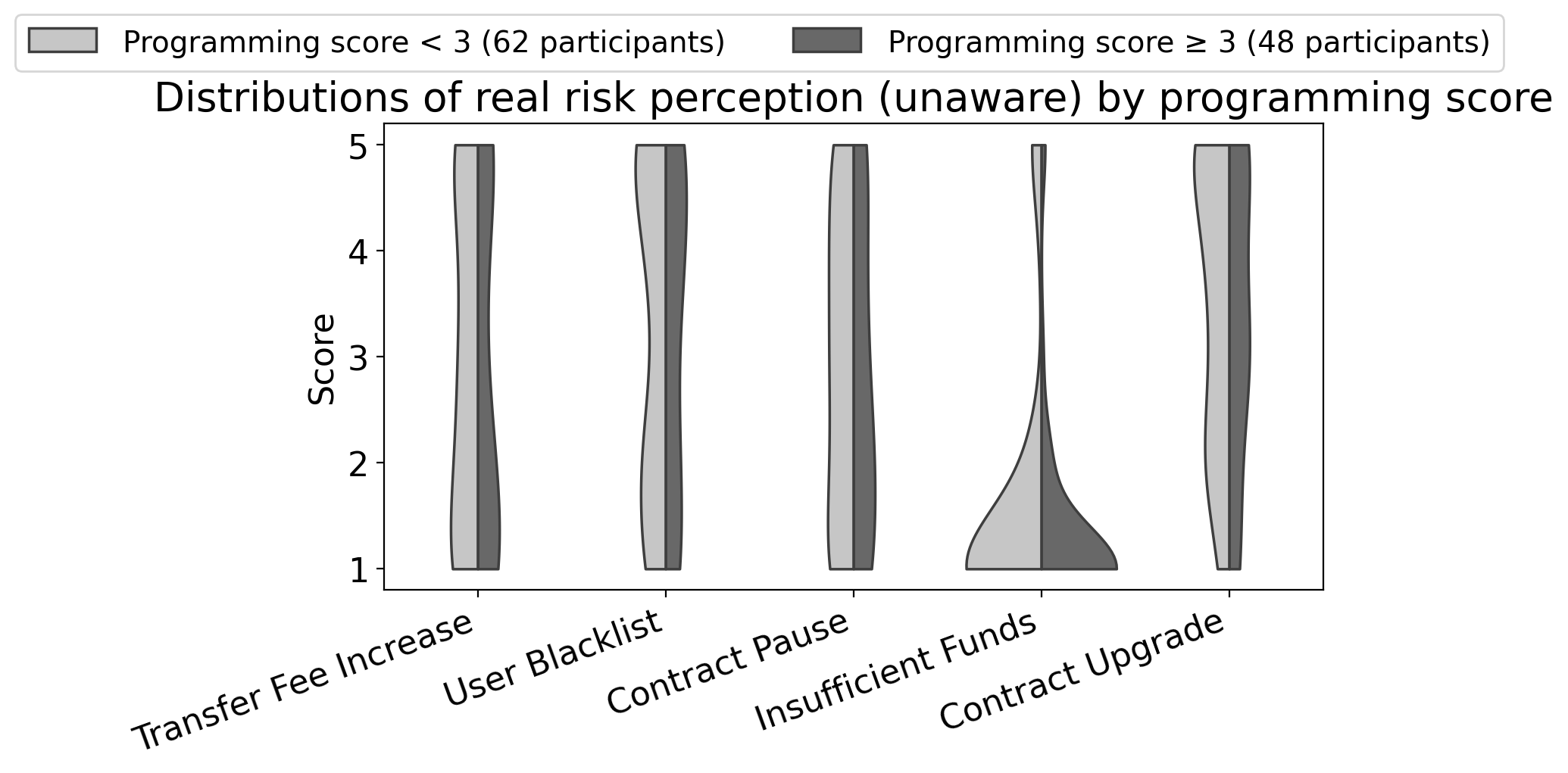}
      \caption{Unawareness for real risks}
    \end{subfigure}
    \begin{subfigure}{0.48\textwidth}
      \includegraphics[width=\textwidth]{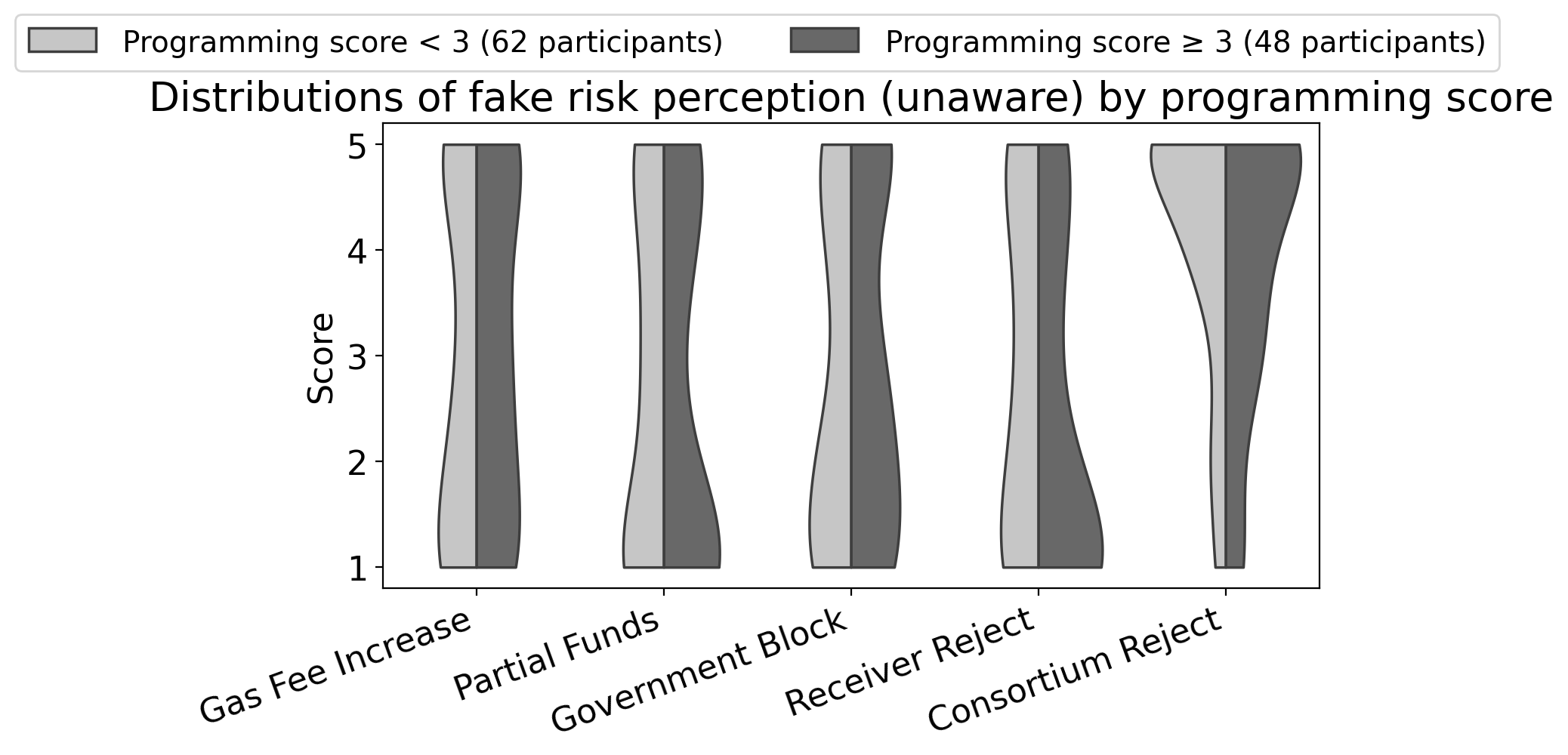}
      \caption{Unawareness for fake risks}
    \end{subfigure}
    \begin{subfigure}{0.48\textwidth}
      \includegraphics[width=\textwidth]{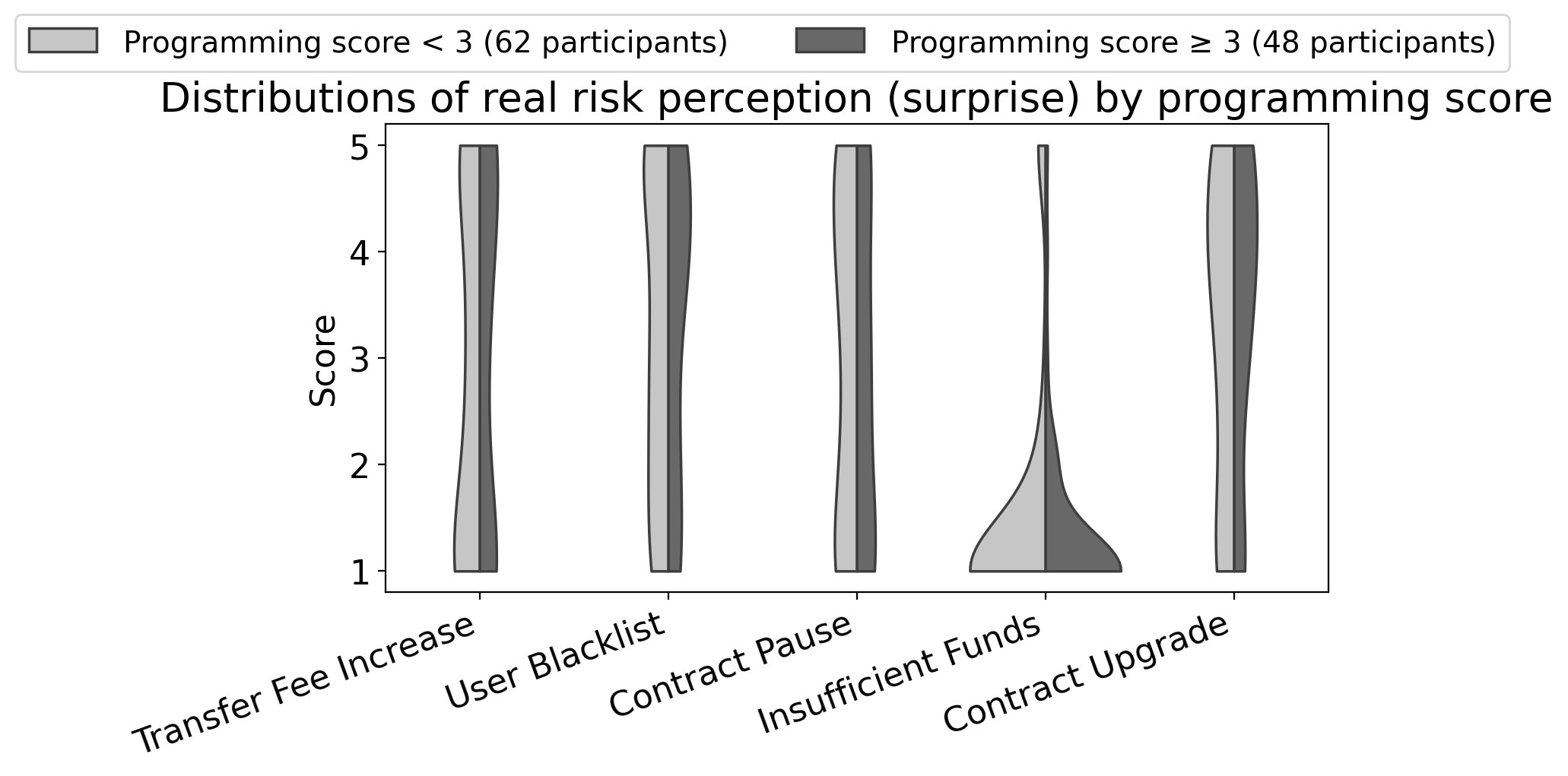}
      \caption{Surprisingness for real risks}
    \end{subfigure}
    \begin{subfigure}{0.48\textwidth}
      \includegraphics[width=\textwidth]{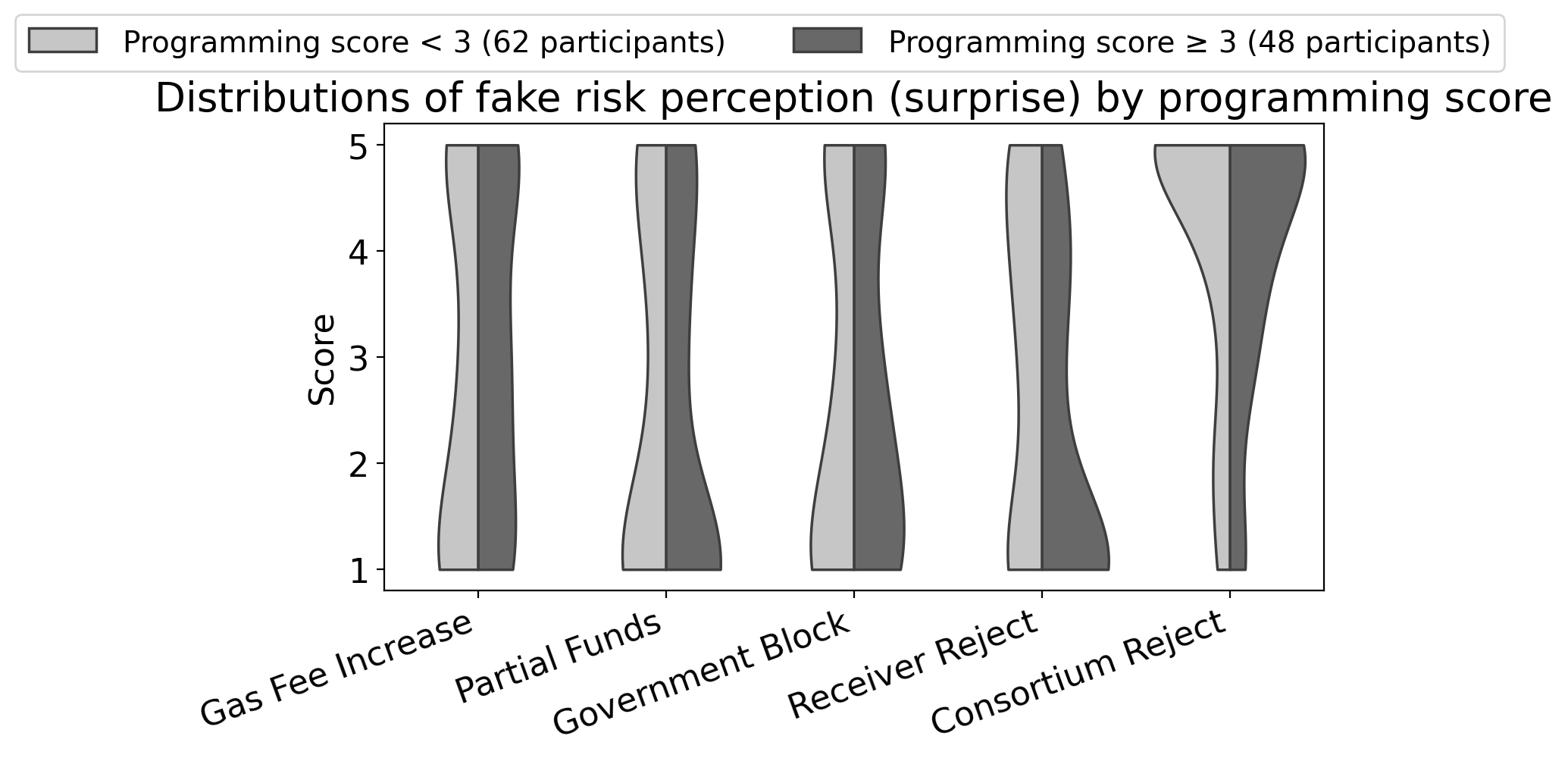}
      \caption{Surprisingness for fake risks}
    \end{subfigure}
    \begin{subfigure}{0.48\textwidth}
      \includegraphics[width=\textwidth]{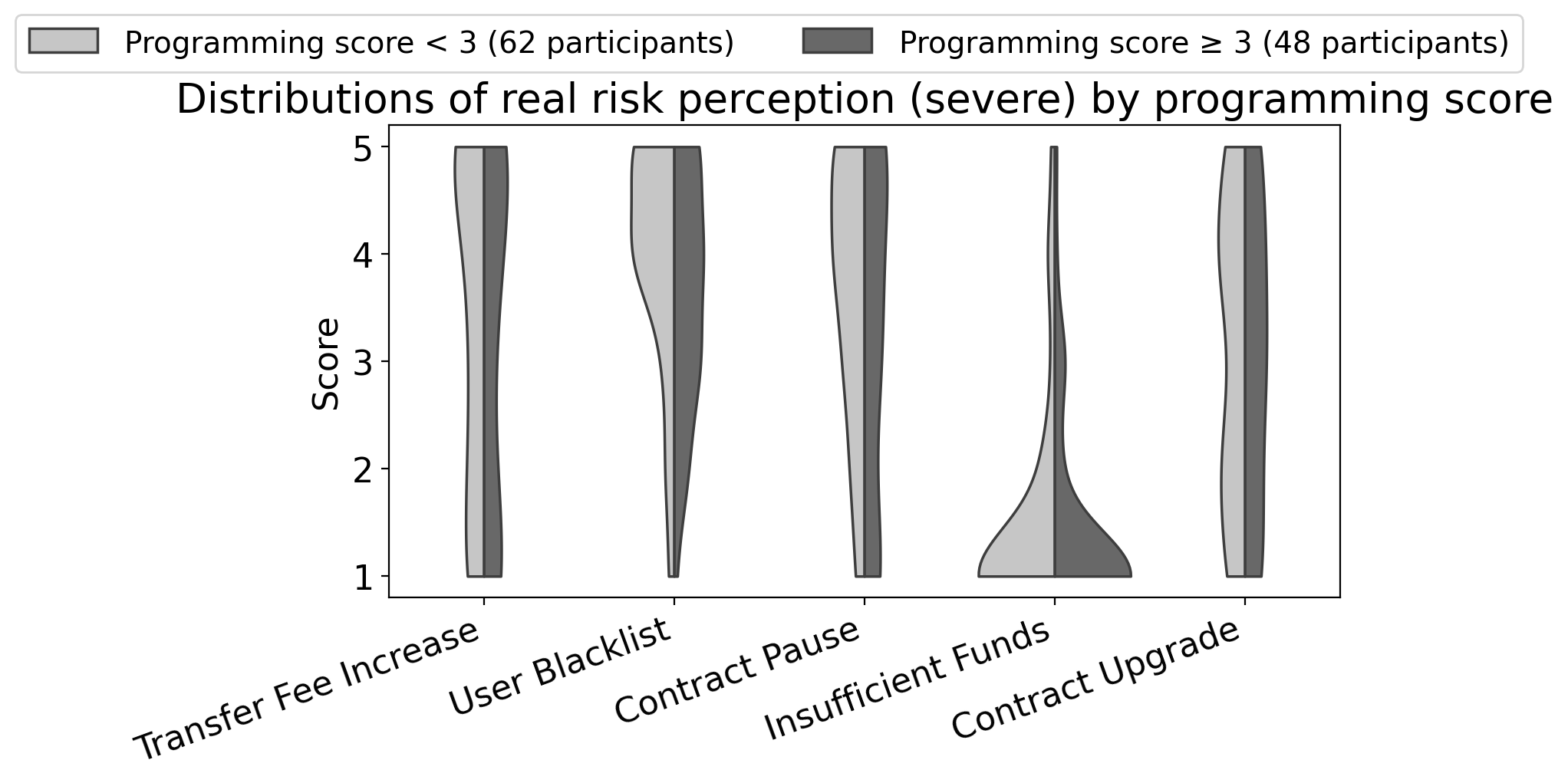}
      \caption{Severity for real risks}
    \end{subfigure}
    \begin{subfigure}{0.48\textwidth}
      \includegraphics[width=\textwidth]{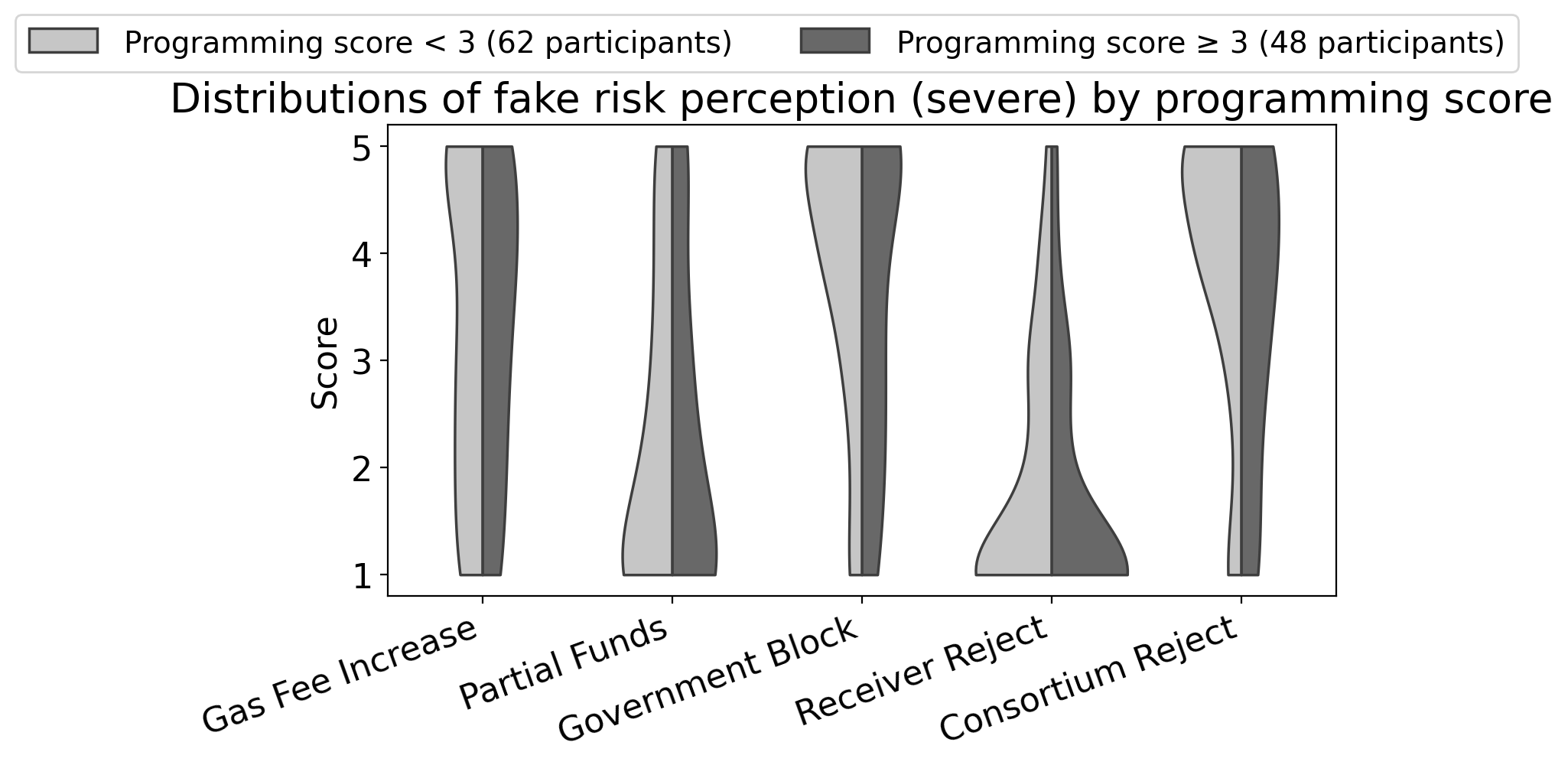}
      \caption{Severity for fake risks}
    \end{subfigure}
    \caption{\centering }Risk perception distributions split by programming proficiency, with the self-rated skill threshold set to three out of five.
    \Description{This details several plots which showcase the distribution of scores for risk perception of real and fake risks, split by self-rated skill proficiency into a proficient group and a non-proficient group. Generally speaking, the distributions look similar despite different skill proficiencies.}
    \end{newbox1}
    \end{figure}

    \begin{figure}[h!]
      \begin{newbox1}
    \centering
    \begin{subfigure}{0.48\textwidth}
      \includegraphics[width=\textwidth]{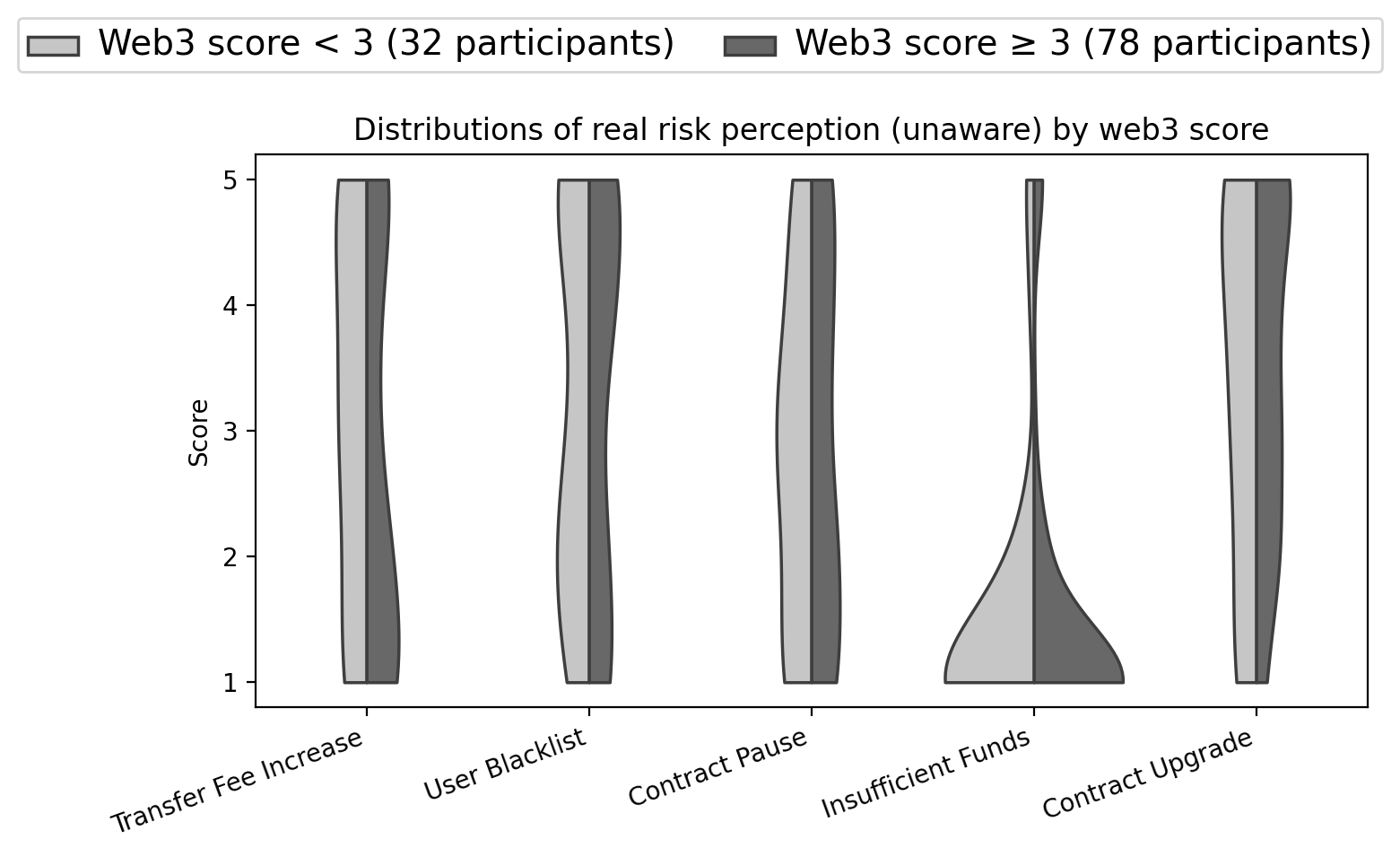}
      \caption{Unawareness for real risks}
    \end{subfigure}
    \begin{subfigure}{0.48\textwidth}
      \includegraphics[width=\textwidth]{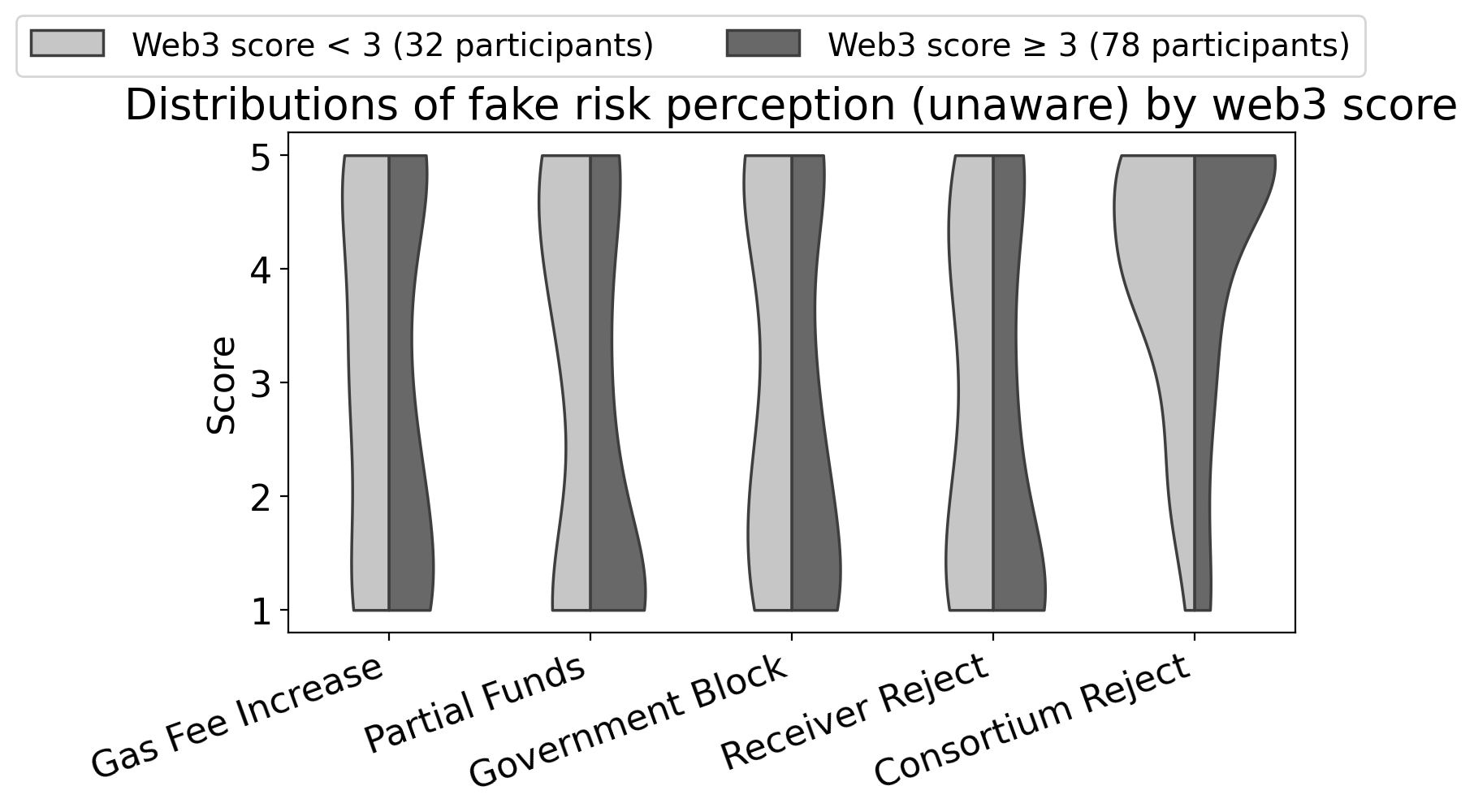}
      \caption{Unawareness for fake risks}
    \end{subfigure}
    \begin{subfigure}{0.48\textwidth}
      \includegraphics[width=\textwidth]{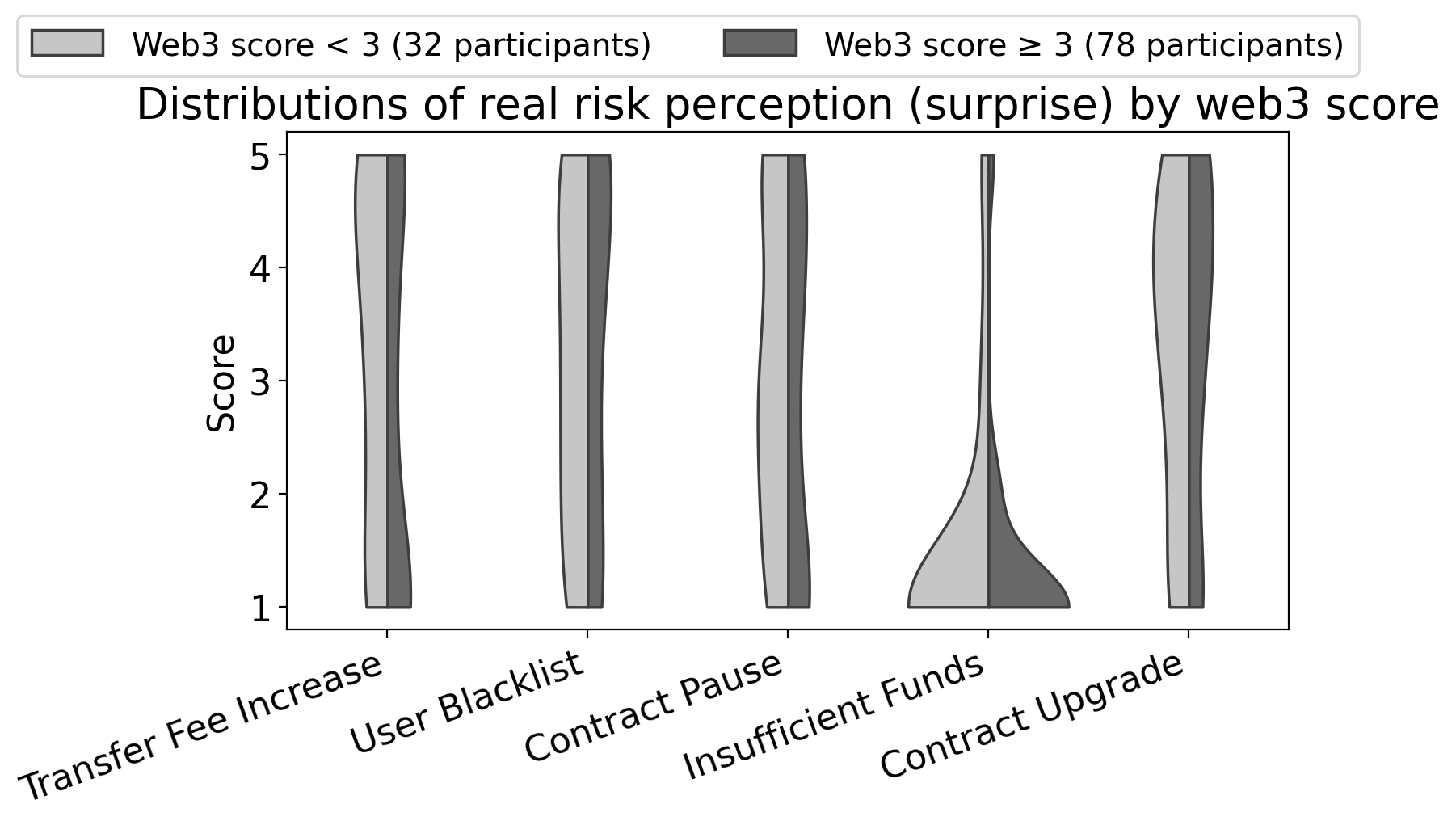}
      \caption{Surprisingness for real risks}
    \end{subfigure}
    \begin{subfigure}{0.48\textwidth}
      \includegraphics[width=\textwidth]{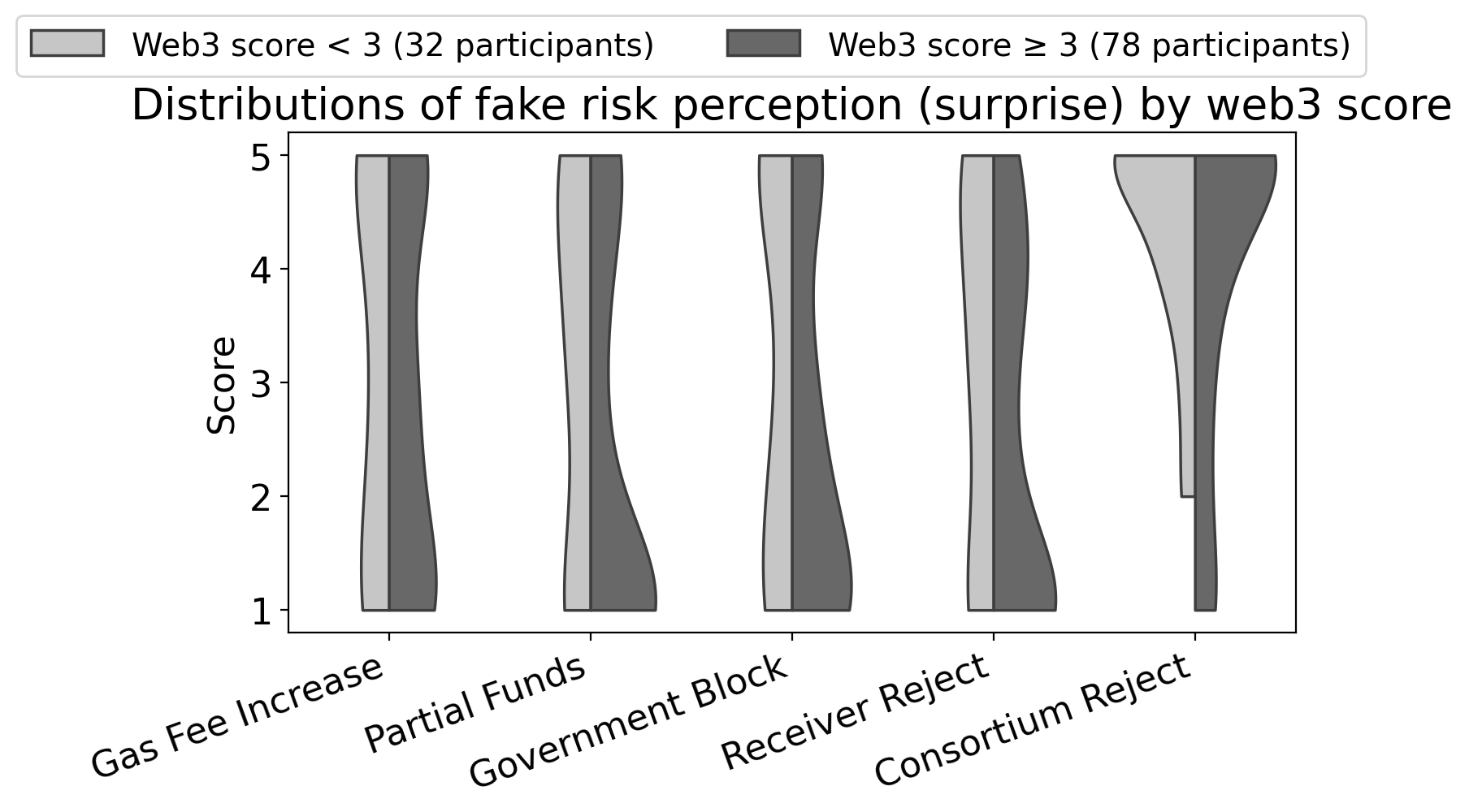}
      \caption{Surprisingness for fake risks}
    \end{subfigure}
    \begin{subfigure}{0.48\textwidth}
      \includegraphics[width=\textwidth]{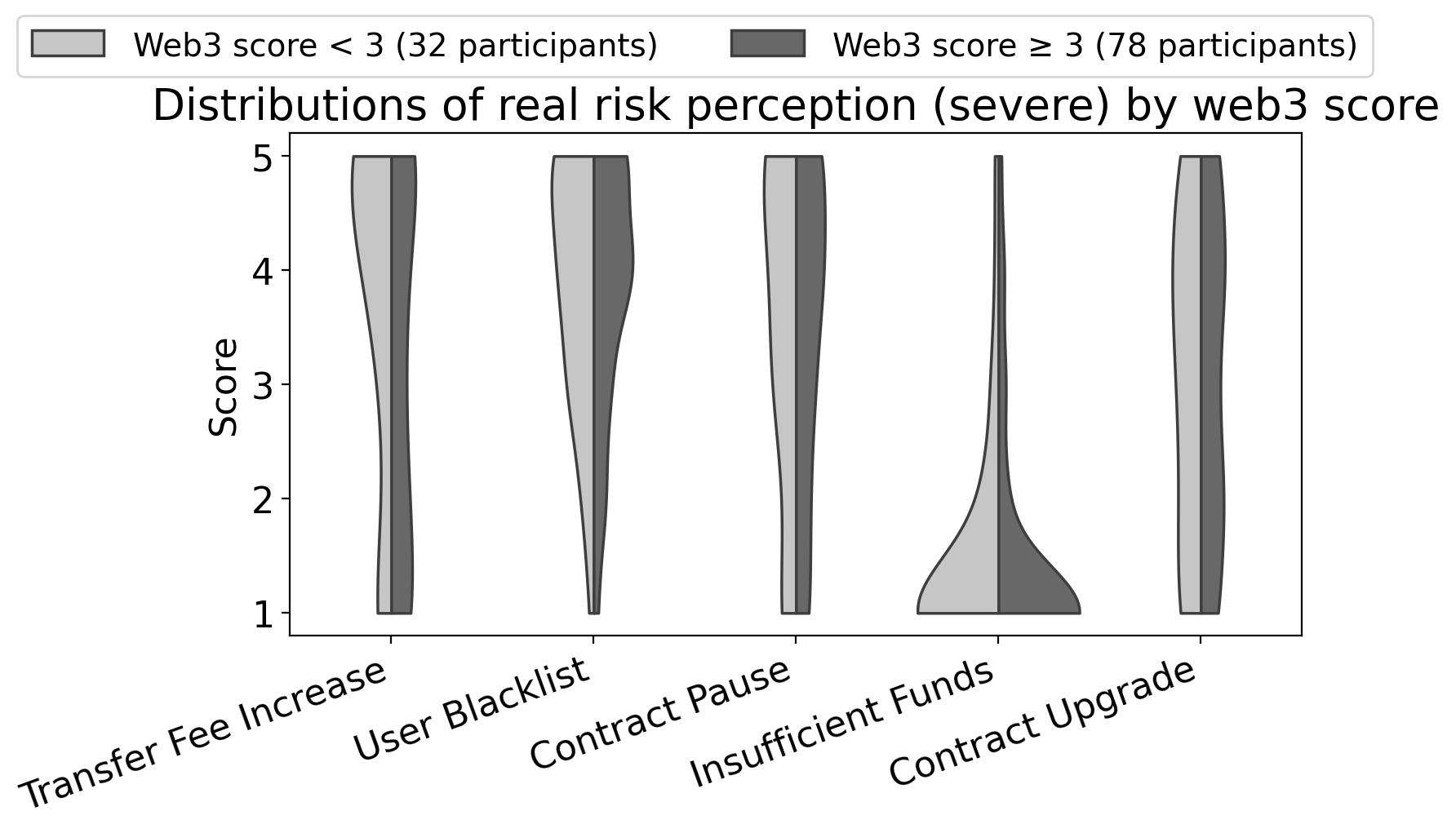}
      \caption{Severity for real risks}
    \end{subfigure}
    \begin{subfigure}{0.48\textwidth}
      \includegraphics[width=\textwidth]{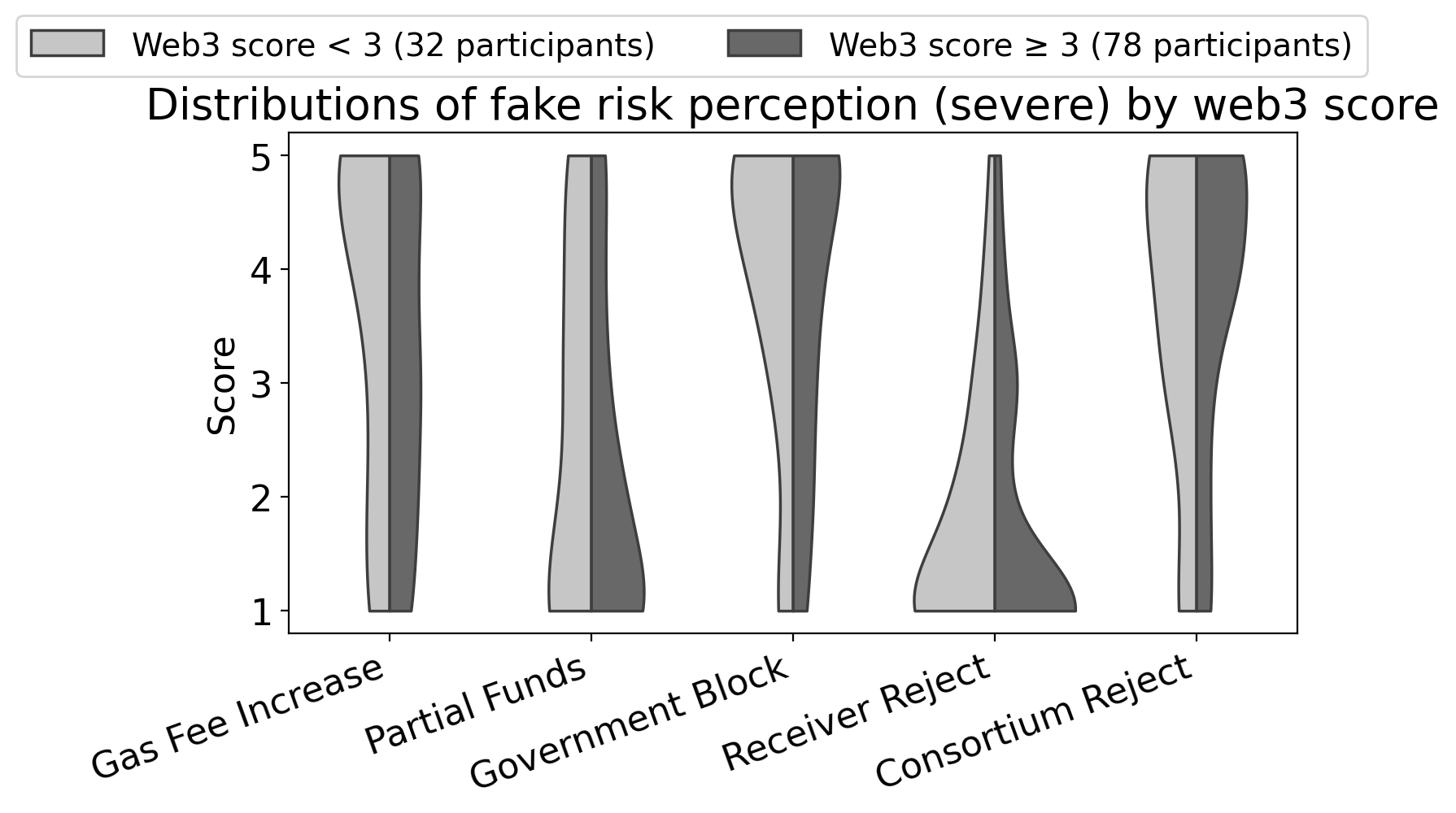}
      \caption{Severity for fake risks}
    \end{subfigure}
    \caption{\centering }Risk perception distributions split by Web3 proficiency, with the self-rated skill threshold set to three out of five.
    \Description{This details several plots which showcase the distribution of scores for risk perception of real and fake risks, split by self-rated skill proficiency into a proficient group and a non-proficient group. Generally speaking, the distributions look similar despite different skill proficiencies.}
        
      \end{newbox1}
    \end{figure}

    \begin{figure}[h!]
      \begin{newbox1}
    \centering
    \begin{subfigure}{0.48\textwidth}
      \includegraphics[width=\textwidth]{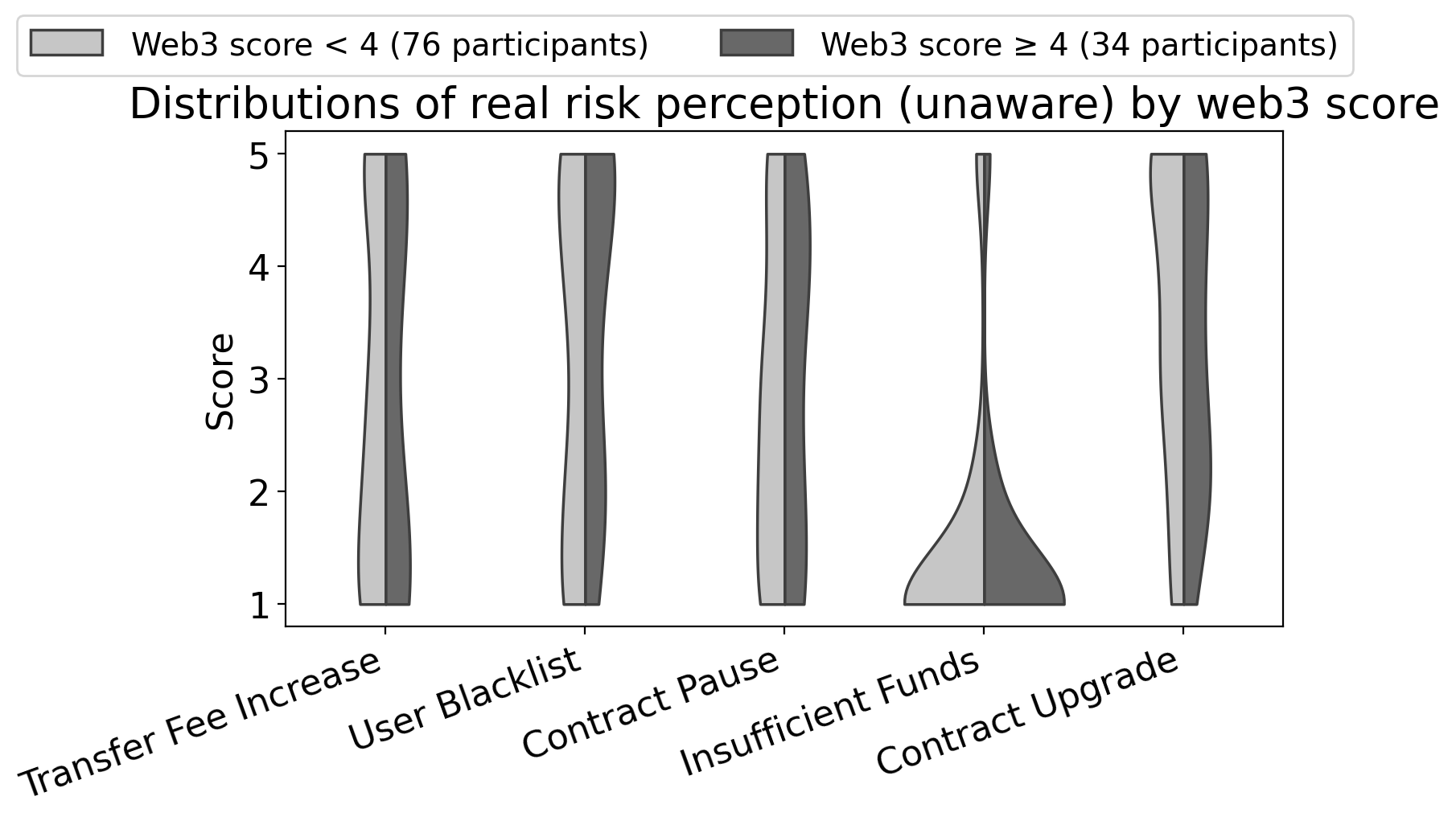}
      \caption{Unawareness for real risks}
    \end{subfigure}
    \begin{subfigure}{0.48\textwidth}
      \includegraphics[width=\textwidth]{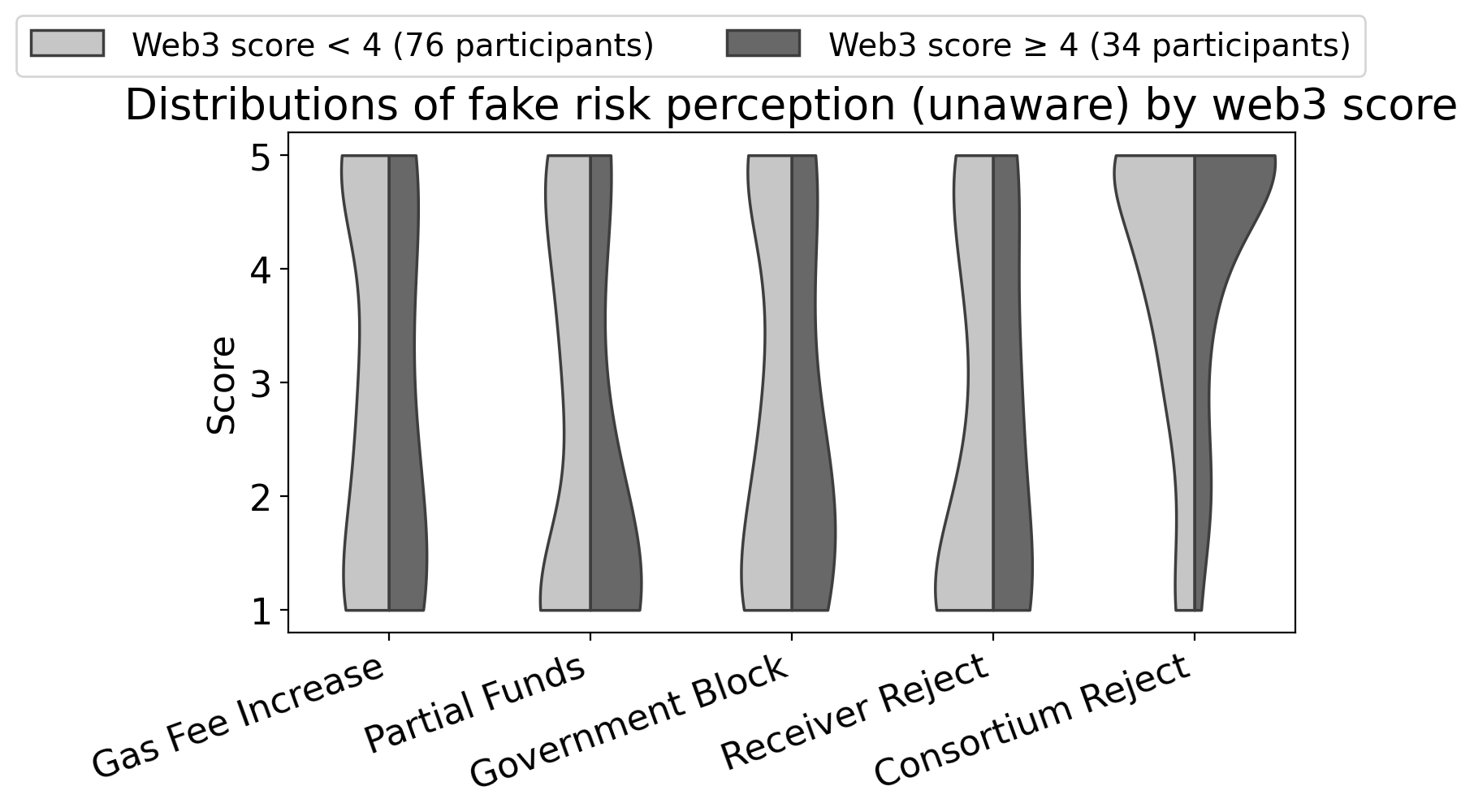}
      \caption{Unawareness for fake risks}
    \end{subfigure}
    \begin{subfigure}{0.48\textwidth}
      \includegraphics[width=\textwidth]{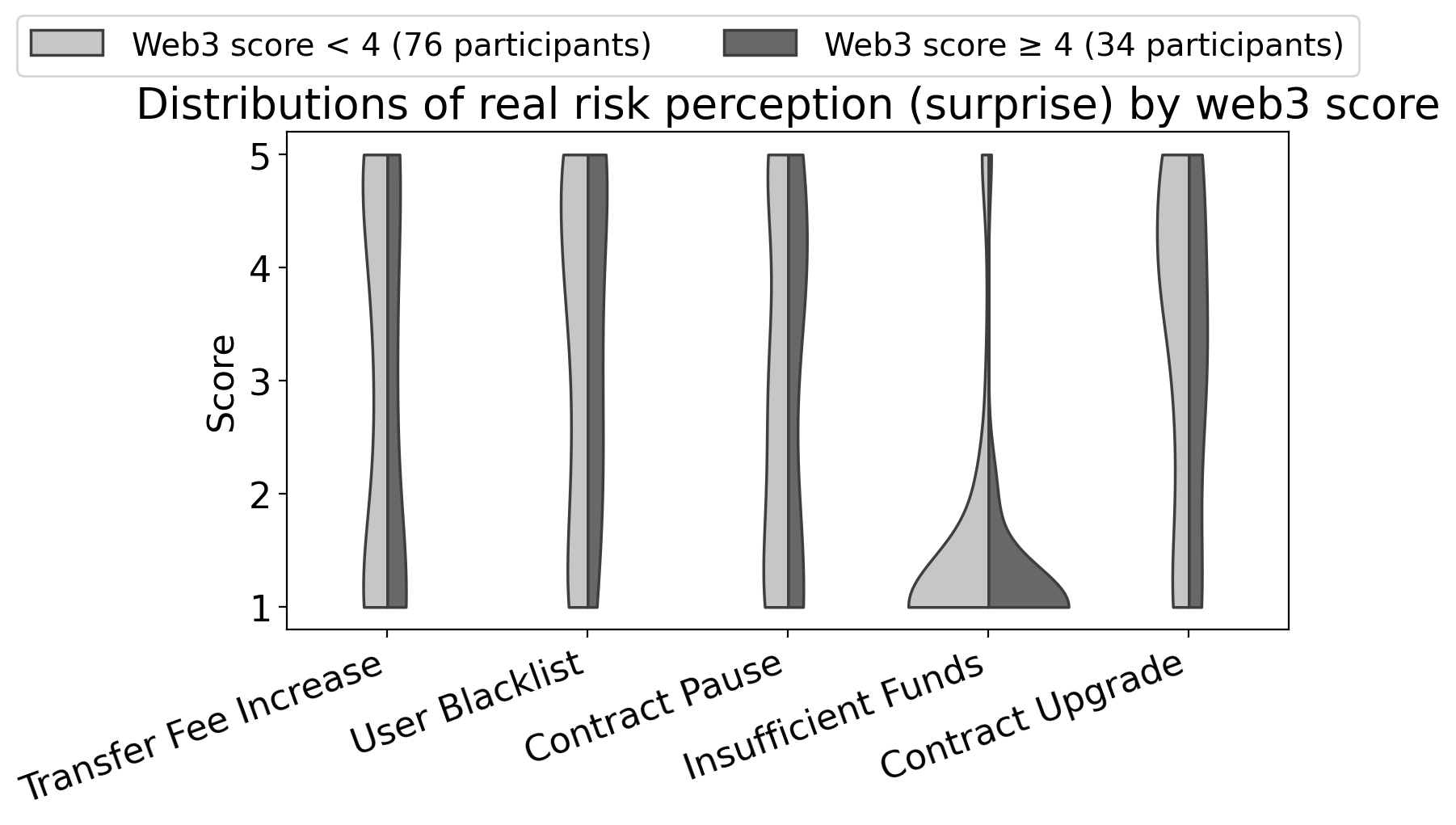}
      \caption{Surprisingness for real risks}
    \end{subfigure}
    \begin{subfigure}{0.48\textwidth}
      \includegraphics[width=\textwidth]{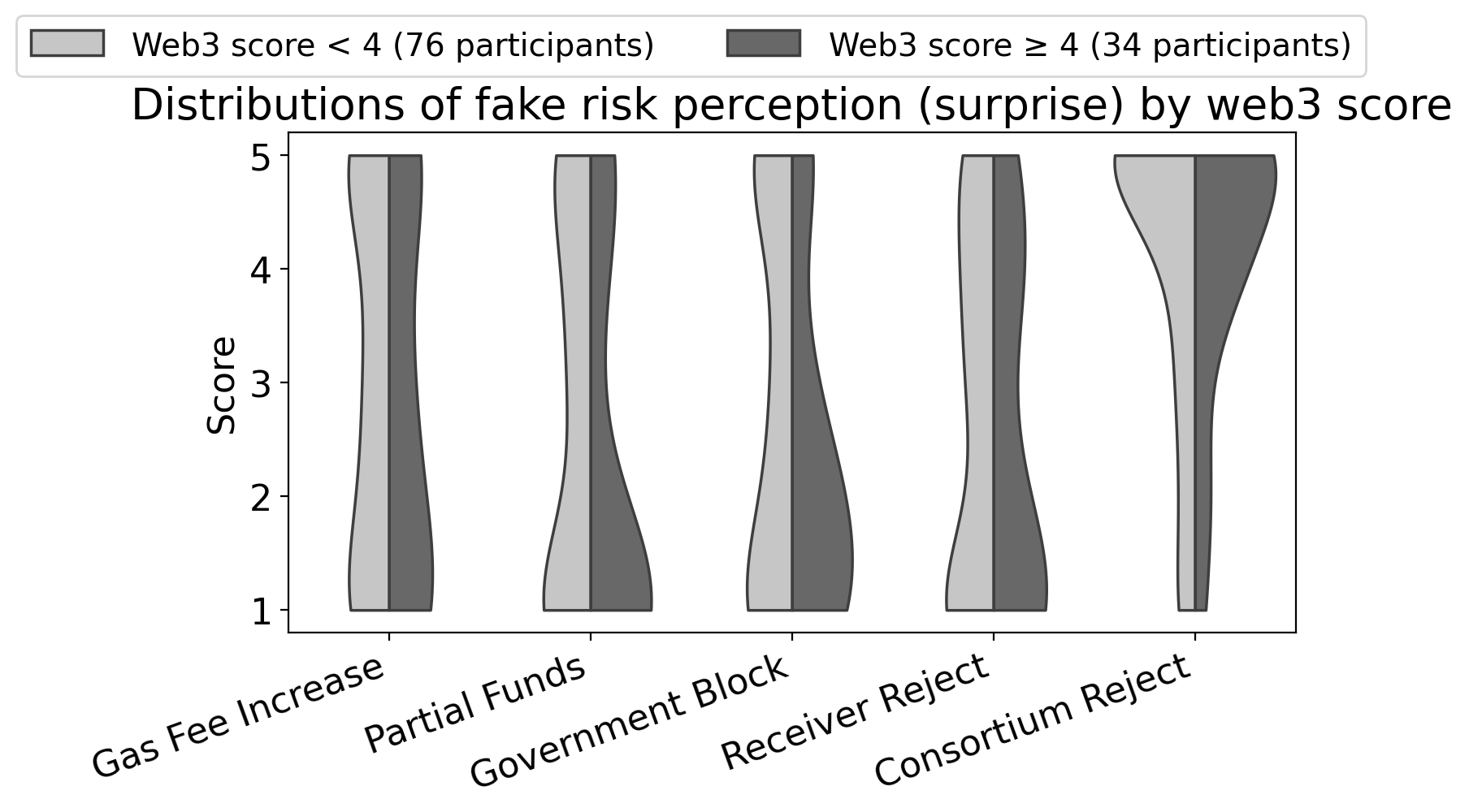}
      \caption{Surprisingness for fake risks}
    \end{subfigure}
    \begin{subfigure}{0.48\textwidth}
      \includegraphics[width=\textwidth]{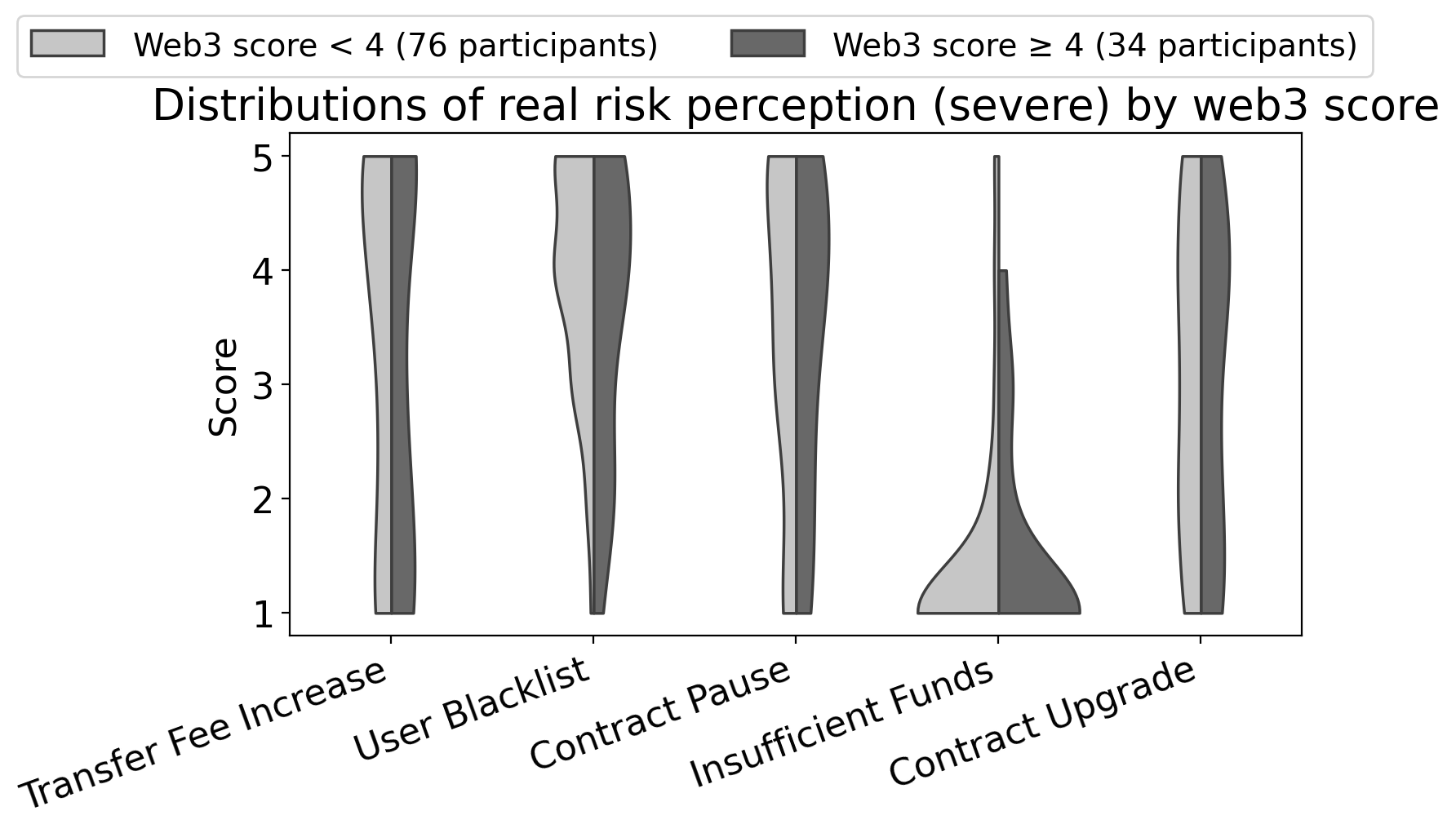}
      \caption{Severity for real risks}
    \end{subfigure}
    \begin{subfigure}{0.48\textwidth}
      \includegraphics[width=\textwidth]{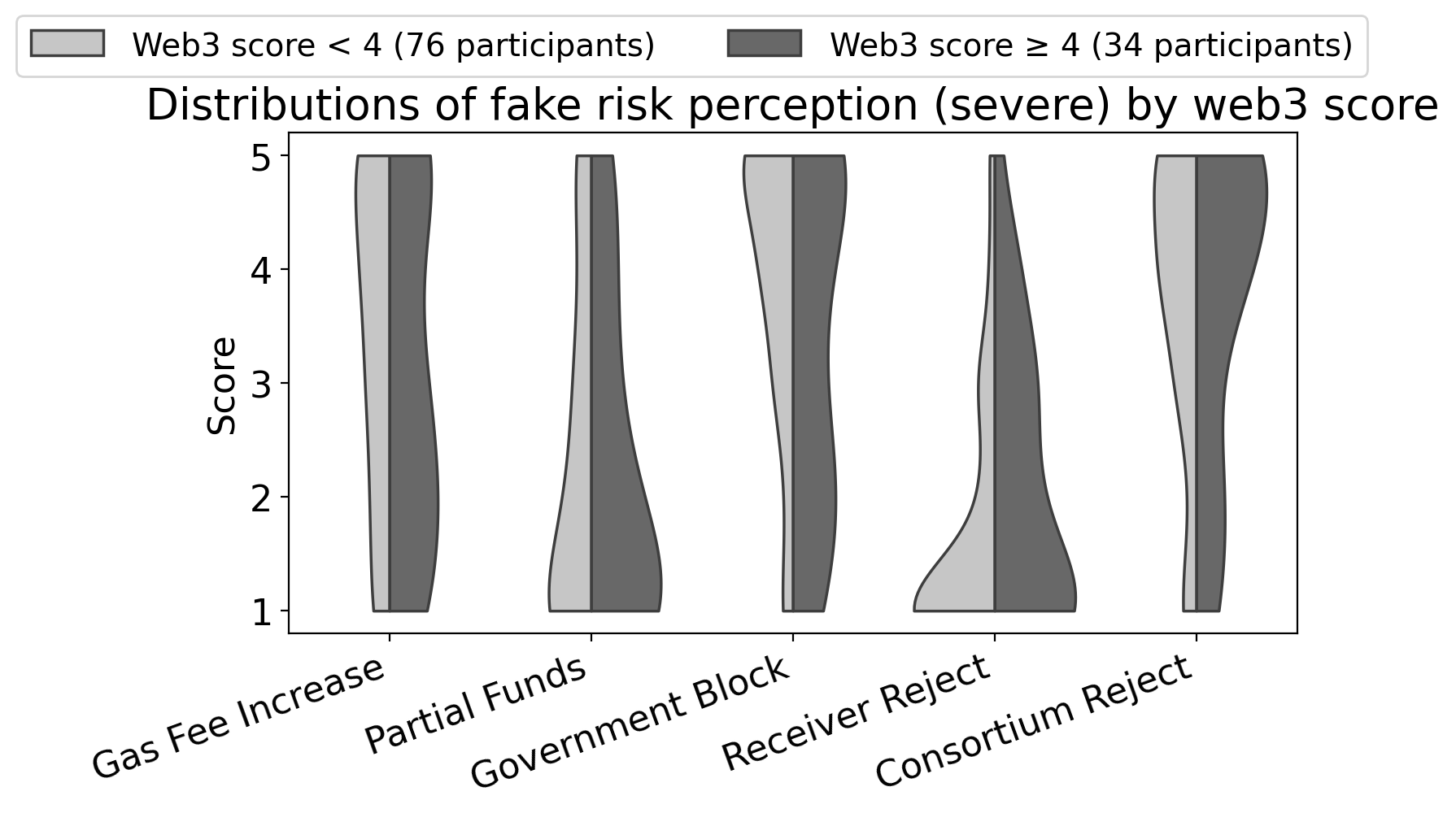}
      \caption{Severity for fake risks}
    \end{subfigure}
    \caption{\centering }Risk perception distributions split by Web3 proficiency, with the self-rated skill threshold set to four out of five.
    \Description{This details several plots which showcase the distribution of scores for risk perception of real and fake risks, split by self-rated skill proficiency into a proficient group and a non-proficient group. Generally speaking, the distributions look similar despite different skill proficiencies.}
      \end{newbox1}
    \end{figure}